\documentclass[12pt]{article}
\usepackage{latexsym}
\usepackage{babel}
\usepackage{amstex}
\begin{document}
\newfont{\fraktgm}{eurm10 scaled 1728}
\newcommand{\graktur}{\baselineskip12.5pt \fraktgm}
\newfont{\fraktfm}{eurm10 scaled 1440}
\newcommand{\frakture}{\baselineskip12.5pt\fraktfm}
\newfont{\fraktrm}{eurm10}
\newcommand{\fraktur}{\baselineskip12.5pt\fraktrm}
\newfont{\fraktem}{eurm6}
\newcommand{\fraktr}{\baselineskip12.5pt\fraktem}
\protect\newtheorem{principle}{Principle}
\newcommand{\dist} {{\rm dist}}
\newcommand{\fini} [1] {{\cal H}_{#1}(\HH)}
\newcommand{\fin} {\fini{fin}} 
\newcommand{\finki} [1] {{\cal K}_{#1}(\HH)}
\newcommand{\fink} {\finki{fin}}
\newcommand{\seta} [1] {I\!\!\!\,#1}
\newcommand{\RR} {{\Bbb R}}
\newcommand{\RRR} {\seta{R}}
\newcommand{\Norm} [1] {\vline \hspace{0.05cm} \vline #1 \vline 
\hspace{0.05cm} \vline \hspace{0.05cm}}
\newcommand{\Betrag} [1] {\vline #1 \vline}
\newcommand{\setb} [1] {I\!\!\!\!\!\:#1}
\newcommand{\setc} [1] {#1\!\!\!#1}
\newcommand{\DD} {\seta{D}}
\newcommand{\NNN} {\seta{N}}
\newcommand{\NN} {{\Bbb N}}
\newcommand{\HH} {{\Bbb H}}
\newcommand{\HHH} {\seta{H}}
\newcommand{\ZZ} {{\Bbb Z}}
\newcommand{\XX} {{\zeta}}
\newcommand{\CC} {{\Bbb C}}
\newcommand{\QQ} {{\Bbb Q}}
\newcommand{\MM} {\seta{M}}
\newcommand{\LL} {\seta{L}}
\newcommand{\FF} {\seta{F}}
\newtheorem{theo}{Theorem}
\newtheorem{nt}{Note}
\newtheorem{lem}{Lemma}
\newtheorem{co}{Corollary}
\newtheorem{de}{Definition}
\newtheorem{fo}{Consequence}
\newtheorem{rle}{Rule}
\newtheorem{rem}{Remark}
\newcommand{\tr} {\mbox{{\rm tr}}}
\newcommand{\norm} [1] {\parallel \hspace{-0.15cm} #1
\hspace{-0.15cm} \parallel}
\newcommand{\betrag} [1] {\mid \hspace{-0.13cm} #1
\hspace{-0.15cm} \mid}
\newcommand{\rsta} [1] {\mid \hspace{-0.1cm} #1 \hspace{-0.04cm}
\rangle}
\newcommand{\lsta} [1] {\langle \hspace{-0.04cm} #1
\hspace{-0.12cm} \mid} 
\newcommand{\pro} {{\cal P}(\HH)} 
\newcommand{\topp} {{\cal T}(\HH)^+_1}
\newcommand{\bop} {{\cal B}(\HH)^+}
\newcommand{\bou} {{\cal B}(\HH)}
\newcommand{\sef} {{\frak E}(\HH)}
\newcommand{\ten} [1] {\otimes_{t \in #1} \HH}
\newcommand{\bout} [1] {{\cal B}^{\otimes}_{#1}(\HH)}
\newcommand{\pout} [1] {{\cal P}^{\otimes}_{#1}(\HH)}
\newcommand{\U} {{\cal U}}
\newcommand{\SSS} {{\cal S}}
\newcommand{\UP} {{\frak U}}
\newcommand{\df} {decoherence functional }
\newcommand{\dfs} {decoherence functionals }
\newcommand{\Rea} {\mbox{\em Re }}
\begin{titlepage}
\centerline{\normalsize DESY 95 - 238 \hfill ISSN 0418 - 9833} 
\centerline{\normalsize December 1995 \hfill} 
\vskip.6in 
\begin{center} 
{\graktur Consistent Histories and Operational Quantum 
Theory} 
\vskip.6in 
{{\Large \frakture Oliver Rudolph} $^*$} 
\vskip.3in 
{\normalsize \sf II.~Institut f\"ur Theoretische Physik, 
Universit\"at Hamburg} 
\vskip.05in 
{\normalsize \sf Luruper Chaussee 149} 
\vskip.05in 
{\normalsize \sf D-22761 Hamburg, Germany}
\vskip.7in
\end{center}
\normalsize
\vfill
\begin{center}
{ABSTRACT}
\end{center}
\smallskip
\noindent
In this work a generalization of the consistent histories 
approach to quantum mechanics is presented. 
We first critically review the consistent 
histories approach to nonrelativistic quantum mechanics in a 
mathematically rigorous way and 
give some general comments about it. We 
investigate to what extent the consistent histories scheme is 
compatible with the results of the operational formulation of 
quantum mechanics. According to the operational approach 
nonrelativistic quantum mechanics is most 
generally formulated in terms of effects, states and operations. 
We formulate a generalized consistent 
histories theory using the concepts and the terminology which have 
proven useful in the operational formulation of quantum mechanics. 
The logical rule of the logical interpretation of quantum 
mechanics is generalized to the present context. The algebraic 
structure of the generalized theory is studied in detail. \\ \\
\bigskip \noindent
\centerline{\vrule height0.25pt depth0.25pt width4cm \hfill}
\noindent
{\footnotesize $^*$ Internet: rudolph@@x4u2.desy.de} 
\end{titlepage}
\newpage
\section{Introduction}
\indent This article is the outcome of an attempt to understand 
the relation of two seemingly completely different modern 
formulations of quantum mechanics. On the one hand we consider 
the insightful modern operational reformulation and 
generalization of quantum mechanics due to many authors. 
We refer the reader to the monographs by Davies (1976), Kraus 
(1983), Ludwig (1970, 1972-79, 1983), Busch, Lahti and 
Mittelstaedt (1991), Busch, Grabowski and Lahti (1995) and 
references therein. On the other hand, we study the consistent 
histories approach and particularly Omn\`es' logical 
interpretation of quantum mechanics. \\ The operational approach 
is based on an analysis of the quantum mechanical measuring 
process and is formulated in terms of accompanying purely 
operational concepts: 
according to the operational approach nonrelativistic 
quantum mechanics is most generally formulated in 
terms of effects, states and operations and the accompanying 
notions like `positive operator valued measures.' The operational 
approach incorporates many ideas from the older orthodox 
interpretations (cf.~(Jammer, 1974; Primas, 1983; Scheibe, 1973)), 
in particular 
the measuring process is considered to be of central importance in 
the foundations of quantum mechanics. \\ 
In contrast, the history approach is the outcome of an effort 
to deal with quantum mechanics of single closed systems, i.e., 
systems which do neither interact with their environment nor are 
exposed to measurements. In this approach the measuring process is 
not considered to be a fundamental notion but rather a derived 
one. \\ In this work we adopt a realistic, individual 
interpretation of quantum mechanics. Further, 
we will adopt a somewhat conservative point of view in that we 
will consider the operational approach as a physically fundamental 
and meaningful approach in the sense that its results and its 
range of applications must be contained in (or derivable from) 
every generalized 
formulation of quantum mechanics based on a realistic 
interpretation, in particular in the consistent histories 
formulation of quantum mechanics. Thus we will 
investigate whether the history approach to quantum mechanics can 
be formulated and generalized in terms of the notions and the 
concepts of the operational formulation of quantum mechanics. \\ 

The history approach to nonrelativistic quantum mechanic 
has received much attention in the last decade. The consistent 
histories approach to nonrelativistic 
quantum mechanics has been inaugurated in a seminal paper by 
Griffiths (1984). 
Nonrelativistic quantum mechanics in its standard formulation is 
not a theory which describes dynamical processes in time, but it 
is a theory which gives probabilities to various possibilities. 
The history approach to quantum mechanics can be looked upon as an 
attempt to remedy this situation by introducing time sequences of 
possibilities as a rough substitute for dynamical processes. 
Nowadays there are two mainstreams of research in this field. The 
first one is mainly due to Roland Omn\`es, who uses histories as a 
framework for constructing a realistic and individual 
interpretation of quantum mechanics. The resulting logical 
interpretation is based on some simple logical rules which we will 
review below. It is thus fair to say that the logical 
interpretation is a purely epistemological interpretation. 
Histories are not necessarily considered to represent features 
belonging to physical reality and are not considered to be 
ontological models for the quantum system (most generally the 
universe) at hand, but are simply considered as a 
useful tool to talk about quantum systems (more precisely, in the 
logical interpretation histories are thought to represent the {\em 
only} meaningful assertions about a quantum mechanical system in 
the language of physics). \\

On the other hand, Gell-Mann and Hartle consider the history 
approach as a convenient setting to discuss the quantum mechanics 
of closed systems, i.e., most generally the universe (Gell-Mann 
and Hartle, 1990a-c, 1993, 1994, 1995; Hartle, 1991, 1994). In 
this way they hope to circumvent some of the conceptual and 
mathematical difficulties inherent in quantum cosmology. They 
propose to use histories to describe the events that have taken 
place and that will take place in the universe. They pay special 
attention to what they call {\em quasiclassical 
domains}. [They prefer now the term {\em 
quasiclassical realm} (Gell-Mann and Hartle, 1995).] Loosely 
speaking a quasiclassical domain is an exhaustive set of 
mutually exclusive coarse-grained 
alternative histories which can be ascribed probabilities and 
whose individual histories are described largely by alternative 
values of a limited set of quasiclassical variables (at a 
succession of times) which exhibit patterns of classical 
correlation 
in time subject to quantum mechanical fluctuations. In order to 
make pre- and retrodiction one needs 
information about the initial state of the universe and about past 
events. Since the human sensory perceptions are rather limited, we 
can deal in general only 
with sets of alternative coarse-grained histories. Furthermore, it 
is only possible to ascribe probabilities to suitably 
coarse-grained histories provided some consistency conditions are 
satisfied. 
The research in this direction is faced with severe difficulties, 
as discussed by Dowker and Kent (1996) and Kent (1995). 
In this work we will not be concerned with the 
Gell-Mann$-$Hartle approach. \\

Griffiths has recently proposed an interesting modification of 
these two approaches (Griffiths, 1995). Griffiths essentially 
claims (amongst others) that inconsistent histories can be 
considered to be `objectively' true separately even 
though inconsistent histories cannot be combined, either in 
constructing descriptions or in making logical inferences about 
them, i.e., even though assertions involving inconsistent 
histories are meaningless. \\ 

Many of the central concepts of the modern operational 
formulation are foreign to the standard consistent 
histories approach and therefore the standard consistent histories 
approach is considerably and unnecessarily restricted in its range 
of applications. In the present work we formulate a generalized 
consistent histories theory using the terminology and the concepts 
which have proven useful in the operational formulation of quantum 
mechanics. The starting points of the present work are the fact 
that the observables in quantum mechanics have to be identified 
with {\em positive operator valued measures} and the reasonable 
claim that the consistent histories approach should take into 
account the full set of quantum mechanical observables. \\

This work is organized as follows. In Section 2 we rewrite the 
formalism of the standard consistent histories approach to 
quantum mechanics in full generality in a mathematically precise 
way. Moreover, this section contains a brief (and inevitably 
incomplete) introduction to the operational formulation of quantum 
mechanics based on the notion of effect and the concept of 
generalized observables. The first part of Section 2 contains a 
discussion of the 
interpretation of quantum mechanics underlying our generalized 
history approach, i.e., we discuss the meaning of concepts like 
`observable,' `property,' and the like. The reader should not 
expect a thorough philosophical analysis. On the contrary, we 
adopt philosophically somewhat naive but physically pragmatic 
points of view. In Section 3 some aspects of the logical 
interpretation of quantum mechanics are discussed. However, the 
treatment in Section 3 is by no means complete and only the aspect 
relevant for this work are discussed. An extensive discussion of 
the logical interpretation can be found in Omn\`es' original work 
(Omn\`es, 1988a-c, 1989, 1990, 1992, 1994, 1995). 
In Section 4 we describe a generalized history theory, which 
generalizes the standard consistent histories approach. In Section 
4 the algebraic and order structures of our generalized history 
approach are investigated. The notion of decoherence functional is 
extended to the generalized framework 
and some of its elementary properties are discussed.
It will turn out that so-called effect algebras 
or difference posets (D-posets) are the basic algebraic 
structure in our formulation. We 
explicitly construct the tensor product of two sets of effects on 
some Hilbert space in the category of effect algebras. The central 
result of Section 4.3 is the identification of the consistent sets 
of generalized histories on which the decoherence functional 
induces a probability measure and the formulation of the 
generalized logical rule of interpretation. 
Finally, in Section 4.4 we use our results to study the possible 
algebraic structures of more general quantum mechanical history 
theories not necessarily restricted to nonrelativistic theories. 
This can be viewed as a generalization of Isham's 
temporal quantum logic (Isham, 1994). In Section 5 we discuss our 
results and present our conclusions. \\

It must be emphasized that the representation and the 
interpretation of the consistent histories approach in this work 
might not be accepted by the authors cited. 
The present work solely reflects the inclination and the 
views of this author. 
\section{Basic Facts about Consistent Histories and 
Operational Quantum Mechanics} 

In this section 
we first review formal aspects of the consistent 
histories approach to quantum theory initiated by Griffiths and 
further developed by Omn\`es, Gell-Mann and Hartle and Isham and 
others. This section contains essentially 
known material. It can be read as a commentary on the standard 
texts. There is no general agreement in the literature, however, 
concerning the interpretational issues discussed in this section. 
Our treatment is based on a realistic and individual 
interpretation of probabilities in quantum mechanics as outlined, 
e.g., in (Busch et al., 1995; Omn\`es, 1992, 1994; Popper, 1982). 
[The views adopted in the present work differ, however, in 
some aspects from those in (Busch et al., 1995; Omn\`es, 1992, 
1994; Popper, 1982).] This interpretation has as one of its basic 
assumptions that there exists a definite physical reality, which 
exists independently of and changes independently of 
(human) observers. 
\begin{rem} Quantum mechanics as a probabilistic theory provides 
no solution of the problem of the 
`actualization of facts.' This is the problem why some events 
take place in the real world while others do not. In quantum 
measurement theory this problem occurs as `objectification 
problem' (Busch et al., 1991). \end{rem} In our 
interpretation we carefully distinguish between the mathematical 
formalism of a theory (the syntactical part of a theory) and the 
semantical part of a theory (the interpretation), which relates 
some abstract concepts of the mathematical formalism to the 
objects of physical reality which they represent. Only those parts 
of the mathematical formalism can be thought of as corresponding 
to elements of physical reality which are 
interpreted in the semantical part of the theory. \\ 

A physical system is a part of the physical reality which has to 
be regarded (in at least one respect) as a physical unit. The 
relevant aspects characterizing a physical system should not be 
affected by the interaction with other parts of physical reality 
(at least to a certain degree of accuracy). Every 
physical theory (in particular quantum mechanics) is concerned 
with the description (and understanding) of some physical systems. 
In this work we do not aim to give a precise definition of the 
notion `physical system,' but rather 
adopt an abstract (but pragmatic) point of view and represent a 
physical system in the mathematical formalism by a collection of 
observables (`the observables of the system'). Observables are 
part of the semantical language of physical theories; they 
represent and systematize in the mathematical formalism possible 
events which may occur in the physical systems and which can be 
described by the theory at hand. \\ 

There is a longstanding debate in quantum 
physics which entities in the formalism can be identified with 
{\sf properties} of a system and what structure the space of all 
properties is supposed to have. No agreement has been achieved 
yet, for some different points of view see, e.g., 
(Omn\`es, 1994; Ludwig, 1983; von Neumann, 1932; Giuntini and 
Greuling, 1989; Cattaneo and Laudisa, 1994). Most authors use the 
terminology 
introduced by von Neumann, who has identified the 
projection operators with the possible `properties' of the quantum 
mechanical system (von Neumann, 1932). However, Ludwig (1970, 
1983) has stressed that the space of all projection operators does 
not satisfy some conditions which may intuitively be associated 
with the notion of `property.' In contrast, the author of the 
present work believes that the problem of identifying the 
properties of a quantum mechanical system is a pseudo-problem. 
We simply identify the possible properties of a system with the 
propositions specifying the domain of values for some observable. 
That is, we 
view a system as a carrier of properties and as a bearer of 
dispositions. However, in the sequel we will use the phrase {\sf 
`proposition about a system'} or following 
partly Birkhoff and von Neumann (1936) the term {\sf physical 
quality} for this concept (actually this term has been 
used in a different sense by Birkhoff and von Neumann) and not 
the term 'property' and also not the term {\sf `pseudoproperty'} 
coined by Ludwig. What exactly we mean by {\sf physical 
quality} and {\sf physical property} will be specified below. 
Whether or not the physical qualities and 
physical properties can be measured ideally or repeatedly 
(Busch et al., 1995) and 
whether or not the set of all qualities and the set of all 
properties satisfies some more or less intuitive 
axioms as claimed, e.g., by Ludwig, is of no 
concern to us. \label{rem2} {\em The notions of observable, 
physical quality and even the features which characterize a 
physical system as such are inferred on theoretical grounds; that 
is, which features characterize a physical system and what a 
physical quality is depends upon the theory and upon the 
interpretation we use.} (However, it is clear that the question 
whether a given theory with its accompanying notions of system and 
physical quality is a `good' theory cannot be decided on purely 
theoretical grounds.) \\ 

In this work we encounter the following point of view which is 
essentially due to Griffiths (1995):
We assert that for a quantum mechanical system there are several 
incompatible (or complementary) frameworks for its theoretical 
description in terms of physical qualities and that there are 
several incompatible (or complementary) frameworks for making 
logical inferences about physical qualities or about time 
sequences of physical qualities. We do not even claim that the 
possible frameworks for the theoretical description coincide with 
the possible frameworks for making logical inferences. (As we will 
see below in our generalized history approach this is nevertheless 
the case.) All different frameworks (or in Griffiths' terminology 
{\em topics of conversation}) are similarly objective. That is, 
the symmetrical treatment of several incompatible frameworks in 
the mathematical formalism of quantum mechanics is not broken in 
the interpretation and (as is asserted in the interpretation) also 
not in the physical reality in the following sense: it is the 
whole objective physical situation (for instance an experimental 
arrangement) which determines the framework that should be used 
for the description and reasoning. \\ 
Thus we assert that for every physical system there are elements 
of physical reality which cannot be combined either in 
constructing a theoretical description or in making logical 
inferences about them. Such complementary elements of reality are 
not independent. The exact form of the framework for the 
theoretical description and for making logical inferences is 
specified below in Rule \ref{rle1} for the standard logical 
interpretation of quantum mechanics and in Rule \ref{rle2} for the 
generalized logical interpretation developed in this work. \\ 

We consider a quantum mechanical system $\cal S$ without 
superselection rules represented by a separable complex Hilbert 
space $\HH$ and a Hamiltonian operator $H$. Every physical state 
of the considered system is mathematically represented by a 
density operator on $\HH$, i.e., a linear, positive, trace-class 
operator on $\HH$ 
with trace 1. We denote the set of all trace-class operators on 
$\HH$ by ${\cal T}(\HH)$ and the set of all density operators on 
$\HH$ by ${\cal T}(\HH)^+_1$. The time 
evolution is governed by the unitary operator $U(t', t) = 
\exp(-i(t'-t) H / \hbar)$ which maps states at time $t$ into 
states at time $t'$ and satisfies $U(t'', t') U(t',t) = U(t'',t)$ 
and $U(t,t)=1$. \\

In the familiar formulations of quantum mechanics the observables 
are identified with (and represented by) the self-adjoint 
operators on 
$\HH$ and according to the spectral theorem observables can be 
identified with projection operator valued (PV) measures on the 
real line; 
that is, there is a one-to-one correspondence between self-adjoint 
operators on $\HH$ and maps ${\cal O} : {\cal B}(\RR) \to {\cal 
P}(\HH)$, 
such that ${\cal O}(\RR) = 1$ and ${\cal O}(\cup_i K_i) = 
\sum_i {\cal O}(K_i)$ for every pairwise disjoint sequence $\{ K_i 
\}_i$ 
in ${\cal B}(\RR)$ (the series converging in the ultraweak 
topology). Here ${\cal B}(\RR)$ denotes the Borel $\sigma$-algebra 
of $\RR$ and ${\cal P}(\HH)$ denotes the set of projection 
operators on $\HH$, i.e., self-adjoint operators $P$ with $P=PP$. 
PV measures are also called {\sc spectral measures.} \\ 
We adopt the following physical interpretation of the so defined 
observables: all meaningful propositions about the 
considered system specify that the value of some observable $\cal 
O$ lies in some set $B \in {\cal B}(\RR)$. We also say that such a 
specification of the value of some observable represents a {\em 
physical quality} of the system or a {\em proposition 
about a system}; in one word a proposition is a {\em speak-able}. 
All 
other propositions about the system in the formalism of the theory 
are considered as representing only syntactical statements. The 
same idea is often expressed in a modern language by saying that 
only observables represent the {\em be-ables} in the 
theory and that all other objects in the mathematical formalism 
have no beable status. We can state 
our assertion also as follows: beables represent the possible {\sf 
events} that may occur in the physical system (Bell, 1987). \\ 

{\small \begin{nt} Beables as defined above are surely the 
pre-eminent concepts in the theory that may be thought of as 
corresponding to some elements of reality. We make no attempt to 
define what the term {\sf reality} exactly means and we do not 
claim that beables are the only elements in the formalism which 
may be considered real. The term {\sf beable} is, however, 
introduced to stress that there are several degrees of reality in 
a physical theory. By way of an example, the possible values of 
the coordinates of an elementary particle describe possible events 
in physical reality and have thus beable status in the quantum 
mechanical description of the particle, whereas, say, the number 
of degrees of freedom of the same elementary particle is a 
somewhat abstract characteristic of the particle which has no 
beable status in the theory but which may nevertheless be 
considered real. In particular the state of the 
system has no beable status in the interpretation of quantum 
mechanics outlined here. The state 
of a system collects all necessary information of the past history 
of the system (e.g., preparation procedures) to compute 
probabilities for future events. In the words of Popper 
(1982) {\em \sf `the real state of a physical system, at any 
moment, may be 
conceived as the sum total of its dispositions$-$or its 
potentialities, or possibilities, or propensities.'} Thus we 
think of the state as an abstract concept in the syntactical part 
of the theory (i.e., in the mathematical formalism) characterizing 
the real physical system. To be not misunderstood, states are 
completely objective; even though we do not think of states as 
real properties of the physical system, we may nevertheless 
consider states as {\sf properties of the integral physical 
situation.} Since quantum mechanics is a probabilistic theory, 
Gleason's theorem fixes the notion of state in the formalism of 
quantum mechanics (see below). \label{rem1} \end{nt}}

The probability of a physical quality represented by the 
projection operator $P$ is in the state represented by the density 
operator $\varrho$ given by ${\tr}(\varrho P)$, where $\tr$ 
denotes the trace in $\HH$. \\ 

There is no agreement in the literature concerning the 
interpretation of this probability. According to a minimal 
operational interpretation it represents the probability that a 
physical quality is found in a measurement. \\

At this point it should be stressed that one has to carefully 
distinguish between physical qualities and the associated 
projection operators. Several physical qualities may correspond to 
the same projection operator. However, from the operational 
approach to quantum mechanics (Kraus, 1983; Ludwig, 1983) it 
follows, that 
different physical qualities corresponding to the same projection 
operator cannot be distinguished by performing yes$-$no 
experiments and thus can only 
collectively be considered to be true or false. For this reason 
in standard quantum mechanics the projection operators are said to 
represent {\em properties} of the system. \\ 

\begin{rem} In this work we adopt Popper's {\sf propensity 
interpretation of probabilities} (Popper, 1959/60, 1982). In the 
propensity 
interpretation the classes of experimentally indistinguishable 
physical qualities (represented by some projection operator) 
represent the ``beables'' and the probabilities are thought to 
express the tendencies in the behaviour of the system. 
They are measures of the propensity or of the tendency of a 
possibility (i.e., a beable) to realize itself upon repetition. 
Probabilities are associated with a single system in contrast 
to the usual frequency interpretation of probabilities where 
probabilities are numbers characterizing ensembles of (similarly 
prepared) systems. \label{rem4} Nevertheless the probabilistic 
statements in the theory can be tested by actual sequences of 
measurements. Popper ascribes the priority for the 
propensity interpretation to a large extend to Land\'e 
(1965). \end{rem} {\small \begin{nt} A 
comprehensive 
and critical discussion of different possibilities to consistently 
interpret the term `probability' can be found in the monographs by 
K.~Popper (1982, 1994). In his discussion of quantum 
mechanics Popper discusses besides the propensity interpretation 
of probability various issues regarding the interpretation of 
quantum mechanics and other related areas of physics. However, 
although the author of the present work believes that the 
propensity interpretation 
of probability is well-suited for the consistent histories 
approach to quantum mechanics and for the logical interpretation, 
the author does not wish to imply that he accepts all of the 
assertions and claims in (Popper, 1982, 1994). On 
the contrary, this author believes that Popper's 
interpretation of quantum mechanics 
as a whole is physically incomplete and that something like the 
logical rule of interpretation has to be added to it in order to 
make it physically complete. \end{nt}} 

According to standard quantum mechanics physical qualities 
correspond to projection operators and observables to PV measures.
However, as discussed at length by 
Davies (1976), Kraus (1983) and Busch et 
al.~(1991, 1995) there is good reason to generalize 
the usual notion of an observable in quantum mechanics. These 
generalized observables have been developed and applied 
independently in various branches of physics, such as, e.g., 
stochastic quantum mechanics (Prugov\v{e}cki, 1992), quantum 
optics 
(Davies, 1976) or even the (conceptual) foundations of quantum 
mechanics (Ludwig, 1983). For a detailed motivation and physical 
justification for the introduction of generalized observables, we 
refer in particular to the recent lucid monograph by Busch, 
Grabowski and Lahti (1995). \\ 

Positive and bounded operators $F$ on 
$\HH$, satisfying \[ 0 \leq F \leq 1, \]
are commonly called {\sc effects} and the set of 
all effects on the Hilbert space $\HH$ will be denoted by $\sef$. 
We further denote the set of all bounded, linear operators on 
$\HH$ by $\bou$, the set of all positive (and hence 
Hermitean), linear, bounded operators on $\HH$ by $\bop$, and the 
set of all projection operators on $\HH$ by $\pro$. \\
If $\HH$ is an infinite dimensional Hilbert space, then the set of 
all projection operators $\pro$ on $\HH$ is weakly 
dense in $\sef$ (Davies, 1976). \\

{\sc Generalized observables} are now identified with {positive 
operator valued (POV) measures} on some measurable space $(\Omega, 
{\cal F}),$ i.e., maps {\fraktur O} $: {\cal F} \to \sef$ with 
the properties: \begin{itemize} \item {\fraktur O}$(A) \geq$ 
{\fraktur O}$(\emptyset)$, for all $A \in \cal F$; \item Let 
$\{A_i\}$ be a countable set of disjoint sets in $\cal F$, then 
{\fraktur O}$(\cup_i A_i) = \sum_i$ {\fraktur O}$(A_i)$, the 
series converging ultraweakly; \item {\fraktur O}$(\Omega) = 1$. 
\end{itemize}
Generalized observables are also called {\sc effect valued 
measures}. 
Ordinary observables (associated with self-adjoint operators on 
$\HH$) are then identified with the projection valued measures on 
the real line $\RR$. 

{\small \begin{nt} Notice, that in general 
there is no unambiguous representation of generalized observables 
as self-adjoint operators on $\HH$. However, there is a one-one 
correspondence between {\sf maximal symmetric operators on $\HH$} 
and POV measures on $\HH$ (Busch et al., 1995). \end{nt}}

The set $\Omega$ represents the set of 
all possible values of the observable {\fraktur O} and $\cal F$ 
represents the allowed {\em coarse-grainings} in $\Omega$. 
Generalized observables which are not ordinary observables are 
often also called {\sc unsharp observables.} 

{\small \begin{nt}The term unsharp observable is 
used with different meanings in the literature, see e.g., 
(Busch et al., 1995). However, the term {\sf unsharp} is somewhat 
misleading. Admittedly, some (but not all) generalized observables 
arise as smeared versions of ordinary observables. In general, 
however, generalized observables must not be considered as unsharp 
or limited counterparts of some underlying more sharply defined 
observables, but rather as independent entities in their own 
right. \end{nt}}

According to Gleason's theorem, for every positive, 
normalized $\sigma$-additive map $p: \sef \to 
[0,1]$, also called {\em generalized probability measure}, there 
exists a unique $\varrho \in \topp$, such that $p(F) = 
{\tr}(\varrho F)$. (A map $p: \sef \to [0,1]$ is called 
{\sc $\sigma$-additive} if for every countable collection $\{ E_i 
\}$ of elements of ${\frak E}(\HH)$ such that $\sum_i E_i \leq 1$, 
one has $p \left( \sum_i E_i \right) = \sum_i p(E_i)$ (convergence 
in the weak operator topology).) It also follows from Gleason's 
theorem that an effect valued measure represents the most general 
notion of an observable compatible with the probabilistic 
structure of Hilbert space quantum mechanics. Gleason's theorem 
has first been proved in (Gleason, 1957); a short proof can for 
instance be found in (Maeda, 1989). \\

Generalizing our above terminology, we regard all propositions 
specifying the value of some (generalized) 
observable as (generalized) physical qualities . In 
order to discriminate physical qualities corresponding to ordinary 
observables from physical qualities corresponding to generalized 
observables, we will sometimes call the former `ordinary physical 
qualities' and the latter `generalized physical qualities.' In the 
generalized approach to every 
physical quality there is associated one effect operator. \\

Every physical quality represents some dichotomy (or binary 
alternative or yes-no alternative) at every instant of time. 
There are physical qualities which are always true, 
e.g., the physical qualities asserting that the value of some 
observable lies somewhere (without further specification) in the 
space of its possible values. Such physical qualities will be 
pairwise identified and collectively denoted by $\fraktur 1$. 
Similarly, physical qualities which are always false will also be 
pairwise identified and collectively denoted by $\fraktur 0$. We 
will denote the set of all physical qualities (with this 
identification) of a quantum system $\cal S$ with corresponding 
Hilbert space $\HH$ by ${\frak P}(\HH)$. Notice, that a 
(nontrivial) member $\frak p$ of ${\frak 
P}(\HH)$ is a proposition specifying the value of some observable 
and is {\em not} an operator on $\HH$ in some way or other 
related to the proposition $\frak p$. \\ 
The effect operator $p$ associated with a physical quality $\frak 
p$ will be denoted by $p = {\frak Z}({\frak p})$, thus defining a 
map ${\frak Z}: {\frak P}(\HH) \to {\frak E}(\HH)$. \\

It is worthwhile to mention that in the operational 
approach to quantum mechanics 
states are defined as equivalence classes of preparing instruments 
of a system, where two preparing instruments are said to be 
equivalent if they cannot be distinguished by measuring binary 
alternatives (Kraus, 1983; Ludwig, 1983). Moreover, effects can 
also be defined as equivalence 
classes of measuring instruments performing yes$-$no measurements, 
where two measuring instruments are called equivalent if for every 
state both measuring instruments are triggered with equal 
frequencies. Since, however, every measuring process can be 
thought of as being composed of elementary yes$-$no measurements 
of physical qualities (at least in principle) (Beltrametti and 
Cassinelli, 1981; Jauch, 1968) and 
since physical qualities represented by the same effect operator 
cannot be distinguished from each other by performing yes$-$no 
measurements, {\sf it follows that in quantum mechanics every pair 
of physical qualities which are represented by the same effect 
operator must be considered equivalent: that is, in quantum 
mechanics one is not interested in single physical qualities but 
in equivalence classes of physical qualities.} We will call this 
equivalence classes of physical qualities {\sc physical properties 
of the physical system}. It is reasonable to assume that physical 
properties of a quantum mechanical system are in one-to-one 
correspondence with effect operators (Kraus, 1983). Therefore in 
generalized quantum mechanics {\sf the effect operators represent 
the beables of the theory} in the mathematical formalism. In the 
standard formulation of quantum mechanics often only the 
projection operators are referred to as representing `properties' 
of the quantum system (following a terminology introduced by von 
Neumann (1932)). Thus effect operators are often referred to as 
representing so-called `unsharp properties.' However, {\sf in 
general effect operators must not be considered as unsharp or 
smeared counterparts of some underlying more sharply defined 
properties but as independent entities in their own right.} Since 
physical properties represented by effect operators can in general 
neither be measured ideally without disturbing the system nor 
measured repeatedly, one may say, loosely speaking, that 
projection operators represent {\sf stable properties} and that 
all other effect operators represent {\sf unstable properties}. \\ 

Again the probability of some physical property $F$ is in the 
state $\varrho$ given by ${\tr}(F \varrho)$. \\ 

Summing up, throughout this paper we 
adopt the following conventions: by {\em a physically meaningful 
proposition about a system} we mean a statement (associated 
with some effect operator) specifying the value of a generalized 
observable; we also talk about {\em physical qualities} in 
this connection; a {\em proposition 
about a system} is thus a statement in the semantical language of 
physics asserting the truth (or realization) of some physical 
qualities.
By a {\em physical property of a system} we mean an equivalence 
class of physical qualities which cannot be distinguished by 
measuring binary alternatives. Physical properties are in one-one 
correspondence with effect operators. \\ \\

We now turn to the discussion of histories in standard 
nonrelativistic quantum mechanics, which, loosely 
speaking, are defined to be sequences of projection operators on 
$\HH$. In standard nonrelativistic quantum mechanics a history 
could alternatively be defined as time sequence of physical 
qualities since (as already remarked above and as will become 
clearer below) different physical qualities corresponding to the 
same projection operator can only simultaneously be true. Thus we 
stick to the usual definition
\begin{de} \label{b} A {\sc homogeneous history} is a map $h : \RR 
\to \pro, t \mapsto h_t$.
We call $t_i(h) := \min(t \in \RR \mid h_t \neq 1)$ the {\sc 
initial} and $t_f(h) := \max(t \in \RR \mid h_t \neq 1)$ the 
{\sc final time } of $h$ respectively. Furthermore, the {\sc 
support of} $h$ is given by ${\frak s}(h) := \{t \in \RR \mid h_t 
\neq 1 \}$. If ${\frak s}(h)$ is 
finite, countable or uncountable, then we say that $h$ is a {\sc 
finite, countable} or {\sc uncountable history} respectively. The 
space of all homogeneous 
histories will be denoted by ${\cal H}(\HH)$, the space of all 
finite homogeneous histories by ${\cal H}_{fin}(\HH)$ and the 
space of all finite homogeneous histories with support $S$ by 
${\cal H}_S(\HH)$. \end{de} 
In this work we focus attention on finite histories. 
In the following we will identify every homogeneous history $h$ 
with the string of its nontrivial projection operators, i.e., 
we write $h \simeq \{ h_{t_k} \}_{t_k \in {\frak s}(h)}$. \\

For every finite subset $S$ of $\RR$ we can consider the Hilbert 
tensor product $\otimes_{t \in S} \HH$ and the algebra $\bout{S}$ 
of bounded linear operators on $\ten{S}$. It has been pointed out 
by Isham (1994) that for any fixed $S$ there is an injective 
(but not surjective) correspondence $\sigma_S$ between finite 
histories with support $S$ and elements of $\bout{S}$ given by \[ 
\sigma_S : \fini{S} \to \bout{S}, h \simeq \{ h_{t_k} \}_{t_k \in 
S} \mapsto \otimes_{t_k \in S} h_{t_k}. \]
The finite homogeneous histories with support $S$ can therefore be 
identified with projection operators on $\ten{S}$. The set of 
all projection operators on $\ten{S}$ will in the sequel be 
denoted by $\pout{S}$. However, not all projection operators in 
$\pout{S}$ have the form $\sigma_S(h)$ with $h \in \fini{S}$. \\
If a homogeneous history vanishes for some $t_0 \in \RR$, i.e., 
$h_{t_0} =0$, then we say that $h$ is a {\sc zero history}. All 
zero histories are collectively denoted by 0, slightly abusing the 
notation. \begin{de} \label{po1} Let $h,k \in {\cal H}(\HH)$. We 
say that $k$ is {\sc coarser than} $h$ if $h_t \leq k_t$ for all 
$t \in \RR$ and write $h \leq k$. If furthermore $h \neq k$, then 
we write $h < k$. The set ${\cal H}(\HH)$ equipped with the 
relation $\leq$ is a partially ordered set. \end{de}
\begin{de} Two homogeneous histories $h$ and $k$ are said to be 
{\sc disjoint} if there is some $t \in \RR$ such that $h_t k_t 
=0$. \end{de}

The identification of finite homogeneous histories with support 
$S$ with projection operators on $\ten{S}$ allows for the 
introduction of a much broader class of histories. To this end we 
recall the well-known fact that 
the set $\pro$ of projection operators on a Hilbert space $\HH$ 
carries the structure of an orthocomplemented complete 
lattice 
provided for $p_1, p_2 \in \pro$ one defines [a] $p_1 \leq 
p_2$ if $p_1$ projects on a subspace of the range of $p_2$, 
($\leq$ defines a partial order on $\pro$), [b] the join $p_1 \vee 
p_2$ of $p_1$ and $p_2$ to be the projection operator which 
projects on 
the smallest closed subspace of $\HH$ which contains the subspaces 
$p_1 \HH$ and $p_2 \HH$, [c] the meet $p_1 \wedge p_2$ of $p_1$ 
and $p_2$ to be the projection operator 
which projects on the intersection of $p_1 \HH$ and $p_2 \HH$ and 
[d] the orthocomplementation $\neg p_1$ of $p_1$ to be the 
projection operator which projects on the 
complement of $p_1 \HH$ in $\HH$ (Birkhoff and von Neumann, 1936).
\begin{de} Let $S$ be a finite subset of $\RR$, then we call the 
space ${\cal K}_S(\HH) := \pout{S}$ of projection operators on 
$\ten{S}$ the {\sc space of finite inhomogeneous histories with 
support $S$}. The space of all 
finite inhomogeneous histories with arbitrary support will be 
denoted by ${\cal K}_{fin}(\HH)$ or by $\pout{fin}$. \end{de}

The lattice operations on $\pout{S}$ induce corresponding 
operations on the finite homogeneous histories in $\fin$, which 
are explicitly described in the following remarks.
\begin{rem} Let $h,k \in \fin$ be two finite homogeneous 
histories, then the {\sc join} $h \vee k$ of $h$ and 
$k$ is defined to be the unique finite history with support 
${\frak s}(h) \cup {\frak s}(k)$ which is 
represented in $\pout{{\frak s}(h) \cup 
{\frak s}(k)}$ by $\left( \otimes_{t_i \in {\frak s}(h)} 
h_{t_i} \right) \bigvee \left(
\otimes_{s_j \in {\frak s}(k)} k_{s_j} \right)$. 
The history $h \vee k$ may be not homogeneous. In this case 
$h \vee k$ is an inhomogeneous history. The {\sc 
join} $\bigvee_j h_j $ of any finite sequence $\{ h_j \}$ of 
pairwise disjoint 
homogeneous histories is analogously defined to be the unique 
finite history with support $\bigcup_j {\frak s}(h_j)$ 
which is represented in $\pout{\cup_j {\frak s}(h_j)}$ by 
$\bigvee_j \left( 
\otimes_{t_i \in {\frak s}(h_j)} h_{t_i} \right)$. \end{rem}
\begin{rem} Let $h,k \in \fin$ be two finite homogeneous 
histories, then the {\sc meet} $h \wedge k$ of $h$ and 
$k$ satisfies that $(h \wedge k)_t := h_t \wedge 
k_t$ is the projection operator on the intersection of the ranges 
of $h_t$ and $k_t$ for all $t \in \RR$. 
The meet operation maps pairs of finite homogeneous histories to a 
finite homogeneous history. \end{rem}
\begin{rem} Let $h$ be a finite homogeneous history with support 
${\frak s}(h)$, then $\neg h$ 
is the unique history with support ${\frak s}(h)$ 
which in $\pout{fin}$ is 
represented by $1 - \bigotimes_{t \in {\frak s}(h)} h_t$. 
We call $\neg h$ the {\sc negation} of $h$. 
The negation $\neg h$ of a finite homogeneous history $h$ may be 
inhomogeneous. Obviously the negation satisfies $h \vee \neg h = 1 
$ and $h \wedge \neg h =0$. It is clear that $\neg h$ is uniquely 
determined by this two conditions. \end{rem}

\begin{lem} Let $S$ be a finite subset of $\RR$, then the set 
${\cal K}_S(\HH)$ is an orthocomplemented complete lattice. 
\end{lem}
\begin{rem} The join, meet and orthocomplementation operations on 
${\cal K}_S(\HH)$ (where $S$ is a finite subset of $\RR$) and on 
$\pout{fin}$ are denoted by the same 
symbols (slightly abusing the notation). \end{rem}

In (Isham, 1994) it is indicated how to imbed ${\cal 
K}_S(\HH)$ into an infinite tensor product of operator algebras 
and how to furnish the latter with a Hilbert lattice structure. 
\begin{de} Two (possibly inhomogeneous) finite histories $h$ and 
$k$ are said to be {\sc disjoint} if $h \leq \neg k$, where $\leq$ 
is the partial order on ${\cal K}_{{\frak s}(h) \cup {\frak 
s}(k)}(\HH)$. We write $h \perp k$. 
\end{de}
\begin{lem} Let $h$ and $k$ denote two disjoint finite histories, 
then $h \wedge k =0$. \end{lem}
\begin{de} A history $h \in \fini{S}$ is called a {\sc simple 
history} if $h_t$ is a projection operator on a one dimensional 
subspace of $\HH$ for every $t \in S$. \end{de}
\begin{lem} \label{d0} For every finite $S \subset \RR$ the space 
$\pout{S}$ can be generated from 
$\fini{S}$ by the application of a countably infinite number of 
$\vee$, $\wedge$ and $\neg$ operations. \end{lem} 
This follows from the fact that, say, in the case of the tensor 
product $\HH_1 \otimes \HH_2$ the set $\{ e_{n_1} \otimes e_{n_2} 
\} $ forms an orthonormal basis for $\HH_1 \otimes \HH_2$ if and 
only if $\{e_{n_1}\}$ and $\{ e_{n_2}\}$ form orthonormal bases 
for $\HH_1$ and $\HH_2$, respectively. A moments thought shows 
even more, namely 
\begin{lem} For every finite $S \subset \RR$ the set $\pout{S}$ 
can be generated from the set of all 
simple histories in $\fini{S}$ by the application of a countably 
infinite number of $\vee$ operations. \label{L0} \end{lem} 
\begin{rem} For every finite $S \subset \RR$ the meet, join and 
orthocomplementation operations 
on $\pout{S}$ induce a meet, join and an orthocomplementation 
operation on $\pout{fin}$ respectively which will be denoted by 
the same symbols. \label{rem10} \end{rem}
\begin{de} Let ${\cal A}$ denote a finite collection $\{ h_k \}$ 
of histories in ${\cal K}_{fin}(\HH)$. Then $\cal A$ is said to be 
{\sc disjoint} if each pair of histories in $\cal A$ is disjoint. 
$\cal A$ is said to be {\sc complete} if $\bigvee_k h_k =1$. 
\end{de}

Furthermore, to every finite homogeneous history $h \in \fin$ we 
associate its {\sc class 
operator with respect to the fiducial time $t_0$} 
\begin{eqnarray} C_{t_0}(h) & := & U(t_0,t_n) h_{t_n} 
U(t_n,t_{n-1}) h_{t_{n-1}} ... U(t_2,t_1) h_{t_1} U(t_1,t_0) \\
& = & U(t_0,t_i(h)) h_{t_n}(t_n) h_{t_{n-1}}(t_{n-1}) ... 
h_{t_1}(t_1) U(t_i(h),t_0), \end{eqnarray}
where we have defined the Heisenberg picture operators
\[ h_{t_k}(t_k) := U(t_k,t_i(h))^{\dagger} h_{t_k} U(t_k,t_i(h)) 
\] with respect to the initial time $t_i(h)$ of $h$.
The class operators are extended to finite inhomogeneous histories 
by requiring that $C_{t_0}$ is additive in the following sense 
\begin{eqnarray} 
\label{a} C_{t_0}(h \vee k) & := &
C_{t_0}(h) + C_{t_0}(k) \mbox{ whenever } h \perp k \\
\label{a1} C_{t_0}(\neg h) & := & 1 - C_{t_0}(h). \end{eqnarray} 
This definitions are compatible with the lattice theoretical 
identities $\neg (h \vee k) = (\neg h) \wedge (\neg k)$ and $\neg 
(h \wedge k) = (\neg h) \vee (\neg k)$. Notice, that 
Equation 
\ref{a1} is a consequence of Equation \ref{a}. In the language of 
Birkhoff (1967) $C_{t_0}$ is an operator-valued {\sc valuation} 
on $\pout{S}$ for every finite $S \subset \RR$. It follows from a 
lemma in (Birkhoff, 1967, Chapter X.1.) that Equations \ref{a} and 
\ref{a1} are 
equivalent to \begin{equation} \label{a2} C_{t_0}(h \vee k) := 
C_{t_0}(h) + C_{t_0}(k), \mbox{ whenever } h \perp k. 
\end{equation} 
The Equations \ref{a} and \ref{a1} are motivated by the identities 
valid for all disjoint $h,k 
\in \pout{fin}$: \begin{eqnarray*} h \vee k & = & h + 
k, \\ \neg h & = & 1 - h. \end{eqnarray*}
The analogue of Equation \ref{a2} for inhomogeneous histories 
which are not finitely generated is 
\begin{equation} \label{e11} C_{t_0} \left( 
\bigvee_{i=1}^{\infty} h_i \right) := \sum_{i=1}^{\infty} 
C_{t_0}(h_i), \mbox{ if } h_j \perp
\left( \bigvee_{i=1}^{j-1} h_i \right) \mbox{ for every } j, 
\end{equation} the series converging ultraweakly. It is clear that 
the left hand sides of Equation \ref{a} and of Equation \ref{e11} 
are well-defined. This can be easily seen from Lemma \ref{L0} and 
from the definition of the tensor product $\otimes_{t \in S} \HH$. 
We mention that the fiducial time $t_0$ can be chosen completely 
arbitrary. \\

By virtue of Lemma \ref{L0} it is therefore enough to know the 
class operators of every finite simple homogeneous history.
\begin{de} Let the state of a quantum mechanical system at 
time $t_0$ be given by the density operator $\varrho(t_0)$. For 
every pair $h$ and $k$ of finite homogeneous histories we define 
the {\sc decoherence weight of $h$ and $k$} by \begin{equation} 
d_{\varrho} (h,k) := {\mbox{\em tr}} \left(C_{t_0}(h) \varrho(t_0) 
C_{t_0}(k)^{\dagger} \right). \end{equation} 
The functional $d_{\varrho}:\fin \times \fin 
\to \CC, (h,k) \mapsto d_{\varrho}(h,k)$ will be called the {\sc 
decoherence functional associated with the state} $\varrho$. 
The decoherence functional is in an obvious way extended to 
finite inhomogeneous histories using Equations ${\mbox{{\em 3}}}$ 
and ${\mbox{{\em 6}}}$. \end{de}
\begin{lem} Let $h,h'$ and $k$ denote finite histories. The 
decoherence functional $d_{\varrho}$ satisfies
\begin{itemize} \item $d_{\varrho}(h,h) \in \RR$ and 
$d_{\varrho}(h,h) 
\geq 0$. \item $d_{\varrho} (h,k) = d_{\varrho}(k,h)^*$. \item 
$d_{\varrho}(1,1) =1$. \item $d_{\varrho}(h \vee h', k) = 
d_{\varrho}(h,k) + d_{\varrho}(h',k)$, whenever $h \perp h'$. 
\item $d_{\varrho}(0,h) =0$, for all 
$h$. \end{itemize} \end{lem}

Now fix $h$ and vary $\varrho$ in $p_{\varrho}(h) := 
d_{\varrho}(h,h)$. Then $p_{\varrho}(h)$ is a positive, linear, 
bounded functional on $\topp$. Therefore it follows from Gleason's 
theorem that there exists a unique effect $F(h) 
\in {\frak E}(\HH)$ such that \[ d_{\varrho}(h,h) = {\tr}[\varrho 
F(h)]. \] Obviously \[ F(h) = C_{t_0}(h)^{\dagger} C_{t_0}(h). \]
We call $F(h)$ the effect associated with the finite history $h$. 
The map $F : \fin \to {\frak E}(\HH)$ is many-to-one and therefore 
information about $h$ is lost when considering $F(h)$ 
instead of $h$. 
\subsection*{Consistent sets of histories}
\begin{de} Let $h$ and $k$ be two disjoint histories in 
$\pout{fin}$. The 
histories $h$ and $k$ are said to be {\sc preconsistent with 
respect to the state $\varrho$} if $\mbox{{\em Re} } 
d_{\varrho}(h,k) = 0$. Any collection $\cal C$ of histories in 
$\pout{fin}$ is said to be 
{\sc preconsistent with respect to the state $\varrho$} if every 
pair of disjoint histories in $\cal C$ is preconsistent with 
respect to the state $\varrho$. Any 
collection ${\cal C}'$ of histories in $\pout{fin}$ is said to be 
{\sc consistent with respect to the state $\varrho$} if 
${\cal C}'$ is a Boolean algebra (with respect to the 
meet, join and orthocomplementation in $\pout{fin}$, see Remark 
${\mbox{{\em 7}}}$ and with unit $1_{{\cal C}'}$) and if 
${\cal C}'$ is preconsistent with 
respect to the state $\varrho$. \end{de}

Note that our above terminology differs somewhat from the 
terminology used by other authors. Further, some authors call a 
pair $h,k$ of histories {\em weakly 
decoherent} if it satisfies $\mbox{\rm Re } d_{\varrho}(h,k) = 0$ 
and {\em mediumly decoherent} if it satisfies $d_{\varrho}(h,k) = 
0$. There can be found other related notions of decoherence 
and consistency in the literature, see, e.g., (Gell-Mann and 
Hartle, 1995; Finkelstein, 1993; Zeh, n.d.). 
The condition Re $d_{\varrho}(h,k) = 0$ is interpreted 
in physical terms by saying that the events $h$ and $k$ have 
vanishing interference in the state $\varrho$. \\
The notion of consistency is important because it is the key to a 
probability interpretation of the numbers $d_{\varrho}(h,h)$ for 
some (pre-)consistent sets of histories. \\
Let us recall that usually a probability space is defined 
to be a triple $(\Omega, {\cal A}, p)$, where $\Omega$ is an 
arbitrary set, $\cal A$ is a Boolean $\sigma$-algebra of subsets 
of $\Omega$ and $p$ is a probability measure on $\cal A$. 
This can be generalized as follows
\begin{de} Let $\cal L$ be a partially ordered set and ${\cal B} 
\subset \cal L$ be a Boolean 
lattice. A nonnegative valuation $m : {\cal B} \to \RR^+$ on 
$\cal B$ which is additive \[ m \left[ \bigvee_{k=1}^{N} \alpha_k 
\right] = \sum_{k=1}^{N} m[\alpha_k], \mbox{ if } \alpha_k \wedge 
\left( \bigvee_{i=1}^{k-1} \alpha_i \right) =0, \mbox{ for every } 
k < N, \] is called a {\sc  finite measure on} $\cal B$. 
If $\cal B$ is a Borel lattice, then $N$ may be taken to be 
$\infty$. In this case $m$ is $\sigma$-additive. If $\cal B$ is 
not a Borel lattice, then $N$ is always finite. If furthermore 
$m[1_{\cal B}] = 1$, then $m$ is called a {\sc probability 
measure on} $\cal B$ and the triple $({\cal L}, {\cal B}, m)$ is 
called a {\sc probability lattice}. \end{de}
A Borel lattice is a Boolean $\sigma$-lattice (Birkhoff, 1967). 
\begin{theo} \label{th1} Let ${\cal C} \subset \pout{fin}$ be a 
Boolean lattice. 
If $\cal C$ is preconsistent with respect to the state $\varrho$, 
then the triple $(\pout{fin}, {\cal C}, p_{\varrho})$ is a 
probability lattice, where $p_{\varrho}$ is defined by
\begin{equation} \label{pro} p_{\varrho} : {\cal C} \to \RR^+, 
p_{\varrho}(h) := \frac{d_{\varrho}(h,h)}{d_{\varrho}(1_{\cal C}, 
1_{\cal C})}. \end{equation} \end{theo}
The proof is straightforward. \\ \\

In the literature it is often tacitly assumed that the 
preconsistent set of histories under consideration forms (or 
generates) a Boolean 
lattice so that a probability interpretation of the diagonal 
values of the \df makes sense. \\
The probability defined by Equation \ref{pro} can for finite 
homogeneous histories be interpreted as conditional probability, 
namely as the probability of the sequence of the propositions 
$h_{t_f} = h_{t_k}, ..., h_{t_{k-j}}$ given that the sequence of 
propositions $h_{t_{k-j-1}}, ..., h_{t_0}$ is realized. 
\begin{lem} \label{w3} Let $(\pout{fin}, {\cal C}, p_{\varrho})$ 
be a probability lattice, where $p_{\varrho}$ is defined by 
Equation \mbox{{\em \ref{pro}}}, then for all $h,k \in \cal C$ 
\begin{itemize} 
\item $0 \leq p_{\varrho}(h) \leq 1$. \item $p_{\varrho}(h \vee k) 
+ p_{\varrho}(h \wedge k) = p_{\varrho}(h) + p_{\varrho}(k).$ 
\item $p_{\varrho}(h) \leq p_{\varrho}(k) \mbox{ if } h \leq k.$ 
\end{itemize} \end{lem}
\begin{co} \label{CSH} Let ${\cal C} \subset \pout{fin}$ be a 
Boolean lattice. 
Then $\cal C$ is a preconsistent set of histories w.r.t.~the state 
$\varrho$ if and only if every pair $h,k$ of histories in $\cal C$ 
satisfies \begin{equation} \label{md} d_{\varrho}(h \vee k, h \vee 
k) + d_{\varrho}(h \wedge k, h \wedge k) = d_{\varrho}(h,h) + 
d_{\varrho}(k,k). \end{equation} \end{co}
\begin{rem} We notice that $d_{\varrho}$ induces also probability 
functionals on sets of histories which are not Boolean lattices. 
Let $\cal C$ be a preconsistent set of pairwise disjoint 
histories, then $m_{\varrho} : {\cal C} \to \RR^+, 
m_{\varrho}(h) := d_{\varrho}(h,h)/\left( \sum_{k \in {\cal 
C}}d_{\varrho}(k,k) \right)$ is an additive functional on 
$\cal C$ and $m_{\varrho}(h)$ can be interpreted as probability of 
$h \in \cal C$. However, since $\cal C$ generates a Boolean 
sublattice of $\fink$ on which $d_{\varrho}$ induces a probability 
measure extending $m_{\varrho}$, it is enough to consider Boolean 
algebras of histories. \end{rem} 

\section{The logical interpretation of quantum mechanics} 

The logical interpretation of (nonrelativistic) quantum mechanics 
is an epistemological interpretation of quantum mechanics. This 
interpretation is mainly due to Roland Omn\`es. 
In this section we first briefly outline some basic assertions of 
the logical interpretation of quantum mechanics. The 
logical interpretation as discussed in this section and Section 2 
differs in some minor details from that in Omn\`es' original work 
(Omn\`es, 1988a-c, 1989, 1990, 1992, 1994, 1995). \\
The logical interpretation is a realistic interpretation of 
quantum mechanics and thus the discussion of Section 2 applies 
here, see in particular the first part of Section 2. 
In the logical formulation the most general propositions about 
a quantum mechanical system which have a physical meaning are 
identified with finite (or at least countably 
infinite) history propositions. Other statements about a system 
which cannot be casted into the framework of history propositions 
are not considered to be meaningful and hence excluded from 
consideration. A ``probability'' is associated with every 
history proposition in a consistent set. The 
set of all probabilities for all history propositions specifies 
the state of the system. In the Hilbert space formulation of 
quantum mechanics the state of a system is characterized 
by a density operator and the probabilities are given by Equation 
\ref{pro}. The set of all meaningful propositions about some 
system in terms of histories 
together with their corresponding probabilities (given by 
Equation \ref{pro}) is considered to be the most general 
conceivable knowledge about a particular quantum mechanical 
system. The probabilities of histories are considered to be 
objective entities in their own right, i.e., numbers 
associated with history propositions describing single systems, 
and not only as quantities approximately equal to frequencies in a 
series of measurements. We adopt again Popper's {\em propensity 
interpretation} for this probabilities, compare Remark \ref{rem4}. 
\\ 

In our terminology introduced above the history propositions 
represent the (temporal) beables (in the sense of 
propensities) and the probabilities are therefore thought to 
express the (temporal) tendencies in the behaviour of the system. 
The probability measure on a consistent Boolean algebra of history 
propositions induced by 
the decoherence functional according to Theorem \ref{th1} defines 
in this consistent Boolean algebra two logical relations, namely 
an implication and an equivalence relation between histories. A 
history proposition $h$ is said to {\sc imply} a history 
proposition $k$ if the conditional probability $p_{\varrho}(k 
{\mid} h) \equiv \frac{p_{\varrho}(h \wedge k)}{p_{\varrho}(h)}$ 
is well-defined and equal 
to one. Two history propositions $h$ and $k$ are said to be {\sc 
equivalent} if $h$ implies $k$ and vice versa. \\
The universal rule of interpretation of quantum mechanics can now 
be formulated as 
\begin{rle} Propositions about quantum mechanical 
systems should \label{rle1} solely be expressed in terms of 
history propositions. Every description of an isolated quantum 
mechanical system should be expressed in terms of finite history 
propositions belonging to a common consistent Boolean algebra of 
histories. Every reasoning relating several propositions should be 
expressed in terms of the logical relations induced by the 
probability measure from Theorem $\mbox{\em \ref{th1}}$ in that 
Boolean algebra. \end{rle}

This logical rule has to be understood as a semantical 
rule which systematizes the language of quantum mechanics. It 
once and for all makes sure, whether a reasoning or an implication 
is allowed or not. This is exactly what one expects from a rule 
building the basis of a complete interpretation. The logical rule 
can also be applied to the hitherto highly problematic 
retro\-dictive reasoning in quantum mechanics. \\
The causal relationship between different histories is 
coded into a logical relationship. This has nothing to do with 
introducing `a new empirical kind of metalogic,' as is sometimes 
claimed (Zeh, n.d.), but simply introduces a convenient way of 
speaking: the logical rule is neither regarded as a scientific law 
of human thought nor intended to modify the logical structure of 
our mathematical reasoning, but it is rather confined to the 
systems of quantum mechanical propositions to which probabilities 
may be ascribed by the theory. \\ 

It should be noticed that in the above formulation there is 
neither a preferred history nor a preferred Boolean algebra in the 
theory, which may be associated in some way with the `actual 
facts' observable in the real world. Omn\`es does not claim to 
have solved the problem of actualization of facts or the 
objectification problem in the quantum measurement 
process. On the contrary, Omn\`es argues in (Omn\`es, 1992, 1994) 
that it is 
neither feasible nor necessary that quantum mechanics provides an 
explanation for the process of actualization of facts. Therefore 
the criticism in this direction raised by Zeh (n.d.) is unfounded. 
\\ 

A few remarks are in order here. Firstly, this author does neither 
claim nor presuppose that quantum mechanics is the ultimate 
universal theory in terms of 
which every natural phenomenon can eventually be described. This 
may or may 
not be the case. No conclusive decision of this question can be 
made at the present. \begin{nt} Some authors explicitly or 
implicitly postulate the universality of quantum mechanics casted 
into the framework of consistent histories. According 
to this postulate the formalism 
of nonrelativistic quantum mechanics (casted into the framework of 
consistent histories) can without significant changes be applied 
to the whole universe and 
every (approximately) isolated part of it. A globally defined 
notion of `time translation' is needed simply to formulate such a 
theory. However, in general curved spacetimes there is no such 
globally defined preferred notion of `time translation.'
Thus this postulate needs further justification 
before it can be accepted as a generally valid fundamental 
principle of physics. Intimately related to this question is the 
question whether the notion of the initial state of the universe 
is meaningful at all. All these nontrivial problems can 
presumably be decided only on the basis of a not yet existing 
quantum theory of gravity. \end{nt} Accordingly, we 
formulate the above Rule \ref{rle1} modestly only for quantum 
mechanical systems, i.e., systems which can indeed be described 
by quantum mechanics. \\ Secondly, 
in his work Omn\`es identifies the possible `properties' of a 
system at a fixed instant of time with projection operators. As 
already discussed in Section 2 we do not use this terminology in 
the present work. \\ Thirdly, Omn\`es has a good deal to say about 
decoherence by the external environment, about recovering 
classical physics and `common sense' from his approach and about 
the notion of truth in quantum mechanics and related questions. It 
is beyond the scope of the present work to discuss these issues 
and thus the reader is referred to (Omn\`es, 1990, 1991, 1992, 
1994, 1995; Dowker and Kent, 1996; Zeh, n.d.). Concerning the 
notion of truth, we recall that in the present work we adopt 
essentially Griffiths' point of view (Griffiths, 1995). \\ 
Fourthly, it is sometimes claimed that in realistic and individual 
interpretations of quantum mechanics it is possible to say that 
complementary physical qualities of a quantum system have at every 
instant of time definite values (Popper, 1982). This claim can be 
shown to have paradoxical consequences and is indeed forbidden by 
Rule \ref{rle1}. \\ 

\section{Generalization of the Consistent Histories 
Approach}

\subsection{Motivation} 
One drawback of the consistent histories formalism 
in its standard formulation is 
that the possible physical qualities of a physical system at any 
particular instant of time are restricted to ordinary physical 
qualities represented by projection operators. \\
The introduction of histories is usually motivated by saying that 
everything that can meaningfully be said about a quantum 
mechanical system can be expressed in terms of time sequences of 
one-time propositions about the system. As discussed above, in the 
usual formulation of quantum mechanics one-time propositions are 
identified with one-time physical qualities, which are 
represented by projection operators, so that histories 
are thought of as being correctly represented by time sequences of 
projection operators. However, as a matter of fact general 
physical qualities in quantum mechanics have to be represented by 
effects. Projection operators represent only special 
physical qualities in quantum mechanics. \\

Omn\`es asserts that the results of measurement theory 
can be deduced from the consistent history approach. 
Similar assertions can also be found in (Griffiths, 1984). 
In particular, 
he argues that in a series of measurements of some proposition 
(represented by a projection operator or a 
history proposition) the empirical frequencies of the results 
are approximately given by the probabilities associated with that 
proposition. 
However, Omn\`es restricts his discussion to a very limited class 
of measurement situations, see Chapter 8 in (Omn\`es, 1994). 
Firstly, he considers only measurements of ordinary observables 
with discrete eigenvalue spectrum and 
secondly only unitary measurement 
interactions which transform eigenstates of the measured 
observable into eigenstates of the same eigenvalue. In the 
language of (Busch et al., 1991) Omn\`es considers only normal 
unitary von Neumann (pre-)measurements. However, the modern 
quantum theory of measurement covers much more general measurement 
situations, specifically measurements of generalized observables, 
see (Busch et al., 1991). 

{\small 
\begin{nt} It follows from a theorem 
by Ozawa (1984) that as a matter of principle the consistent 
history approach in its standard formulation is unable to cover 
measurements of continuous observables. The restriction of the 
discussion of measurement theory to ordinary discrete observables 
in (Omn\`es, 1994) is therefore not only a matter of convenience. 
Continuous observables can only be dealt with after introducing a 
discrete {\sf reading scale}. That is, continuous observables have 
to be replaced by discrete coarse-grained observables. However, 
such a procedure does in general manifestly destroy invariance 
properties of the continuous variables. \end{nt}} 
The interaction between the measurement apparatus and the 
measuring object in real measurements usually takes place during a 
finite time interval and cannot be associated with a fixed point 
of time. Further, many 
measurements have only a finite precision (`unsharp 
measurements') and cannot be considered 
to be measurements of physical qualities represented by projection 
operators or to be sequences of measurements of physical qualities 
represented by history propositions. As discussed by Ludwig 
(1983), Davies (1976), and by Busch, Lahti and Mittelstaedt (1991) 
the physical quality measured in general measurement 
situations has to be represented by some effect operator. 
However, the above formulated universal logical Rule \ref{rle1} 
forbids to make predictions for or even to talk about results from 
such unsharp measurements and also forbids to draw any conclusions 
from them. This state of affairs is clearly 
unsatisfactory. Thus we feel that 
Omn\`es' logical rule has to be extended to cover also more 
realistic measurement situations. \\
The question arises whether the logical 
interpretation can be generalized such that generalized physical 
qualities can be dealt with and such that the more general 
measurement situations can be described by the consistent history 
approach to quantum mechanics. \\ 

Since projection valued measures 
represent observables which can be measured ideally and repeatedly 
(at least in principle) and since effect valued 
measures represent observables which in general cannot be measured 
ideally and repeatedly, one might be tempted to reject this 
argument and to argue apologetically that in the consistent 
history approach a quantum system 
is described by (and only by) the set of its `properties' (which 
are usually identified with projection operators) which are 
asserted to be the fundamental ingredients in the description of 
quantum systems. According to this line of thought other 
(unsharp) observables are not fundamental and 
need thus not to be taken into account. However, we have 
already argued against this attitude in Section 2 and will not 
repeat our arguments here. It has 
been pointed out by Busch et al.~(1989)  that unsharp 
observables are {\em \sf `far from being mere imperfections'} and 
that this point of view {\em \sf `amounts to a severe restriction 
of the measurement theoretic possibilities of quantum mechanics. 
For instance one could not interpret the Stern-Gerlach experiment 
as a measurement of a spin observable at all.'} Here of course a 
{\em real} Stern-Gerlach experiment is meant and not one of its 
idealized textbook versions. The reader is 
referred to the lucid monograph by Busch, Grabowski and Lahti 
(1995). 
\\ One may also argue that Naimark's theorem (as stated, e.g., in 
(Busch et al., 1995)) implies that only PV measures need to be 
taken into 
account since POV measures can be replaced by PV measures on a 
larger Hilbert space $\widetilde{\HH}$. [Naimark's theorem states 
that every maximal 
symmetric operator on a Hilbert space $\HH$ is the restriction (to 
$\HH$) of some self-adjoint operator on a larger Hilbert space 
$\widetilde{{\HH}} \supset \HH$.] However, the Hilbert space 
$\widetilde{{\HH}}$ has in general no direct physical 
interpretation, and even if it has, then it typically represents 
the environment of the considered system or a measuring apparatus. 
In any case the PV measures on the Hilbert space 
$\widetilde{{\HH}}$ can in general not be interpreted as 
describing solely the system under consideration and the 
projection operators in the range of the PV measures cannot in 
general be thought of as representing properties of the system 
[e.g., when $\widetilde{{\HH}}$ describes a measuring situation on 
$\HH$, then typically the PV measures arising from POV measures on 
$\HH$ represent pointer observables of the measuring apparatus]. 
\\ 
Therefore POV measures have to be considered as the observables in 
the theory and the question 
of the status of generalized observables and of the corresponding 
generalized physical qualities in the consistent history approach 
cannot be avoided. \\ 

{\sf In summary of the above discussion we conclude that 
a generalization of the standard consistent histories approach is 
needed for the following reasons: Firstly, the notion of 
observable in the standard consistent histories scheme is 
restricted to the class of ordinary observables and secondly, it 
is not possible to describe realistic measurements in the language 
provided by the standard consistent histories theory.}
It is the target of this section to generalize the consistent 
histories approach appropriately and thereby get rid of these 
drawbacks. In the first part of this section we construct the 
generalization of the consistent histories theory and of the 
logical interpretation for nonrelativistic quantum mechanics. At 
the end of this section we study the consequences of our results 
for the structure of more general temporal history theories in the 
spirit of 
(Isham, 1994). In Section 5 below we will furthermore briefly 
argue that our generalized history approach incorporates in a 
natural way histories of `quasi-projectors' and that we can 
dispose of using approximate consistency conditions. \\
\subsection{The Space of Effect Histories} 
As we have discussed above the equivalence classes of physical 
qualities of a system are 
the objects in the theory which are interpreted to represent 
possible events of physical reality. Thus in order to 
make pre- and retrodiction one has to study  {\em effect 
histories}, i.e., sequences of effects on $\HH$. 
\begin{de} \label{b2} A {\sc homogeneous effect 
history} is a map $u : \RR \to {\frak E}(\HH), t \mapsto u_t$.
The {\sc support of} $u$ is given by ${\frak s}(u) := \{t \in \RR 
\mid u_t \neq 1 \}$. If ${\frak s}(u)$ is 
finite, countable or uncountable, then we say that $u$ is a {\sc 
finite, countable} or {\sc uncountable effect history} 
respectively. The space of all homogeneous effect 
histories will be denoted by ${\Bbb E}(\HH)$, the space of all 
finite homogeneous effect histories by ${\Bbb E}_{fin}(\HH)$ and 
the space of all finite homogeneous effect histories with support 
$S$ by ${\Bbb E}_S(\HH)$. All homogeneous effect histories for 
which there exists at least one $t \in \RR$ such that $u_t = 0 $ 
are collectively denoted by $0$, slightly abusing the notation. 
\end{de} 

\begin{rem} Let $G_r$ be a finite set with $r$ elements and 
supplied with two binary operations 
(free meet and free join) which will be denoted by 
$\wedge_{fr}$ and $\vee_{fr}$ respectively.
{\sc Finite polynomials} in $G_r$ can be built from elements of 
$G_r$ by at most finitely many applications of $\wedge_{fr}$ and 
$\vee_{fr}$. On the space of all finite polynomials in $G_r$ a 
{\sf congruence relation} is defined by imposing the algebraic 
identities valid in every lattice, see (Birkhoff, 1967). The {\sc 
free lattice ${\cal L}(G_r)$ 
generated by} $G_r$ is the quotient space of the space of all 
finite polynomials in $G_r$ by this congruence relation. 
Now let $G_{\aleph}$ be a set with $\aleph$ elements, $\aleph$ any 
cardinal number. The free lattice ${\cal L}(G_{\aleph})$ 
generated by $G_{\aleph}$ can be constructed as follows: for any 
finite subset $T \subset 
G_{\aleph}$ one has constructed ${\cal L}(T)$. Let $T$ and $S$ be 
two finite subsets of $G_{\aleph}$ satisfying $T \subset S$, then 
the canonical embedding $\eta_{TS}: T \to S$ can be extended to a 
monomorphism of free lattices $\bar{\eta}_{TS} : 
{\cal L}(T) \to {\cal L}(S)$ (Birkhoff, 1967, Theorem VI.16). 
Define \[ \widetilde{\cal L} := \bigcup_{T \subset G_{\aleph} 
\atop T \hspace{0.2em} finite} {\cal L}(T). \] Two elements $g_1 
\in {\cal L}(T_1)$ and 
$g_2 \in {\cal L}(T_2)$ ($T_1$ and $T_2$ finite) are {\sc 
equivalent} if there is a $T_{12} \subset G_{\aleph}$ such that 
$T_1 \subset T_{12}$ 
and $T_2 \subset T_{12}$ and $\bar{\eta}_{T_1 T_{12}} (g_1) = 
\bar{\eta}_{T_2 T_{12}} (g_2)$. The {\sc free lattice 
${\cal L}(G_{\aleph})$ generated by $G_{\aleph}$} is now the 
quotient space of $\widetilde{\cal L}$ modulo this equivalence 
relation. It is easy to show that the lattice operations on ${\cal 
L}(T)$ induce the structure of a lattice on ${\cal 
L}(G_{\aleph})$. The lattice operations on ${\cal L}(G_{\aleph})$ 
will be denoted by $\wedge_{\cal L}$ and $\vee_{\cal L}$. 
More information about free lattices and free Boolean algebras can 
be found, e.g., in the monograph by Birkhoff (1967). \end{rem}

It is not clear at all how to 
define the meet and the join of two effect histories. Therefore 
the notion of inhomogeneous effect history cannot be defined in a 
natural way compatible with the partial ordering on ${\frak 
E}(\HH)$. But we can define 
\begin{de} \label{b4} 
The free lattice generated by 
${\Bbb E}_{fin}(\HH)$ will be denoted by ${\cal L}\left({\Bbb 
E}_{fin}(\HH) \right)$. 
The meet and 
the join operations in ${\cal L} \left({\Bbb 
E}_{fin}(\HH) \right)$ will be denoted by $\wedge_{\cal L}$ and 
$\vee_{\cal L}$ respectively. \end{de}

\begin{rem} \label{cgr} We can define a partial order on ${\Bbb 
E}_{fin}(\HH)$. For $u_1, u_2 \in {\Bbb E}_{fin}(\HH)$ we set $u_1 
\leq u_2$ if $(u_1)_t \leq (u_2)_t$ for all $t \in \RR$ and say 
that $u_2$ is {\sc coarser} than $u_1$. This 
partial ordering induces partially defined meet and join 
operations (denoted by $\wedge$ and $\vee$) on ${\Bbb 
E}_{fin}(\HH)$. Every finite 
polynomial in ${\cal L} \left({\Bbb E}_{fin}(\HH)\right)$ can be 
transformed in {\sc equivalent} polynomials by inserting the 
identifications $u_1 \vee_{\cal L} u_2 = 
u_1 \vee u_2$ and $u_1 \wedge_{\cal L} u_2 = 
u_1 \wedge u_2$ whenever the right hand sides are well-defined in 
${\Bbb E}_{fin}(\HH)$. This defines an equivalence relation 
$\sim_{\Bbb E}$ on ${\cal L} \left({\Bbb E}_{fin}(\HH) \right)$. 
It is clear that the physically interesting 
objects are the $\sim_{\Bbb E}$-equivalence classes in the 
quotient space ${{\cal L} \left({\Bbb E}_{fin}(\HH) 
\right)}/{\sim_{\Bbb E}}$. 
By construction $\sim_{\Bbb E}$ is a {\sf 
congruence relation} on ${\cal L} \left({\Bbb E}_{fin}(\HH) 
\right)$. (For the notion of congruence relation see, e.g., 
(Birkhoff, 1967; Skornjakov, 1977).) 
The $\sim_{\Bbb E}$-congruence 
class of $u \in {\Bbb E}_{fin}(\HH)$ will be denoted by 
$\XX(u)$. For $u \in {\frak A} \subset 
{\cal L} \left({\Bbb E}_{fin}(\HH) \right)$ we write $\XX^{\frak 
A}(u) := \{ u' 
\in {\frak A} {\mid} u' \sim_{\Bbb E} u \}$ and $\XX({\frak A}) := 
\{ \XX(a) {\mid} a \in {\frak A} \}$. 
\end{rem}
\begin{de} \label{IH} The quotient space ${{\cal L} \left({\Bbb 
E}_{fin}(\HH) \right)}/{\sim_{\Bbb E}}$ is called {\sc the 
space of finite inhomogeneous effect histories} and will be 
denoted by $\widehat{\Bbb E}_{fin}(\HH)$. A finite inhomogeneous 
effect history induces a map $\hat{u} : \RR \to {{\cal L} 
\left({{\frak E}}(\HH) \right)}/{\sim_{\Bbb E}}, t \mapsto 
\XX(u)_t$, but not vice versa. 
The {\sc support of} 
$\hat{u}$ is given by ${\frak s}(\hat{u}) := 
\{t \in \RR \mid \hat{u}_t \neq 1 \}$. The space 
$\widehat{{\Bbb E}}(\HH)$ of {\sc general inhomogeneous effect 
histories} and the space $\widehat{{\Bbb E}}_S(\HH)$ of {\sc 
inhomogeneous effect histories with finite support} $S$ are 
defined analogously. \end{de} 

For homogeneous finite effect histories $u$ we define 
the class operator by  \begin{eqnarray} \label{cop}
C_{t_0}(u) & := & U(t_0,t_n) \sqrt{u_{t_n}} 
U(t_n,t_{n-1}) \sqrt{u_{t_{n-1}}} ... U(t_2,t_1) 
\sqrt{u_{t_1}} U(t_1,t_0) \\ & = & U(t_0,t_i(u)) 
\sqrt{u_{t_n}}(t_n) \sqrt{u_{t_{n-1}}}(t_{n-1}) ... 
\sqrt{u_{t_1}}(t_1) U(t_i(u),t_0), \end{eqnarray} where we 
have defined the Heisenberg picture operators \[ 
\sqrt{u_{t_k}}(t_k) := U(t_k,t_i(u))^{\dagger} 
\sqrt{u_{t_k}} U(t_k,t_i(u)) \] with respect to the initial 
time $t_i(u)$ of $u$. \\ 

For every pair 
$u$ and $v$ of finite homogeneous effect histories we define the 
{\sc decoherence weight} of $u$ and $v$ by \begin{equation} 
d_{\varrho} (u,v) := {\tr} \left(C_{t_0}(u) \varrho(t_0) 
C_{t_0}(v)^{\dagger} \right). \end{equation} The functional 
$d_{\varrho}: 
{\Bbb E}_{fin}(\HH) \times {\Bbb E}_{fin}(\HH) 
\to \CC, (u,v) \mapsto d_{\varrho}(u,v)$ will be called the {\sc 
decoherence functional associated with the state} $\varrho$. 
There immediately arises a serious difficulty with this 
decoherence functional. At first sight it seems difficult (if not 
impossible) to construct a natural mathematical structure on the 
space of effect histories such that the decoherence functional is 
additive in both arguments. Without this structure a consistency 
condition generalizing Equation \ref{md} cannot even be formulated 
and an interpretation of $d_{\varrho}(u,u)$ as probability seems 
to be impossible. (We will see below that the \df induces a 
probability functional on some sets of effect histories on which 
the \df is not additive.) These questions are investigated in the 
next subsection. {\small \begin{nt} It is straightforward to 
generalize the notion of {\sf linearly positive history} 
introduced by Goldstein and Page (1995) and to introduce {\sf 
linearly positive effect histories} and the like. However, since 
the physical significance of the Goldstein-Page condition remains 
somewhat elusive, we will not consider it in this work. 
\end{nt}} 

\subsection{Consistent Effect Histories and the 
Generalized Logical Rule of Interpretation} 
Isham (1994) has studied the logico-algebraic structure of 
the standard consistent histories approach and has discussed 
generalizations of this structure as models for more general 
history theories which may have applications to quantum space-time 
theories. It is the aim of this subsection to discuss this line of 
thought from the point of view of our generalized histories 
involving 
effects and to use our results to generalize the logical 
interpretation to the present framework. We start with some 
important definitions
\begin{de} A {\sc difference poset} or {\sc D-poset} is a 
partially ordered set $D$ with greatest element 1 and with a 
partial binary operation $\ominus : D_2 \to D$, where $D_2 \subset 
D \times D$, such that 
\begin{itemize} \item $b \ominus a$ is defined if and only if $a 
\leq b$ for all $a,b \in D$, \item $b \ominus a \leq b$ for all $a 
\leq b$, \item $b \ominus ( b \ominus a ) = a$, for all $a \leq 
b$, \item $a \leq b \leq c \Rightarrow c \ominus b \leq c \ominus 
a$ and $(c \ominus a) \ominus (c \ominus b) = b \ominus a$.
\end{itemize} \end{de}

Difference posets have been introduced by K\^opka and Chovanec 
(1994) and have been further studied in (K\^opka, 1992; 
Dvure\v{c}enskij and Pulmannov\'a 1994a-c; Dvure\v{c}enskij 1995; 
Pulmannov\'a, 1995). 
\begin{de} A set $D$ with two special elements $0,1 \in D$ 
supplied with a partially defined associative and commutative 
operation $\oplus : D_2' \to D$, where $D_2' \subset D \times D$, 
is called an {\sc effect 
algebra} if \begin{itemize} \item For every $a \in D$ there exists 
a unique $a' \in D$ such that $a \oplus a'$ is defined and $a 
\oplus a' =1$, 
\item If $1 \oplus b$ is defined, then $b=0$, for all $b \in D$.
\end{itemize} 
An effect algebra $D$ is called an {\sc orthoalgebra} if 
furthermore \begin{itemize} \item If $b \oplus b$ is defined, then 
$b=0$, for all $b \in D$. \end{itemize} \end{de}

Effect algebras have been introduced by Foulis and Bennett (1994). 
Whenever $a \oplus b$ is well-defined for $a,b 
\in D$, then we write $a \perp b$. 
Let $(D, \ominus)$ be a D-poset. Define
\[ a \oplus b := 1 \ominus ( (1 \ominus a) \ominus b), \]
whenever the right hand side is well-defined. Then $\oplus$ is a 
well-defined partial binary operation on $D$ and $(D,\oplus)$ is 
an effect algebra. Conversely, let $(D, \oplus)$ be an effect 
algebra. Define 
\[ b \ominus a := (a \oplus b')', \] whenever the right hand side 
is well-defined. Then $\ominus$ is a well-defined partially binary 
operation on $D$. Further, define $a \leq b$ for $a,b \in D$ if 
there exists $c \in D$ such that $c \perp a$ and $a \oplus c =b$. 
Then $(D,\ominus)$ is a D-poset.
Therefore the notions of D-poset and effect algebra are 
equivalent and we will use both terms synonymously in the 
following. 
\begin{de} A finite subset $\{ a_1, a_2, ..., a_n \}$ of a D-poset 
$(D, \oplus)$ is said to be $\bigoplus$-{\sc orthogonal} if 
$\bigoplus_{i=1}^n a_i := a_1 \oplus a_2 \oplus \cdot \cdot \cdot 
\oplus a_n$ can be defined recursively. In this case 
$\bigoplus_{i=1}^n a_i$ is called the $\bigoplus$-{\sc sum} of 
$\{ a_1, a_2, ..., a_n \}$. In particular, for every $n \in \NN$, 
we define $a n := n a := \bigoplus_{i=1}^n a$, for all $a \in D$ 
for which the right hand side is well-defined. A 
$\bigoplus$-{\sc orthogonal} subset $\{ a_1, a_2, ..., a_n \}$ of 
a D-poset $D$ is said to be {\sc complete} if $\bigoplus_{i=1}^n 
a_i = 1$. The family of all $\bigoplus$-sums of subsets of $\{ 
a_1, a_2, ..., a_n \}$ will be denoted by $\bigoplus \{ a_1, a_2, 
..., a_n \}$. \\ Furthermore, let $D_0 \subset D$, then the set of 
all well-defined finite $\bigoplus$-sums of elements of $D_0$ will 
be denoted by $\bigoplus D_0$. \end{de}

Since $\oplus$ is commutative and associative, $\bigoplus_{i=1}^n 
a_i$ is indeed well-defined. 
\begin{rem} If $\{ a_1, a_2, ..., a_n \}$ is a finite, complete, 
$\bigoplus$-orthogonal subset of a D-poset $(D, \oplus)$, then 
$\left( \bigoplus \{ a_1, a_2, ..., a_n \}, \oplus \right)$ is 
itself a D-poset. \end{rem}
\begin{rem} The set ${\frak E}(\HH)$ of all effect operators on a 
Hilbert space $\HH$ ${\mbox{\em [}}$with the scalar product 
denoted by $(\cdot, \cdot){\mbox{\em ]}}$ can be organized into a 
D-poset by defining a partial 
ordering and a partial binary operation $\ominus$ on $D$ by
$A \leq B$ if $(Ax,x) \leq (Bx,x)$ for all $x \in \HH$ and $C = B 
\ominus A$ if $(Bx,x) - (Ax,x) = (Cx,x)$ for all $x \in 
\HH$. \end{rem} \begin{de} Let $(D, \oplus)$ be a D-poset. A {\sc 
probability measure on} $D$ is a map $p : D \to \RR^+$ satisfying 
$p(1) = 1$ and $p(a \oplus b) = p(a) + p(b)$, whenever $a \oplus 
b$ is well-defined. \end{de}

We next summarize the general axioms for a 
generalized quantum theory based on histories as stated by Isham 
(1994).
According to Isham the algebraic structure underlying the 
generalized quantum mechanics of histories consists of the 
following ingredients:
\begin{enumerate}
\item The space of history propositions $\UP$. \begin{itemize} 
\item Isham and Linden (1994) suggest that the minimal 
useful mathematical structure on $\UP$ is that of an orthoalgebra. 
One may also consider the case that 
$\UP$ is an orthocomplemented lattice or has an even stronger 
structure (e.g., a complete orthocomplemented lattice). 
(Pulmannov\'a (1995) proposed that the space of all histories 
in a generalized quantum theory may admit the structure of a 
D-poset. In the present work a physical justification of this 
proposal is given.) 
\end{itemize}
\item The space of decoherence functionals. \\
A decoherence functional is a map $d : \UP \times \UP \to \CC$ 
which satisfies for all $\alpha, \alpha',\beta \in \UP$
\begin{itemize} \item $d(\alpha,\alpha) \in \RR$ and 
$d(\alpha,\alpha) 
\geq 0$. \item $d(\alpha,\beta) = d(\beta,\alpha)^*$. \item 
$d(1,1) =1$. \item $d(0,\alpha) =0$, for all 
$\alpha$. \item \begin{itemize} \item If $\UP$ is an orthoalgebra 
(or more generally an effect algebra), then $d(\alpha_1 \oplus 
\alpha_2, \beta) = d(\alpha_1,\beta) + d(\alpha_2,\beta)$ for all 
$\alpha_1, \alpha_2, \beta \in \UP$ with $\alpha_1 \perp 
\alpha_2$. \item If $\UP$ is a lattice, then $d(\alpha_1 \vee 
\alpha_2, \beta) = 
d(\alpha_1,\beta) + d(\alpha_2,\beta) - d(\alpha_1 \wedge 
\alpha_2,\beta)$ for all $\alpha_1, \alpha_2, \beta \in \UP$. 
\end{itemize} \end{itemize} 
\item The physical interpretation of the above axioms is the same 
as in the case of the standard consistent history formulation. One 
can define (pre)consistent sets of histories and interpret 
$d(\alpha, \alpha)$ as the probability of the history $\alpha$. \\ 
\\ 
Besides the structures of the space of histories and the space of 
decoherence functionals which are believed to be present also in 
more general history theories than nonrelativistic quantum 
mechanics, Isham (1994) also describes structures arising in the 
history formulation of nonrelativistic quantum mechanics which may 
be an artefact of that theory and may be meaningless in more 
general quantum history theories. 
\item The space $\U$ of {\em history filters} or {\em homogeneous 
histories}. \begin{itemize} \item $\U$ is a partial semigroup with 
composition law $\circ$. It satisfies $\alpha \circ 1 = \alpha$ 
and $\alpha \circ 0 = 0$ for all $\alpha \in \U$. If $\alpha \circ 
\beta$ is defined, then $\alpha \circ \beta = \alpha \wedge 
\beta$. \item $\U$ is a partially ordered set 
with a unit $1$ and a null history $0$. The partial order
is denoted by $\leq$. \item On $\U$ an operation of meet, denoted 
by $\wedge$, is defined which satisfies $1 \wedge \alpha = \alpha$ 
and $0 \wedge \alpha = 0$ for all $\alpha \in \U$. \item 
There exists an embedding $\tau : \U \to \UP$, i.e., 
$\tau(U) \subset \UP$. \item $\UP$ can be 
generated from $\U$ by the application of a finite or countably 
infinite number of the algebraic operations defined on $\frak U$. 
\end{itemize}
Clearly, the axioms for $\cal U$ have to be modified when taking 
into account homogeneous effect histories. {We will 
return to this point below.} The definition of a semigroup will 
also be given below. \item The space of temporal supports $\SSS$. 
\begin{itemize} \item $\SSS$ is a partial semigroup with unit. 
\item There exists a semigroup homomorphism $\kappa : \U \to \SSS$ 
such that $\kappa(1) = \kappa(0) =1 \in \SSS$. \end{itemize} 
$\SSS$ may contain only one element $\SSS = \{ 1 \}$.
\end{enumerate}

In the standard history approach to quantum mechanics we 
have the following identifications: $\SSS$ is the 
space of all finite subsets of $\RR$ and $\U$ $:= \fin$.
Isham's axioms, as stated above, are not the most general 
structure possible. There is great freedom to add further axioms 
to the list or to remove some (note that we have already omitted 
some of Isham's axioms).
As discussed in Section 2, finite 
homogeneous histories can be identified with projection operators 
on $\ten{S}$ in a natural way. One aim of this subsection is to 
examine to what extent Isham's results can be extended to our 
generalized effect histories and to explore its consequences for 
more general history theories. To this end the concept of the 
tensor product of effect algebras is of utmost importance.
The tensor product of a pair of D-posets is defined by a universal 
mapping property, cf.~(Dvure\v{c}enskij, 1995).
\begin{de} Let $D_1$, $D_2$, ..., $D_n$ and $L$ denote D-posets. 
\begin{itemize} \item A mapping
$\phi : D_1 \to L$ is said to be a {\sc morphism} if (i) 
$\phi(1)=1$, (ii) $a \perp b$ implies 
$\phi(a) \perp \phi(b)$ and $\phi(a \oplus b) = \phi(a) \oplus 
\phi(b)$, for all $a,b \in D_1$. \item A mapping
$\phi : D_1 \to L$ is said to be a {\sc monomorphism} if (i) 
$\phi$ is a morphism and (ii) $\phi(a) \perp 
\phi(b)$ implies $a \perp b$, for all $a,b \in D_1$. \item A 
mapping $\phi : D_1 \to L$ 
is said to be an {\sc isomorphism} if $\phi$ is a surjective 
monomorphism. \item A mapping $\theta : D_1 \times D_2 \to L$ is 
said to be a {\sc semi-bimorphism} if \begin{itemize} \item $a,b 
\in D_1$ with $a \perp b$ imply $\theta(a,c) \perp \theta(b,c)$ 
and $\theta(a \oplus b, c) = \theta(a,c) \oplus \theta(b,c)$ for 
all $c \in D_2$; \item $c,d \in D_2$ with $c \perp d$ imply 
$\theta(a,c) \perp \theta(a,d)$ and $\theta(a, c \oplus d) = 
\theta(a,c) \oplus \theta(a,d)$ for all $a \in D_1$; \end{itemize} 
\item A mapping $\theta : D_1 \times D_2 \to L$ is 
said to be a {\sc bimorphism} if it is a semi-bimorphism and 
$\theta(1,1) =1$. \item A mapping $\theta : D_1 \times D_2 \times 
\cdot \cdot \cdot \times D_n \to L$ is 
said to be a {\sc multimorphism} if it is a semi-bimorphism in 
every pair of its arguments and if $\theta(1,...,1) = 1.$ 
\end{itemize} \end{de}
\begin{de} Let $D_1, ..., D_n$ denote a family of $n$ D-posets. We 
say that a 
pair $({\frak D}, \Theta)$ consisting of a D-poset $\frak D$ and a 
multimorphism $\Theta : D_1 \times \cdot \cdot \cdot \times D_n 
\to \frak D$ is a {\sc tensor 
product} of the family $D_1, ..., D_n$ if the following conditions 
are satisfied \begin{itemize} \item If $L$ is a D-poset and 
$\theta_L : D_1 \times \cdot \cdot \cdot \times D_n \to L$ a 
multimorphism, then there exists a morphism 
$\beta : {\frak D} \to L$ such that $\theta_L = \beta \circ 
\Theta$; \item Every element of $\frak D$ is a finite orthogonal 
sum of elements of the form $\Theta(a_1,..., a_n)$ with $a_i \in 
D_i$, for all $i$. \end{itemize} We also write ${\frak D} = 
\bigotimes_{i=1}^n D_i = D_1 \otimes \cdot \cdot \cdot \otimes 
D_n$ and $\Theta = \otimes$. \end{de} 

Dvure\v{c}enskij (1995) has proven that two D-posets $D_1$ 
and $D_2$ admit a tensor product if and only if there is a 
difference poset $L$ for which there is a bimorphism $\theta_L : 
D_1 \times D_2 \to L$. The definition of the tensor product of 
D-posets and Dvure\v{c}enskij's theorem can straightforwardly be 
extended to any finite number of D-posets, see (Pulmannov\'a, 
1995). \\

Let in the sequel $\Bbb N$ denote the set of positive integers 
including 0. We adopt the following definitions \\ \\
{\bf Definition} {\em A set $\Omega$ with an operation $\top: 
\Omega \times \Omega \to \Omega$ is called a {\sc semigroup} if 
the associative law is satisfied.} \\ \\
{\bf Definition} {\em A triple 
$(\Omega, \top, \odot)$ is called a {\sc semiring} if 
\begin{itemize} \item $(\Omega, \top)$ is an abelian semigroup 
with neutral element, \item $(\Omega, \odot)$ is a semigroup with 
unit $1$, \item the distributive laws are satisfied. 
\end{itemize}} 
\noindent {\bf Example}: Let $\alpha \in \RR^+\backslash \{0\}$ 
and define ${\Bbb N}_{\alpha} := {\{ n^{\alpha}}{\mid}{n \in {\Bbb 
N}\}}$. A semigroup operation $+_{\alpha} : {\Bbb N}_{\alpha} 
\times {\Bbb N}_{\alpha} \to {\Bbb N}_{\alpha}$ is defined by $u 
+_{\alpha} v := \left( 
u^{1/{\alpha}} + v^{1/{\alpha}} \right)^{\alpha}$. Further, define
${\odot}_{\alpha} : {\Bbb N}_{\alpha} \times {\Bbb N}_{\alpha} \to 
{\Bbb N}_{\alpha}$ by $u {\odot}_{\alpha} v := \left( 
u^{1/{\alpha}} \cdot v^{1/{\alpha}} \right)^{\alpha} = u \cdot v$. 
Then the triple $({\Bbb N}_{\alpha}, +_{\alpha}, 
{\odot}_{\alpha})$ is a semiring. For all $\alpha \in 
\RR^+\backslash\{0\}$ there is a semiring isomorphism 
$\iota_{\alpha} : {\Bbb N}_{\alpha} \hookrightarrow {\Bbb N}, 
\iota_{\alpha}(u) := u^{1/{\alpha}}$. We call $({\Bbb N}_{\alpha}, 
+_{\alpha}, {\odot}_{\alpha})$ the semiring of 
{\sc $\alpha$-scaled natural numbers.} \\ 

\noindent {\bf Definition} {\em Let $(M, +)$ be an abelian 
semigroup and 
$(\Omega, +, \odot)$ be a semiring and $\cdot : \Omega \times M 
\to M$ be a `scalar multiplication.' Then the system $(M, +; 
\Omega, \odot; \cdot)$ is called a {\sc semimodule over the 
semiring} $(\Omega, +, \odot)$ if for all $\omega, \omega_1, 
\omega_2 \in \Omega$ and $m, m_1, m_2 \in M$ \begin{itemize} \item 
$\omega \cdot ( m_1 + m_2) = \omega \cdot m_1 + \omega m_2,$ \item 
$(\omega_1 + \omega_2) \cdot m = \omega_1 \cdot m + \omega_2 \cdot 
m,$ \item $(\omega_1 \odot \omega_2) \cdot m = \omega_1 \cdot 
(\omega_2 \cdot m)$, \item $1 \cdot m=m$. \end{itemize}} 

Consider now two complex Hilbert spaces $\HH_1$ and $\HH_2$ and 
the corresponding sets of effects ${\frak E}(\HH_1)$ and ${\frak 
E}(\HH_2)$. An immediate consequence of Dvure\v{c}enskij's theorem 
is that the tensor product of ${\frak E}(\HH_1)$ and ${\frak 
E}(\HH_2)$ exists. To gain insight into the structure of the space 
of effect histories it is necessary to explicitly construct the 
tensor product of ${\frak E}(\HH_1)$ and ${\frak E}(\HH_2)$. 
To this end we regard $\HH_1$ as right semimodule over $\Bbb N$ 
and 
$\HH_2$ as left semimodule over $\Bbb N$ and denote by $\HH_1 
\otimes_{\Bbb N} 
\HH_2$ the tensor product of the $\Bbb N$-semimodules $\HH_1$ and 
$\HH_2$ in the category of semimodules over $\Bbb N$ , that is, 
$\HH_1 
\otimes_{\Bbb N} \HH_2$ is the free abelian semigroup generated by 
elements of the form $\psi \otimes_{\Bbb N} \varphi$ with $\psi 
\in \HH_1$ and $\varphi \in \HH_2$ subject to the relations
\begin{enumerate} \item $(\psi_1 + \psi_2) \otimes_{\Bbb N} 
\varphi = (\psi_1 \otimes_{\Bbb N} \varphi) + (\psi_2 
\otimes_{\Bbb N} \varphi),$ for all $\psi_1, \psi_2 \in \HH_1$ and 
$\varphi \in \HH_2$; \item  $\psi \otimes_{\Bbb N} (\varphi_1 + 
\varphi_2) = (\psi \otimes_{\Bbb N} \varphi_1) + (\psi 
\otimes_{\Bbb N} \varphi_2),$ for all $\psi \in \HH_1$ and 
$\varphi_1, \varphi_2 \in \HH_2$; \item $n (\psi \otimes_{\Bbb N} 
\varphi) \equiv (\psi n) \otimes_{\Bbb N} 
\varphi = \psi \otimes_{\Bbb N} (n \varphi),$ for all $\psi \in 
\HH_1, \varphi \in \HH_2$ and $n \in \Bbb N$. \end{enumerate}
To every $E_1 \in {\frak E}(\HH_1)$ and $E_2 \in {\frak E}(\HH_2)$ 
one can define a map $\HH_1 \otimes_{\Bbb N} \HH_2 
\to \HH_1 \otimes_{\Bbb N} \HH_2$ by \[ E_1 \otimes_{\Bbb N} E_2 
(\psi \otimes_{\Bbb N} \varphi) := E_1 \psi \otimes_{\Bbb N} E_2 
\varphi. \]
${\frak E}(\HH_1)$ and ${\frak E}(\HH_2)$ can also be regarded as 
right and left $\Bbb N$-semimodules respectively. 
We can thus consider the tensor product ${\frak E}(\HH_1) 
\otimes_{\Bbb N} {\frak E}(\HH_2)$ of ${\frak E}(\HH_1)$ and 
${\frak E}(\HH_2)$ in the category of $\Bbb N$-semimodules.
${\frak E}(\HH_1) \otimes_{\Bbb N} {\frak E}(\HH_2)$ is the free 
abelian semigroup generated by elements of the form $E_1 
\otimes_{\Bbb N} E_2$, where $E_1 \in {\frak E}(\HH_1)$ and $E_2 
\in {\frak E}(\HH_2)$ subject to the relations \begin{enumerate} 
\item $n(E_1 \otimes_{\Bbb N} E_2) \equiv (E_1 n) 
\otimes_{\Bbb N} E_2 = E_1 \otimes_{\Bbb N} (n E_2)$, for all $E_1 
\in {\frak E}(\HH_1), E_2 \in {\frak E}(\HH_2)$ and $n \in \Bbb 
N$; \item $(E_1 \otimes_{\Bbb N} E_2) + (E_1 \otimes_{\Bbb N} 
\tilde{E}_2) = E_1 \otimes_{\Bbb N} (E_2 + \tilde{E}_2),$ for all 
$E_1 \in {\frak E}(\HH_1)$ and $E_2, \tilde{E}_2 \in {\frak 
E}(\HH_2)$; \item $(E_1 \otimes_{\Bbb N} E_2) + (\tilde{E}_1 
\otimes_{\Bbb N} E_2) = (E_1 + \tilde{E}_1) \otimes_{\Bbb N} E_2,$ 
for all $E_1, \tilde{E}_1 \in {\frak E}(\HH_1)$ and $E_2 \in 
{\frak E}(\HH_2)$. \end{enumerate}

The scalar products on $\HH_1$ and $\HH_2$ allow to define a 
scalar product on $\HH_1 \otimes_{\Bbb N} \HH_2$ given by 
\begin{eqnarray*} \langle \Phi \mid A \mid \Psi \rangle := \langle 
\phi \mid E_1 
\mid \psi \rangle \langle \omega \mid E_2 \mid \varphi \rangle, 
& & \mbox{ if } \Phi = \phi \otimes_{\Bbb N} \omega, \Psi = \psi 
\otimes_{\Bbb N} \varphi \in \HH_1 \otimes_{\Bbb N} \HH_2 \\ & & 
\mbox{ and } A = E_1 \otimes_{\Bbb N} E_2 \in {\frak E}(\HH_1) 
\otimes_{\Bbb N} {\frak E}(\HH_2). \end{eqnarray*} This 'scalar 
product' is extended by linearity to arbitrary $A \in {\frak 
E}(\HH_1) \otimes_{\Bbb N} {\frak E}(\HH_2)$ and $\Phi, \Psi \in 
\HH_1 \otimes_{\Bbb N} \HH_2$. A partial order on ${\frak 
E}(\HH_1) \otimes_{\Bbb N} {\frak E}(\HH_2)$ can now be defined. 
For $A, B \in {\frak E}(\HH_1) \otimes_{\Bbb N} {\frak E}(\HH_2)$ 
we set $A \leq B$ if $ \langle \Psi \mid A \mid \Psi \rangle \leq 
\langle \Psi \mid B \mid \Psi \rangle$ for all $\Psi \in \HH_1 
\otimes_{\Bbb N} \HH_2$. \\ 
A finite family $\{ E_i \otimes_{\Bbb N} F_i \}_i$ of elements of 
${\frak E}(\HH_1) \otimes_{\Bbb N} {\frak E}(\HH_2)$ is a {\sc 
decomposition} of $1 \otimes_{\Bbb N} 1$ if $\sum_i E_i 
\otimes_{\Bbb N} F_i = 1 \otimes_{\Bbb N} 1$ in ${\frak 
E}(\HH_1) \otimes_{\Bbb N} {\frak E}(\HH_2)$. \\

In order to establish the effect algebra structure on ${\frak 
E}(\HH_1) \otimes_{\Bbb N} {\frak E}(\HH_2)$ we define a partial 
binary operation $\oplus_{\Bbb N}$ on ${\frak E}(\HH_1) 
\otimes_{\Bbb N} {\frak E}(\HH_2)$ by 
\[ A \oplus_{\Bbb N} B := A + B, \mbox{ if } A + B \leq 1, \]
for all $A, B \in {\frak E}(\HH_1) \otimes_{\Bbb N} {\frak 
E}(\HH_2)$. 
However, $({\frak E}(\HH_1) \otimes_{\Bbb N} {\frak E}(\HH_2), 
\oplus_{\Bbb N})$ is not yet the tensor product of ${\frak 
E}(\HH_1)$ and ${\frak E}(\HH_2)$. Indeed, $({\frak E}(\HH_1) 
\otimes_{\Bbb N} {\frak E}(\HH_2), \oplus_{\Bbb N})$ is not even a 
D-poset. \\ 
We next define another partial binary operation $\oplus_{{\frak 
D}}$ on ${\frak E}(\HH_1) \otimes_{\Bbb N} {\frak E}(\HH_2)$. 
We say that a $\oplus_{\frak D}$-sum of the form $(E_1 
\otimes_{\Bbb N} F_1) \oplus_{\frak D} (E_2 \otimes_{\Bbb N} F_2) 
\oplus_{\frak D} \cdot \cdot \cdot \oplus_{\frak D} (E_n 
\otimes_{\Bbb N} F_n)$ of finitely many elements $E_i 
\otimes_{\Bbb N} F_i \in {\frak E}(\HH_1) \otimes_{\Bbb N} {\frak 
E}(\HH_2)$ exists if there exists a decomposition 
$\bigoplus_{{\Bbb N}, i} e_i 
\otimes_{\Bbb N} f_i$ of $1 \otimes_{\Bbb N} 1$ in ${\frak 
E}(\HH_1) \otimes_{\Bbb N} {\frak E}(\HH_2)$ such that every term 
of the form $E_i \otimes_{\Bbb N} F_i$ occurs also in the 
decomposition $\bigoplus_{{\Bbb N}, i \in I} e_i \otimes_{\Bbb N} 
f_i$ of $1 \otimes_{\Bbb N} 1$, i.e., if there for every $i \in 
\{1, ..., n \}$ there exists a $j \in I$ such that $E_i 
\otimes_{\Bbb N} F_i = e_j \otimes_{\Bbb N} f_j$. \\
In this case we also adopt the following way of speaking: we say 
that the $\oplus_{\frak D}$-sum $(E_1 \otimes_{\Bbb N} F_1) 
\oplus_{\frak D} (E_2 \otimes_{\Bbb N} F_2) \oplus_{\frak D} \cdot 
\cdot \cdot \oplus_{\frak D} (E_n \otimes_{\Bbb N} F_n)$ is a {\sc 
$\oplus_{\frak D}$-part of} $1 \otimes_{\Bbb N} 1$. \\
In this case we set \begin{equation} (E_1 \otimes_{\Bbb N} F_1) 
\oplus_{\frak D} (E_2 \otimes_{\Bbb N} F_2) \oplus_{\frak D} \cdot 
\cdot \cdot \oplus_{\frak D} (E_n \otimes_{\Bbb N} F_n) :=(E_1 
\otimes_{\Bbb N} F_1) \oplus_{\Bbb N} (E_2 \otimes_{\Bbb N} F_2) 
\oplus_{\Bbb N} \cdot \cdot \cdot \oplus_{\Bbb N} (E_n 
\otimes_{\Bbb N} F_n). \end{equation} 
We further say that a $\oplus_{\frak D}$-sum $(E_1 \otimes_{\Bbb 
N} F_1) \oplus_{\frak D} (E_2 \otimes_{\Bbb N} F_2) \oplus_{\frak 
D} \cdot \cdot \cdot \oplus_{\frak D} (E_n \otimes_{\Bbb N} F_n)$ 
is a {\sc minimal representation} of its associated 
$\oplus_{\frak D}$-part of $1 \otimes_{\Bbb N} 1$ if the number of 
terms in the sum cannot be reduced by applying the defining 
relations 1., 2.~and 3.~of ${\frak E}(\HH_1) \otimes_{\Bbb N} 
{\frak E}(\HH_2)$ (see above). \\ 
The $\oplus_{\frak D}$-sum $G_{\alpha} \oplus_{\frak D} G_{\beta}$ 
of two arbitrary elements $G_{\alpha}$ and $G_{\beta} \in {\frak 
E}(\HH_1) \otimes_{\Bbb N} {\frak E}(\HH_2)$ is defined if 
$G_{\alpha}$ possesses a well-defined decomposition 
\[ G_{\alpha} = (E_{\alpha_1} 
\otimes_{\Bbb N} F_{\alpha_1}) \oplus_{\frak D} \cdot \cdot \cdot 
\oplus_{\frak D} (E_{\alpha_n} \otimes_{\Bbb N} F_{\alpha_n}), \]
and if $G_{\beta}$ possesses a well-defined decomposition \[ 
G_{\beta} = 
(E_{\beta_1} \otimes_{\Bbb N} F_{\beta_1}) \oplus_{\frak D} \cdot 
\cdot \cdot \oplus_{\frak D} (E_{\beta_m} \otimes_{\Bbb N} 
F_{\beta_m}), \] (i.e., $G_{\alpha}$ and $G_{\beta}$ are both 
well-defined $\oplus_{\frak D}$-parts of $1 \otimes_{\Bbb N} 1$) 
such that \[ (G_{\alpha} 
\oplus_{\frak D} G_{\beta}) := (E_{\alpha_1} \otimes_{\Bbb N} 
F_{\alpha_1}) \oplus_{\frak D} \cdot \cdot \cdot \oplus_{\frak D} 
(E_{\alpha_n} \otimes_{\Bbb N} F_{\alpha_n}) \oplus_{\frak D} 
(E_{\beta_1} \otimes_{\Bbb N} F_{\beta_1}) \oplus_{\frak D} \cdot 
\cdot \cdot \oplus_{\frak D} (E_{\beta_m} \otimes_{\Bbb N} 
F_{\beta_m}) \] is well-defined. It is clear that 
$(G_{\alpha} \oplus_{\frak D} G_{\beta})$ such defined is 
independent of the particular decompositions considered. 
We say that two $\oplus_{\frak D}$-parts of $1 \otimes_{\Bbb N} 
1$, say, $G_{\delta}$ and $G_{\gamma}$ are {\sc equivalent} if for 
all $\oplus_{\frak D}$-parts $G$ of $1 \otimes_{\Bbb N} 1$ the 
following holds: $G \oplus_{\frak D} G_{\delta}$ is well-defined 
and equal to $1 \otimes_{\Bbb N} 1$ if and only if $G 
\oplus_{\frak D} G_{\gamma}$ is well-defined and equal to $1 
\otimes_{\Bbb N} 1$. We also write $G_{\delta} \sim_{\frak D} 
G_{\gamma}$. We write for the $\sim_{\frak D}$-equivalence class 
of $G$ either $[G]$ or (abusing the notation) simply $G$. It is 
clear that the equivalence class of a homogeneous element $E 
\otimes_{\Bbb N} F$ contains only the element $E \otimes_{\Bbb N} 
F$, i.e., $[E \otimes_{\Bbb N} F] = \{ E \otimes_{\Bbb N} F \}$. 
The partial addition $\oplus_{\frak D}$ defined above for 
$\oplus_{\frak D}$-parts of $1 \otimes_{\Bbb N} 1$ induces a 
partial addition on the equivalence classes of 
$\oplus_{\frak D}$-parts of $1 \otimes_{\Bbb N} 1$ (denoted by the 
same symbol) by \[ \left[G_{\delta} \right] \oplus_{\frak D} 
\left[G_{\gamma} \right] := \left[G_{\delta} \oplus_{\frak D} 
G_{\gamma} \right], \] for all $G_{\delta}$ and $G_{\gamma}$ for 
which the right hand side is well-defined. 
It is clear that every equivalence class $[G]$ contains at most 
two elements. Further, if $[G]$ contains more than one element, 
then the complementary class $[G]' := [G']$ of $[G]$ (defined by 
$[G] \oplus_{\frak D} [G]' = [1]$) contains only one element but 
not vice versa. 
For well-defined $\oplus_{\frak D}$-parts $G \in {\frak E}(\HH_1) 
\otimes_{\Bbb N} {\frak E}(\HH_2)$ of $1 \otimes_{\Bbb N} 1$ with 
a minimal representation 
of the form $G = (E_1 \otimes_{\Bbb N} F_1) 
\oplus_{\frak D} (E_2 \otimes_{\Bbb N} F_2) \oplus_{\frak D} \cdot 
\cdot \cdot \oplus_{\frak D} (E_n \otimes_{\Bbb N} F_n)$ there are 
two candidates both of which may serve as complements in case they 
are well-defined, namely \begin{eqnarray*} \bullet \hspace{0.5em}
G^* & := & \left( E_1 \otimes_{\Bbb N} \left( 1-F_1 \right) 
\right) \oplus_{\frak D} \left( E_2 \otimes_{\Bbb N} \left( 1-F_2 
\right) \right) \oplus_{\frak D} \cdot \cdot \cdot \oplus_{\frak 
D} \left( E_n \otimes_{\Bbb N} \left( 1-F_n \right) \right) 
\oplus_{\frak D} \\ & & \oplus_{\frak D} \left( 
\left( 1 - \left( E_1 \oplus E_2 \oplus \cdot \cdot \cdot \oplus 
E_n \right) \right) \otimes_{\Bbb N} 1 \right) \end{eqnarray*} 
\begin{eqnarray*} \bullet \hspace{0.5em} G^{**} & := & \left( 
\left( 1-E_1 
\right) \otimes_{\Bbb N} F_1 \right) \oplus_{\frak D} \left( 
\left( 1-E_2 \right) \otimes_{\Bbb N} F_2 \right) \oplus_{\frak D} 
\cdot \cdot \cdot \oplus_{\frak D} \left( \left( 1-E_n \right) 
\otimes_{\Bbb N} F_n \right) \oplus_{\frak D} \\ & & 
\oplus_{\frak D} \left( 1 \otimes_{\Bbb N} \left( 1 - \left( F_1 
\oplus F_2 \oplus \cdot \cdot \cdot \oplus F_n \right) \right) 
\right). \end{eqnarray*} 
However, if $G$ is well-defined, it is clear that either $G^*$ or 
$G^{**}$ is well-defined. (If they are both well-defined, then 
they are equivalent of course since also $G \oplus_{\frak D} G^*$ 
and $G \oplus_{\frak D} G^{**}$ are both well-defined and equal to 
$1 \otimes_{\Bbb N} 1$.) Define ${\frak E}(\HH_1) 
\widehat{\otimes}_{\frak D} {\frak E}(\HH_2)$ as the effect 
algebra with partial binary operation $\oplus_{\frak D}$ as the 
set consisting of all $\sim_{\frak D}$-equivalence classes of 
well-defined finite $\oplus_{\frak D}$-sums $G$ of elements 
of the form $E_1 \widehat{\otimes}_{\frak D} E_2 := E_1 
\otimes_{\Bbb N} E_2 \in {\frak E}(\HH_1) \otimes_{\Bbb N} {\frak 
E}(\HH_2)$, i.e., consisting of all $\sim_{\frak D}$-equivalence 
classes of $\oplus_{\frak D}$-parts of $1 \otimes_{\Bbb N} 1$, 
subject to the relations \begin{enumerate} \item $n(E_1 
\widehat{\otimes}_{\frak D} E_2) = (n E_1)  
\widehat{\otimes}_{\frak D} E_2 = E_1 \widehat{\otimes}_{\frak D} 
(n E_2)$, for all $E_1 \in 
{\frak E}(\HH_1), E_2 \in {\frak E}(\HH_2)$ and $n \in {\Bbb N},$ 
\\ whenever the expressions are well-defined;  \item $(E_1 
\widehat{\otimes}_{\frak D} E_2) \oplus_{\frak D} (E_1 
\widehat{\otimes}_{\frak D} \tilde{E}_2) = E_1 
\widehat{\otimes}_{\frak D} (E_2 \oplus \tilde{E}_2),$ for all 
$E_1 \in {\frak E}(\HH_1)$ and $E_2, \tilde{E}_2 \in {\frak 
E}(\HH_2)$, \\ whenever the expressions are well-defined; \item 
$(E_1 \widehat{\otimes}_{\frak D} E_2) \oplus_{\frak D} 
(\tilde{E}_1 
\widehat{\otimes}_{\frak D} E_2) = (E_1 \oplus \tilde{E}_1) 
\widehat{\otimes}_{\frak D} E_2,$ 
for all $E_1, \tilde{E}_1 \in {\frak E}(\HH_1)$ and $E_2 \in 
{\frak E}(\HH_2)$, \\ whenever the expressions are well-defined. 
\end{enumerate} 
\begin{theo} The pair $({\frak E}(\HH_1) \widehat{\otimes}_{\frak 
D} {\frak E}(\HH_2), \oplus_{\frak D})$ is the tensor product of 
${\frak E}(\HH_1)$ and ${\frak E}(\HH_2)$. \end{theo}
{\bf Proof:} That every element $G$ of ${\frak E}(\HH_1) 
\widehat{\otimes}_{\frak D} {\frak E}(\HH_2)$ is a finite 
$\oplus_{\frak D}$-orthogonal sum of elements of the form $E_1 
\otimes_{\Bbb N} E_2$ 
with $E_1 \in {\frak E}(\HH_1)$ and $E_2 \in {\frak E}(\HH_2)$
and that to every $G \in {\frak E}(\HH_1) \widehat{\otimes}_{\frak 
D} {\frak E}(\HH_2)$ there is a unique $G^* \in {\frak E}(\HH_1) 
\widehat{\otimes}_{\frak D} {\frak E}(\HH_2)$ such that $G 
\oplus_{\frak D} G^* = 1 \otimes_{\Bbb N} 1 \in {\frak E}(\HH_1) 
\widehat{\otimes}_{\frak D} {\frak E}(\HH_2)$ is clear by 
construction. Let $(({\frak D}, \widehat{\oplus}_{\frak D}); 
\Theta)$ denote the 
tensor product of ${\frak E}(\HH_1)$ and ${\frak E}(\HH_2)$. Then 
there exists a morphism $\theta : {\frak D} \to {\frak E}(\HH_1) 
\widehat{\otimes}_{\frak D} {\frak E}(\HH_2)$ such that 
$\widehat{\otimes}_{\frak D} = \theta \circ \Theta$. 
Every element in ${\frak E}(\HH_1) 
\widehat{\otimes}_{\frak D} {\frak E}(\HH_2)$ is an equivalence 
class of $\oplus_{\frak 
D}$-parts of decompositions of $1 \otimes_{\Bbb N} 1 
\in {\frak E}(\HH_1) 
\widehat{\otimes}_{\frak D} {\frak E}(\HH_2)$. The map $\Theta 
\circ \widehat{\otimes}_{\frak D}^{-1} : \widehat{\otimes}_{\frak 
D} \left({\frak E}(\HH_1) \times {\frak E}(\HH_2) \right) \to 
\frak D$ maps the 
collection of all homogeneous terms (i.e., those of the form $e_i 
\otimes_{\Bbb N} f_j$) in every particular 
decomposition of $1 \otimes_{\Bbb N} 1$ to a complete, 
$\widehat{\oplus}_{\frak D}$-orthogonal subset of $\frak D$.
Thus $\Theta \circ \widehat{\otimes}_{\frak D}^{-1}$ can be 
extended to a map which maps every equivalence class of 
$\oplus_{\frak D}$-parts of $1 \otimes_{\Bbb N} 1$ to a 
$\widehat{\oplus}_{\frak D}$-part of $1_{\frak D} = \Theta \circ 
\widehat{\otimes}_{\frak D}^{-1}(1 \otimes_{\Bbb N} 1)$. We denote 
this extension also by $\Theta \circ \widehat{\otimes}_{\frak 
D}^{-1}$. Since then 
$\theta \circ \Theta \circ \widehat{\otimes}_{\frak D}^{-1} = 
id_{{\frak E}(\HH_1) \widehat{\otimes}_{\frak D} {\frak 
E}(\HH_2)}$, it follows that $\theta$ is surjective. 
We show that $\theta$ is also 
injective. To this end consider ${\frak d}_1, {\frak d}_2 \in 
\frak D$, such that 
$\theta({\frak d}_1) = {\theta}({\frak d}_2)$. 
${\theta}({\frak d}_1)$ and ${\theta}({\frak d}_2)$ are 
both equivalence classes of $\oplus_{\frak 
D}$-parts of $\widehat{\otimes}_{\frak D}(1, 1)$. Thus $\Theta 
\circ \widehat{\otimes}_{\frak D}^{-1}({\theta}({\frak 
d}_1)) = \Theta \circ \widehat{\otimes}_{\frak D}^{-1}
({\theta}({\frak d}_2))$. 
Since $\frak D$ is a D-poset, ${\frak d}_1$ and ${\frak d}_2$ are 
$\widehat{\oplus}_{\frak D}$-sums of elements of the form ${\frak 
d}_1^i = \Theta(e_{11}^i, e_{12}^i)$ and ${\frak d}_2^j = 
\Theta(e_{21}^j, e_{22}^j)$, i.e., ${\frak d}_1 = 
\widehat{\oplus}_{\frak D} {\frak d}_1^i = \widehat{\oplus}_{\frak 
D} \Theta(e_{11}^i, e_{12}^i)$ and ${\frak d}_2 = 
\widehat{\oplus}_{\frak D} {\frak d}_2^j = \widehat{\oplus}_{\frak 
D} \Theta(e_{21}^j, e_{22}^j)$. With $\theta \circ \Theta = 
\widehat{\otimes}_{\frak D}$ it follows $\Theta \circ 
\widehat{\otimes}_{\frak D}^{-1} \left( \theta({\frak d}_1^i) 
\right) = {\frak d}_1^i $ and thus $\Theta \circ 
\widehat{\otimes}_{\frak D}^{-1} \left( \theta({\frak d}_1) 
\right) = {\frak d}_1$ and similarly $\Theta \circ 
\widehat{\otimes}_{\frak D}^{-1} \left( \theta({\frak d}_2) 
\right) = {\frak d}_2$. Thus 
${\frak d}_1 = {\frak d}_2$. It follows from the above argument 
that ${\theta}^{-1}$ is also a morphism. Thus ${\theta}$ 
is an isomorphism between D-posets and therefore $({\frak D}, 
\widehat{\oplus}_{\frak D}) = ({\frak E}(\HH_1) 
\widehat{\otimes}_{\frak D} {\frak E}(\HH_2), \oplus_{\frak D})$.
\hfill $\Box$ \\

This result can immediately be generalized to the tensor product 
of finitely many such effect algebras. With an obvious 
generalization of our above notation we have
\begin{co} The pair $({\frak E}(\HH_1) \widehat{\otimes}_{\frak D} 
{\frak E}(\HH_2) \widehat{\otimes}_{\frak D} \cdot \cdot \cdot 
\widehat{\otimes}_{\frak D} {\frak E}(\HH_n) , \oplus_{\frak D})$ 
is the tensor product of the family ${\frak E}(\HH_1), ..., {\frak 
E}(\HH_n)$. \end{co} 

In the sequel we denote by $\Theta$ the map defined in the proof 
of Theorem 2 given by \[ \Theta: {\frak E}(\HH_1) \times \cdot 
\cdot \cdot \times {\frak E}(\HH_n) \to {\frak E}(\HH_1) 
\widehat{\otimes}_{\frak D} \cdot \cdot \cdot 
\widehat{\otimes}_{\frak D} {\frak E}(\HH_n), \Theta(u_{t_1}, ..., 
u_{t_n}) := u_{t_1} \widehat{\otimes}_{\frak D} \cdot \cdot \cdot 
\widehat{\otimes}_{\frak D} u_{t_n}. \] Remember that the 
equivalence class $\Theta(u_{t_1}, ..., 
u_{t_n})$ contains only the element $u_{t_1} 
\widehat{\otimes}_{\frak D} \cdot \cdot \cdot 
\widehat{\otimes}_{\frak D} u_{t_n}$. However, general equivalence 
classes $[G] \in {\frak E}(\HH_1) \widehat{\otimes}_{\frak D} 
\cdot \cdot \cdot \widehat{\otimes}_{\frak D} {\frak E}(\HH_n)$ 
contain at most $n!$ elements.  \\ 

Now we consider homogeneous 
effect histories with fixed finite support $S = \{ t_1, ..., t_n 
\}$. The set of all such histories can be identified with the 
Cartesian product 
${\frak E}(\HH)_{t_1} \times \cdot \cdot \cdot \times {\frak 
E}(\HH)_{t_n}$. The class operator defined by Equation \ref{cop} 
factorizes according to {\small 
\[ u = (u_{t_1}, ..., u_{t_n}) 
\stackrel{\Theta}{\longrightarrow} 
\Theta(u) = u_{t_1} \widehat{\otimes}_{\frak D} \cdot \cdot \cdot 
\widehat{\otimes}_{\frak D} u_{t_n} \stackrel{\sqrt{ 
}}{\longrightarrow} \sqrt{\Theta(u)} = u^{1/2}_{t_1} 
\widehat{\otimes}_{\frak D} \cdot \cdot \cdot 
\widehat{\otimes}_{\frak D} u^{1/2}_{t_n} 
\stackrel{C_{t_0}'}{\longrightarrow} C_{t_0}'\left( 
\sqrt{\Theta(u)} \right) = C_{t_0}(u), \]} where $C_{t_0}'$ is 
defined in an obvious way. \\ 
We define a decoherence functional $\widehat{d}_{\varrho,S}$ for 
pairs $(a,b)$ of homogeneous elements in \\ 
$\left( {\frak E}(\HH)_{t_1} \otimes_{\Bbb N} \cdot \cdot \cdot 
\otimes_{\Bbb N} {\frak E}(\HH)_{t_n} \right) \times 
\left( {\frak E}(\HH)_{t_1} \otimes_{\Bbb N} \cdot \cdot 
\cdot \otimes_{\Bbb N} {\frak E}(\HH)_{t_n} \right)$ 
by \begin{equation} \label{df1} \widehat{d}_{\varrho,S}(a,b) := 
{\tr}\left(C_{t_0}' \left(\sqrt{a} \right) \varrho C_{t_0}' 
\left(\sqrt{b} \right)^{\dagger}\right). \end{equation} The such 
defined \df cannot be extended to a 
$\oplus_{\frak D}$-additive functional on $\left( {\frak 
E}(\HH)_{t_1} \widehat{\otimes}_{\frak D} \cdot \cdot \cdot 
\widehat{\otimes}_{\frak D} {\frak E}(\HH)_{t_n} \right) \times 
\left( {\frak E}(\HH)_{t_1} \widehat{\otimes}_{\frak D} \cdot 
\cdot \cdot \widehat{\otimes}_{\frak D} {\frak 
E}(\HH)_{t_n} 
\right)$, that is, the D-poset structure given by $\oplus_{\frak 
D}$ is not the physically interesting one. \\
The decoherence functional $\widehat{d}_{\varrho,S}$ can easily be 
extended to arbitrary finite homogeneous effect histories 
\[ d_{\varrho, {\frak N}} : {\Bbb E}_{fin}(\HH) \times {\Bbb 
E}_{fin}(\HH) \to \CC, d_{\varrho, {\frak N}}(u, v) := 
\widehat{d}_{\varrho, {\frak s}(u) \cup {\frak s}(v)} \left( 
\Theta(u), \Theta(v) \right), \] for all $u, v \in {\Bbb 
E}_{fin}(\HH)$. \\ 

However, before we proceed to define and investigate the algebraic 
structure in terms of which the general consistency condition in 
our generalized history theory can be formulated, we briefly 
mention that there exist special sets of effect 
histories on which the decoherence functional induces a 
probability measure even though in general none of the familiar 
consistency conditions is satisfied for the elements of the 
special sets. \\ To this end fix 
an arbitrary finite homogeneous effect history $u_0$, 
fix a $t^* > t_f(u_0)$ and for every $E \in {\frak E}(\HH)$ denote 
by $u_E$ the finite homogeneous 
effect history defined by ${\frak s}(u_E) := {\frak s}(u_0) \cup 
\{ t^* \}$ and by  \[ (u_E)_t := \left\{ \begin{array}{r@{\quad : 
\quad}l} (u_0)_t & t \neq t^* \\ E & t = t^* \end{array} 
\right. , \] that is, $u_E$ is the extension of $u_0$ by the 
effect $E$. 
The above decoherence functional $\widehat{d}_{\varrho,{\frak 
s}(u_0) \cup \{ t^* \}}$ induces a probability measure on the set 
$\widetilde{{\frak E}} := \{ u_E \mid E \in {\frak E}(\HH) \} 
\simeq {\frak E}(\HH)$ by \begin{equation} \label{E7} p_{\varrho} 
: \widetilde{\frak E} \to \RR, p_{\varrho}(u_E) := 
\widehat{d}_{\varrho,{\frak s}(u_0) \cup \{ t^* \}} \left( 
\Theta(u_E), \Theta(u_E) \right). 
\end{equation} The D-poset structure on ${\frak E}(\HH)$ 
induces a D-poset structure on $\widetilde{{\frak E}}$. It is easy 
to see that $p_{\varrho}$ is additive on $\widetilde{\frak 
E}$ in the sense $p_{\varrho}(u_{E_1 \oplus E_2}) = 
p_{\varrho}(u_{E_1}) + p_{\varrho}(u_{E_2})$, whenever $E_1 \oplus 
E_2$ is well-defined. 
We will say that the D-poset of histories $\widetilde{\frak E}$ 
which can be constructed in this way (and which in particular is 
isomorphic to ${\frak E}(\HH)$) is a {\sc full 
D-poset of effect histories}. 
If $E$ and $F$ are orthogonal projection operators, then 
$\widehat{d}_{\varrho,{\frak s}(u_0) \cup \{ t^* \}}(\Theta(u_E), 
\Theta(u_F)) = 0$. However, if $E$ and $F$ are 
not orthogonal projection operators, then in general 
Re $\widehat{d}_{\varrho,{\frak s}(u_0) \cup \{ 
t^* \}}(\Theta(u_E), \Theta(u_F)) \neq 0$. The probability 
$p_{\varrho}(u_E)$ can be 
interpreted as the conditional probability that the event $E$ 
will be realized at time $t^*$ given that the history $u_0$ is 
realized. Stated differently, the probability of the history 
$u_E$ in the state $\varrho$ equals the probability of the effect 
$E$ in the (reduced) state $C_{t_0}(u_0) \varrho 
C_{t_0}(u_0)^{\dagger}$, i.e., $p_{\varrho}(u_E) = p_{C_{t_0}(u_0) 
\varrho C_{t_0}(u_0)^{\dagger}}(E)$. \\ 
We mention already here that it is possible to construct 
further exceptional D-posets of effect histories such that the 
decoherence functional is additive in both arguments. 
{\nopagebreak We will return to \nopagebreak this point below.} 
\pagebreak[3] 

We now return to our determination and investigation of the 
natural algebraic structure which makes the decoherence functional 
additive in both arguments. \\
It is possible to define a tensor product ${\frak 
E}(\HH)_{t_1} \widetilde{\otimes}_{\frak D} \cdot \cdot \cdot 
\widetilde{\otimes}_{\frak D} {\frak E}(\HH)_{t_n} $
such that the corresponding decoherence functional 
$d_{\varrho,S}$ is additive in both arguments as we will 
show next. To this end we again consider ${\frak 
E}(\HH_1) \otimes_{\Bbb N} {\frak E}(\HH_2)$ and define a partial 
binary operation $\widetilde{\oplus}_{\Bbb N}$ by
\begin{equation} A \widetilde{\oplus}_{\Bbb N} B := \left( A^{1/2} 
+ B^{1/2} \right)^2, \mbox{ if } A^{1/2} + B^{1/2} \leq 1, 
\end{equation} 
for all homogeneous elements $A = A_1 \otimes_{\Bbb N} A_2, B = 
B_1 \otimes_{\Bbb N} B_2 \in {\frak E}(\HH_1) \otimes_{\Bbb N} 
{\frak E}(\HH_2).$ Finite $\widetilde{\oplus}_{\Bbb N}$-sums of 
homogeneous elements are defined recursively and the set of all 
well-defined finite $\widetilde{\oplus}_{\Bbb N}$-sums of 
homogeneous elements will be denoted by ${\frak E}(\HH_1) 
\bar{\otimes}_{\Bbb N} {\frak E}(\HH_2)$. Notice that $A 
\widetilde{\oplus}_{\Bbb N} B$ is a well-defined bounded, linear 
operator on $\HH_1 \otimes_{\Bbb N} \HH_2$, but in general $A 
\widetilde{\oplus}_{\Bbb N} B \notin {\frak 
E}(\HH_1) \otimes_{\Bbb N} {\frak E}(\HH_2)$.

We have defined above a partial operation 
$\ominus$ on the space ${\frak E}(\HH)$ of effect 
operators on the Hilbert space $\HH$ such that $({\frak E}(\HH), 
\ominus)$ is a D-poset. 
However, the D-poset structure on ${\frak E}(\HH)$ is not unique. 
Define $A \widetilde{\oplus} B := \left( A^{1/2} + B^{1/2} 
\right)^2$, for all 
$A,B \in {\frak E}(\HH)$ satisfying $A^{1/2} + B^{1/2} \leq 1$.
To prove that $({\frak E}(\HH), \widetilde{\oplus})$ is a D-poset 
is straightforward. Moreover, $({\frak E}(\HH), 
\widetilde{\oplus})$ is an ${\Bbb N}_2$-semimodule. 
In this work we denote the D-poset $({\frak 
E}(\HH), \oplus)$ briefly by ${\frak E}(\HH)$. When we refer to 
${\frak E}(\HH)$ supplied with the D-poset structure given by 
$\widetilde{\oplus}$, then we explicitly write $({\frak E}(\HH), 
\widetilde{\oplus})$. \\ We can now proceed as above and define a 
partial binary operation $\widetilde{\oplus}_{\frak D}$ and 
construct the tensor product of $({\frak E}(\HH_1), 
\widetilde{\oplus})$ and $({\frak E}(\HH_2), \widetilde{\oplus})$ 
in the category of ${\Bbb N}_2$-semimodules. Everywhere in our 
above description of the 
definition of $\oplus_{\frak D}$ preceding Theorem 2 we simply 
have to replace 
\begin{itemize} \item $\oplus$ by $\widetilde{\oplus}$; \item 
$\oplus_{\Bbb N}$ by $\widetilde{\oplus}_{\Bbb N}$; \item 
$\oplus_{\frak D}$ by $\widetilde{\oplus}_{\frak D}$; 
\item $\sim_{\frak D}$ by $\approx_{\frak D}$. 
\end{itemize} 
{\footnotesize We say that a $\widetilde{\oplus}_{\frak D}$-sum of 
the form $(E_1 \otimes_{\Bbb N} F_1) \widetilde{\oplus}_{\frak D} 
(E_2 \otimes_{\Bbb N} F_2) \widetilde{\oplus}_{\frak D} \cdot 
\cdot \cdot \widetilde{\oplus}_{\frak D} (E_n \otimes_{\Bbb N} 
F_n)$ of finitely many elements $E_i {\otimes}_{\Bbb N} F_i \in 
{\frak E}(\HH_1) {\otimes}_{\Bbb N} {\frak E}(\HH_2)$ exists if 
there exists a decomposition $\widetilde{\bigoplus}_{{\Bbb N}, i} 
e_i \otimes_{\Bbb N} f_i$ of 
$1 \otimes_{\Bbb N} 1$ in ${\frak E}(\HH_1) \bar{\otimes}_{\Bbb N} 
{\frak E}(\HH_2)$ such that every term of the form $E_i 
\otimes_{\Bbb N} F_i$ occurs 
also in the decomposition $\widetilde{\bigoplus}_{{\Bbb N}, i \in 
I} e_i \otimes_{\Bbb N} 
f_i$ of $1 \otimes_{\Bbb N} 1$, i.e., if there for every $i \in 
\{1, ..., n \}$ there exists a $j \in I$ such that $E_i 
\otimes_{\Bbb N} F_i = e_j \otimes_{\Bbb N} f_j$. \\
In this case we also adopt the following way of speaking: we say 
that the $\widetilde{\oplus}_{\frak D}$-sum $(E_1 {\otimes}_{\Bbb 
N} F_1) \widetilde{\oplus}_{\frak D} (E_2 {\otimes}_{\Bbb N} F_2) 
\widetilde{\oplus}_{\frak D} \cdot \cdot \cdot 
\widetilde{\oplus}_{\frak D} (E_n {\otimes}_{\Bbb N} F_n)$ is a 
{\sc $\widetilde{\oplus}_{\frak D}$-part of} $1 {\otimes}_{\Bbb N} 
1$. \\ In this case we set \begin{equation} (E_1 {\otimes}_{\Bbb 
N} F_1) \widetilde{\oplus}_{\frak D} (E_2 {\otimes}_{\Bbb N} F_2) 
\widetilde{\oplus}_{\frak D} \cdot \cdot \cdot 
\widetilde{\oplus}_{\frak D} (E_n {\otimes}_{\Bbb N} F_n) :=(E_1 
{\otimes}_{\Bbb N} F_1) \widetilde{\oplus}_{\Bbb N} (E_2 
{\otimes}_{\Bbb N} F_2) \widetilde{\oplus}_{\Bbb N} \cdot \cdot 
\cdot \widetilde{\oplus}_{\Bbb N} (E_n {\otimes}_{\Bbb N} F_n). 
\end{equation} 
We further say that a $\widetilde{\oplus}_{\frak D}$-sum $(E_1 
\otimes_{\Bbb 
N} F_1) \widetilde{\oplus}_{\frak D} (E_2 \otimes_{\Bbb N} F_2) 
\widetilde{\oplus}_{\frak 
D} \cdot \cdot \cdot \widetilde{\oplus}_{\frak D} (E_n 
\otimes_{\Bbb N} F_n)$ 
is a {\sc minimal representation} of its associated 
$\widetilde{\oplus}_{\frak D}$-part of $1 \otimes_{\Bbb N} 1$ if 
the number of 
terms in the sum cannot be reduced by applying the following 
relations \begin{enumerate} \item $n(E_1 
{\otimes}_{\Bbb N} E_2) \equiv (n E_1) 
{\otimes}_{\Bbb N} E_2 = E_1 {\otimes}_{\Bbb N} (n E_2)$, for all 
$E_1 \in ({\frak E}(\HH_1), 
\widetilde{\oplus}), E_2 \in ({\frak E}(\HH_2), 
\widetilde{\oplus})$ and $n \in {\Bbb N}_2$. \\ 
whenever the expressions are well-defined;  \item $(E_1 
{\otimes}_{\Bbb N} E_2) \widetilde{\oplus}_{\frak D} 
(E_1 {\otimes}_{\Bbb N} \tilde{E}_2) = E_1 
{\otimes}_{\Bbb N} (E_2 \widetilde{\oplus} 
\tilde{E}_2),$ for all $E_1 \in ({\frak E}(\HH_1), 
\widetilde{\oplus})$ and $E_2, \tilde{E}_2 \in ({\frak E}(\HH_2), 
\widetilde{\oplus})$, \\ whenever the expressions are 
well-defined; \item $(E_1 {\otimes}_{\Bbb N} E_2) 
\widetilde{\oplus}_{\frak D} (\tilde{E}_1 
{\otimes}_{\Bbb N} E_2) = (E_1 \widetilde{\oplus} 
\tilde{E}_1) {\otimes}_{\Bbb N} E_2,$ 
for all $E_1, \tilde{E}_1 \in ({\frak E}(\HH_1), 
\widetilde{\oplus})$ and $E_2 \in ({\frak E}(\HH_2), 
\widetilde{\oplus})$, \\ whenever the expressions are 
well-defined. \end{enumerate} 
The $\widetilde{\oplus}_{\frak D}$-sum $(G_{\alpha} 
\widetilde{\oplus}_{\frak D} G_{\beta})$ of two arbitrary 
elements $G_{\alpha}$ and $G_{\beta} \in {\frak E}(\HH_1) 
\bar{\otimes}_{\Bbb N} {\frak E}(\HH_2)$ is defined if 
$G_{\alpha}$ possesses a well-defined decomposition \[ G_{\alpha} 
= (E_{\alpha_1} {\otimes}_{\Bbb N} F_{\alpha_1}) 
\widetilde{\oplus}_{\frak D} 
\cdot \cdot \cdot \widetilde{\oplus}_{\frak D} (E_{\alpha_n} 
{\otimes}_{\Bbb N} F_{\alpha_n}), \] and if $G_{\beta}$ possesses 
a well-defined decomposition \[ G_{\beta} = 
(E_{\beta_1} {\otimes}_{\Bbb N} F_{\beta_1}) 
\widetilde{\oplus}_{\frak D} \cdot \cdot \cdot 
\widetilde{\oplus}_{\frak D} (E_{\beta_m} {\otimes}_{\Bbb N} 
F_{\beta_m}), \] (i.e., $G_{\alpha}$ and $G_{\beta}$ are both 
well-defined $\widetilde{\oplus}_{\frak D}$-parts of $1 
\otimes_{\Bbb N} 1$) such that \begin{equation} (G_{\alpha} 
\widetilde{\oplus}_{\frak D} G_{\beta}) := (E_{\alpha_1} 
{\otimes}_{\Bbb N} F_{\alpha_1}) \widetilde{\oplus}_{\frak D} 
\cdot \cdot \cdot \widetilde{\oplus}_{\frak D} (E_{\alpha_n} 
{\otimes}_{\Bbb N} F_{\alpha_n}) \widetilde{\oplus}_{\frak D} 
(E_{\beta_1} {\otimes}_{\Bbb N} F_{\beta_1}) 
\widetilde{\oplus}_{\frak D} \cdot \cdot \cdot 
\widetilde{\oplus}_{\frak D} (E_{\beta_m} {\otimes}_{\Bbb N} 
F_{\beta_m}) \end{equation} is 
well-defined. It is clear that $(G_{\alpha} 
\widetilde{\oplus}_{\frak D} G_{\beta})$ such defined is 
independent of the particular decompositions considered. 
We say that two $\widetilde{\oplus}_{\frak D}$-parts of $1 
\otimes_{\Bbb N} 
1$, say, $G_{\delta}$ and $G_{\gamma}$ are {\sc equivalent} if for 
all $\widetilde{\oplus}_{\frak D}$-parts $G$ of $1 \otimes_{\Bbb 
N} 1$ the 
following holds: $G \widetilde{\oplus}_{\frak D} G_{\delta}$ is 
well-defined 
and equal to $1 \otimes_{\Bbb N} 1$ if and only if $G 
\widetilde{\oplus}_{\frak D} G_{\gamma}$ is well-defined and equal 
to $1 
\otimes_{\Bbb N} 1$. We also write $G_{\delta} \approx_{\frak D} 
G_{\gamma}$. We write for the $\approx_{\frak D}$-equivalence 
class 
of $G$ either $[G]$ or (abusing the notation) simply $G$. It is 
clear that the equivalence class of a homogeneous element $E 
\otimes_{\Bbb N} F$ contains only the element $E \otimes_{\Bbb N} 
F$, i.e., $[E \otimes_{\Bbb N} F] = \{ E \otimes_{\Bbb N} F \}$. 
The partial addition $\widetilde{\oplus}_{\frak D}$ defined above 
for 
$\widetilde{\oplus}_{\frak D}$-parts of $1 \otimes_{\Bbb N} 1$ 
induces a 
partial addition on the equivalence classes of 
$\widetilde{\oplus}_{\frak D}$-parts of $1 \otimes_{\Bbb N} 1$ 
(denoted by the 
same symbol) by \[ \left[G_{\delta} \right] 
\widetilde{\oplus}_{\frak D} 
\left[G_{\gamma} \right] := \left[G_{\delta} 
\widetilde{\oplus}_{\frak D} 
G_{\gamma} \right], \] for all $G_{\delta}$ and $G_{\gamma}$ for 
which the right hand side is well-defined. 
It is clear that every equivalence class $[G]$ contains at most 
two elements. Further, if $[G]$ contains more than one element, 
then the 
complementary class $[G]' := [G']$ of $[G]$ (defined by $[G] 
\widetilde{\oplus}_{\frak D} [G]' = [1]$) contains only one 
element but not vice versa. 
For well-defined $\widetilde{\oplus}_{\frak D}$-parts $G \in 
{\frak E}(\HH_1) {\otimes}_{\Bbb N} 
{\frak E}(\HH_2)$ of $1 \otimes_{\Bbb N} 1$ with a minimal 
representation of the form $G := (E_1 {\otimes}_{\Bbb N} F_1) 
\widetilde{\oplus}_{\frak D} (E_2 {\otimes}_{\Bbb N} F_2) 
\widetilde{\oplus}_{\frak D} \cdot \cdot \cdot 
\widetilde{\oplus}_{\frak D} (E_n {\otimes}_{\Bbb N} F_n)$ there 
are two candidates both of which may serve as complements in case 
they are well-defined, namely \begin{eqnarray*} \bullet 
\hspace{0.5em} G^* & := & 
\left( E_1 {\otimes}_{\Bbb N} \left( 1-F_1^{1/2} \right)^2 \right) 
\widetilde{\oplus}_{\frak D} \left( E_2 {\otimes}_{\Bbb N} \left( 
1-F_2^{1/2} \right)^2 \right) \widetilde{\oplus}_{\frak D} \cdot 
\cdot \cdot \widetilde{\oplus}_{\frak D} \left( E_n 
{\otimes}_{\Bbb N} \left( 1-F_n^{1/2} \right)^2 \right) 
\widetilde{\oplus}_{\frak D} \\ & & 
\widetilde{\oplus}_{\frak D} \left( \left( 1 - \left( E_1^{1/2} 
\widetilde{\oplus} E^{1/2}_2 \widetilde{\oplus} \cdot \cdot \cdot 
\widetilde{\oplus} E_n^{1/2} \right)^2 \right) {\otimes}_{\Bbb N} 
1 \right) \end{eqnarray*} \begin{eqnarray*} \bullet \hspace{0.5em} 
G^{**} & := & 
\left( \left( 1-E^{1/2}_1 \right)^2 {\otimes}_{\Bbb N} F_1 \right) 
\widetilde{\oplus}_{\frak D} \left( \left( 1-E_2^{1/2} \right)^2 
{\otimes}_{\Bbb N} F_2 \right) \widetilde{\oplus}_{\frak D} \cdot 
\cdot \cdot \widetilde{\oplus}_{\frak D} \left( \left( 1-E_n^{1/2} 
\right)^2 {\otimes}_{\Bbb N} F_n \right) \widetilde{\oplus}_{\frak 
D} \\ & & \widetilde{\oplus}_{\frak 
D} \left( 1 {\otimes}_{\Bbb N} \left( 
1 - \left( F_1^{1/2} \widetilde{\oplus} F_2^{1/2} 
\widetilde{\oplus} \cdot \cdot \cdot \widetilde{\oplus} F_n^{1/2} 
\right)^2 \right) \right). \end{eqnarray*} 
However, if $G$ is well-defined, it is clear that either $G^*$ or 
$G^{**}$ is also well-defined. (If they are both well-defined, 
then they are 
equivalent of course since then also $G \widetilde{\oplus}_{\frak 
D} G^*$ and $G \widetilde{\oplus}_{\frak D} G^{**}$ are both 
well-defined and equal to $1 {\otimes}_{\Bbb N} 1$.) 
Define ${\frak E}(\HH_1) \widetilde{\otimes}_{\frak D} {\frak 
E}(\HH_2)$ as the effect algebra with partial binary operation 
$\widetilde{\oplus}_{\frak D}$ as the set consisting of all 
$\approx_{\frak D}$-equivalence classes of well-defined finite 
$\widetilde{\oplus}_{\frak D}$-sums $G$ of elements of the form
$E_1 \widetilde{\otimes}_{\frak D} E_2 := E_1 \otimes_{\Bbb 
N} E_2 \in {\frak E}(\HH_1) \otimes_{\Bbb N} {\frak E}(\HH_2)$, 
i.e., consisting of all $\approx_{\frak D}$-equivalence classes of 
$\widetilde{\oplus}_{\frak D}$-parts of $1 \otimes_{\Bbb N} 1$,  
subject to the relations \begin{enumerate} \item $n(E_1 
\widetilde{\otimes}_{\frak D} E_2) \equiv (n E_1) 
\widetilde{\otimes}_{\frak D} E_2 = E_1 \widetilde{\otimes}_{\frak 
D} (n E_2)$, for all $E_1 \in ({\frak E}(\HH_1), 
\widetilde{\oplus}), E_2 \in ({\frak E}(\HH_2), 
\widetilde{\oplus})$ and $n \in {\Bbb N}_2$. \\ 
whenever the expressions are well-defined;  \item $(E_1 
\widetilde{\otimes}_{\frak D} E_2) \widetilde{\oplus}_{\frak D} 
(E_1 \widetilde{\otimes}_{\frak D} \tilde{E}_2) = E_1 
\widetilde{\otimes}_{\frak D} (E_2 \widetilde{\oplus} 
\tilde{E}_2),$ for all $E_1 \in ({\frak E}(\HH_1), 
\widetilde{\oplus})$ and $E_2, \tilde{E}_2 \in ({\frak E}(\HH_2), 
\widetilde{\oplus})$, \\ whenever the expressions are 
well-defined; \item $(E_1 \widetilde{\otimes}_{\frak D} E_2) 
\widetilde{\oplus}_{\frak D} (\tilde{E}_1 
\widetilde{\otimes}_{\frak D} E_2) = (E_1 \widetilde{\oplus} 
\tilde{E}_1) \widetilde{\otimes}_{\frak D} E_2,$ 
for all $E_1, \tilde{E}_1 \in ({\frak E}(\HH_1), 
\widetilde{\oplus})$ and $E_2 \in ({\frak E}(\HH_2), 
\widetilde{\oplus})$, \\ whenever the expressions are 
well-defined. \end{enumerate}} 
\begin{lem} The pair $({\frak E}(\HH_1) \widetilde{\otimes}_{\frak 
D} {\frak E}(\HH_2), \widetilde{\oplus}_{\frak D})$ is the tensor 
product of the D-posets $({\frak E}(\HH_1), \widetilde{\oplus})$ 
and $({\frak E}(\HH_2), \widetilde{\oplus})$ in the category of 
${\Bbb N}_2$-semimodules. \label{le8} \end{lem}

Lemma \ref{le8} can easily be extended to any finite family 
${\frak E}(\HH_1), ..., {\frak E}(\HH_n)$. The \df 
$\widehat{d}_{\varrho, S}$ defined above in Equation \ref{df1} for 
pairs $(a,b)$ of homogeneous elements in 
$\left( {\frak E}(\HH)_{t_1} \otimes_{\Bbb N} \cdot \cdot \cdot 
\otimes_{\Bbb N} {\frak E}(\HH)_{t_n} \right) \times 
\left( {\frak E}(\HH)_{t_1} \otimes_{\Bbb N} \cdot \cdot 
\cdot \otimes_{\Bbb N} {\frak E}(\HH)_{t_n} \right)$ can now be 
extended to an additive \df $d_{\varrho, S}$ on $\left({\frak 
E}(\HH)_{t_1} \widetilde{\otimes}_{\frak D} \cdot 
\cdot \cdot \widetilde{\otimes}_{\frak D} {\frak E}(\HH)_{t_n} 
\right) \times \left({\frak E}(\HH)_{t_1} 
\widetilde{\otimes}_{\frak D} 
\cdot \cdot \cdot \widetilde{\otimes}_{\frak D} 
{\frak E}(\HH)_{t_n} \right)$. 
\begin{theo} The decoherence functional $d_{\varrho,S}: 
\left({\frak E}(\HH)_{t_1} \widetilde{\otimes}_{\frak D} \cdot 
\cdot \cdot \widetilde{\otimes}_{\frak D} {\frak E}(\HH)_{t_n} 
\right) \times \\ \times \left({\frak E}(\HH)_{t_1} 
\widetilde{\otimes}_{\frak D} 
\cdot \cdot \cdot \widetilde{\otimes}_{\frak D} 
{\frak E}(\HH)_{t_n} \right) \to \CC, d_{\varrho,S}(a,b) := 
{\mbox{\em tr}} \left(C_{t_0}' \left(\sqrt{a} \right) \varrho 
C_{t_0}' \left(\sqrt{b} \right)^{\dagger} \right), $ 
satisfies for all $a, b \in {\frak E}(\HH)_{t_1} 
\widetilde{\otimes}_{\frak D} \cdot \cdot \cdot 
\widetilde{\otimes}_{\frak D} {\frak E}(\HH)_{t_n}$ 
\begin{itemize} 
\item $d_{\varrho,S}(a,a) \in \RR$ and $d_{\varrho,S}(a,a) \geq 
0$. \item $d_{\varrho,S}(a,b) = d_{\varrho,S}(b,a)^*$. \item 
$d_{\varrho,S}(1,1) =1$. \item $d_{\varrho,S}(0,a) =0$, for all 
$a$. \item $d_{\varrho,S}(a_1 \widetilde{\oplus}_{\frak D} a_2, b) 
= d_{\varrho,S}(a_1,b) + d_{\varrho,S}(a_2,b)$ for all $a_1, a_2, 
b \in {\frak E}(\HH)_{t_1} \widetilde{\otimes}_{\frak D} \cdot 
\cdot \cdot  \widetilde{\otimes}_{\frak D} {\frak 
E}(\HH)_{t_n},$ for which the left hand side is well-defined. 
\end{itemize} \label{T3} \end{theo} 
\begin{de} Let $({\frak T}, \leq)$ be a partially ordered set. A 
${\frak T}$-{\sc directed system} of D-posets is a family 
$D_{\frak T} := \{ D_t, t \in {\frak T} \}$ of D-posets supplied 
with a family 
of morphisms $f_{ts} : D_t \to D_s, t, s \in \frak T$, defined iff 
$t \leq s$, such that \begin{itemize} \item $f_{tt} = id_{D_t}$, 
for all $t \in \frak T$; \item If $t \leq s \leq r$ in $\frak T$, 
then $f_{sr}f_{ts} = f_{tr}$. \end{itemize}
Let $D_{\frak T}$ be a ${\frak T}$-directed system of D-posets. 
Then a D-poset $\frak L$ supplied with a family of morphisms $\{ 
f_t : D_t \to {\frak L} \}_{t \in {\frak T}}$ is called the {\sc 
direct limit} of $D_{\frak T}$ if \begin{itemize} \item If $t \leq 
s$ in $\frak T$, then $f_s f_{ts} = f_t$; \item If $D$ is a 
D-poset supplied with a set of morphisms $\{ g_t : D_t \to D, t 
\in {\frak T} \}$, then there exists a unique morphism $g : {\frak 
L} \to D$, such that $g f_t = g_t$, for all $t \in {\frak T}$. 
\end{itemize} \end{de} 

The direct limit of a directed system of D-posets always exists 
(Pulmannov\'a, 1995). \\

Let in the sequel $\frak T$ denote the set of all finite subsets 
of $\RR$ partially ordered by set inclusion. For every $t \in \RR$ 
set ${\frak E}(\HH)_t := {\frak E}(\HH)$ and for every $T = \{ 
t_1, ..., t_n \} \in \frak T$ set ${\frak E}(\HH)_T := {\frak 
E}(\HH)_{t_1} \widetilde{\otimes}_{\frak D} \cdot \cdot \cdot 
\widetilde{\otimes}_{\frak D} {\frak E}(\HH)_{t_n}$. Then it has 
been shown by Pulmannov\'a (1995) that for every $T \subset S \in 
\frak T$ there 
exists a morphism $f_{TS}: {\frak E}(\HH)_T \to {\frak E}(\HH)_S$ 
such that $\{ {\frak E}(\HH)_T, T \in {\frak T} \}$ supplied with 
$\{ f_{TS}, T \subset S \in {\frak T} \}$ is a $\frak T$-directed 
system. Let, e.g., $T = \{ t_1, t_3 \} \subset S = \{ t_1, t_2, 
t_3 \}$, then $f_{TS}(A \widetilde{\otimes}_{\frak D} B) = A 
\widetilde{\otimes}_{\frak D} 1 \widetilde{\otimes}_{\frak D} B$. 
\\ Therefore the direct limit of $\{ ({\frak E}(\HH)_T, 
\widetilde{\oplus}_{\frak D}), T \in {\frak 
T} \}$ exists and will be denoted by $({\frak E}(\HH)_{{\frak T}}, 
\widetilde{\oplus}_{\frak D})$. 
${\frak E}(\HH)_{{\frak T}}$ can be constructed as follows: 
consider the disjoint union $\cup_{T \in {\frak T}} {\frak 
E}(\HH)_T$ and call two elements $h_1, h_2$ of $\cup_{T \in {\frak 
T}} {\frak E}(\HH)_T$ {\sc equivalent} if there exist $T_1, T_2, 
T_{12} \in {\frak T}$ such that $h_1 \in T_1 \subset T_{12}$, $h_2 
\in T_2 \subset T_{12}$ and such that $f_{T_1 T_{12}}(h_1) = 
f_{T_2 T_{12}}(h_2)$. Then ${\frak E}(\HH)_{{\frak T}}$ is the 
quotient space of $\cup_{T \in {\frak T}} {\frak E}(\HH)_T$ by the 
such defined equivalence relation. It is easy to extend the 
D-poset structures on ${\frak E}(\HH)_T$, for $T \in \frak T$ to a 
D-poset structure on ${\frak E}(\HH)_{{\frak T}}$. Thus the above 
defined decoherence functional can be generalized to a 
decoherence functional on ${\frak E}(\HH)_{{\frak T}} \times 
{\frak E}(\HH)_{{\frak T}}$ by
\[ d_{\varrho, {\frak T}} : {\frak E}(\HH)_{\frak T} \times 
{\frak E}(\HH)_{\frak T} \to \CC, d_{\varrho, {\frak T}} (A, B) 
:= d_{\varrho, {\frak s}(A) \cup {\frak s}(B)} \left(f_{{\frak 
s}(A), {\frak s}(A) \cup {\frak s}(B)} A, f_{{\frak s}(B), {\frak 
s}(A) \cup {\frak s}(B)} B \right), \] 
where ${\frak s}(A)$ denotes the temporal support of $A$ and 
${\frak s}(B)$ denotes 
the temporal support of $B$ and $f_{{\frak s}(A), {\frak s}(A) 
\cup {\frak s}(B)}$ and 
$f_{{\frak s}(B), {\frak s}(A) \cup {\frak s}(B)}$ denote the 
canonical morphisms in the directed system 
$\{ {\frak E}(\HH)_T, T \in {\frak T} \}$ from 
${\frak E}(\HH)_{{\frak s}(A)}$ and from ${\frak E}(\HH)_{{\frak 
s}(B)}$ to ${\frak 
E}(\HH)_{{\frak s}(A) \cup {\frak s}(B)}$ respectively. 
\begin{lem} The so defined decoherence functional $d_{\varrho, 
{\frak T}}$ satisfies all properties listed in Theorem $\mbox{{\em 
\ref{T3}}}$; in particular, $d_{\varrho,{\frak T}}(a_1 
\widetilde{\oplus}_{\frak D} a_2, b) = d_{\varrho,{\frak 
T}}(a_1,b) + d_{\varrho,{\frak T}}(a_2,b)$ for all $a_1, a_2, 
b \in {\frak E}(\HH)_{\frak T}$ for which the left hand side is 
well-defined. \end{lem} 
Now we are ready to formulate the consistency conditions in our 
generalized history formalism. 
\begin{de} Let $\cal G$ be a set of elements of $({\frak 
E}(\HH)_{{\frak T}}, \widetilde{\oplus}_{\frak D})$. Then an 
arbitrary subset $\Delta \cal G$ of $\widetilde{\bigoplus}_{\frak 
D} {\cal G}$ is said to be {\sc 
preconsistent with respect to the state $\varrho$} if 
\[ \mbox{{\em Re }} d_{\varrho, {\frak T}} (a,b) =0, \mbox{ for 
all } a, b \in \Delta {\cal G} \mbox{ for which } a 
\widetilde{\oplus}_{\frak D} b \mbox{ is well-defined in } 
\widetilde{\oplus}_{\frak D} {\cal G}. \] 
Further, $\widetilde{\bigoplus}_{\frak D} {\cal G}$ is said to be 
{\sc consistent with respect to the state $\varrho$} if 
\[ {\mbox{\em Re }} d_{\varrho, {\frak T}} (a,b) =0, \mbox{ 
for all } a, b \in \widetilde{\oplus}_{\frak D} {\cal G} \mbox{ 
for which } a \widetilde{\oplus}_{\frak D} b \mbox{ is 
well-defined}. \]
An arbitrary subset ${\cal G}_0$ of ${\frak E}(\HH)_{{\frak T}}$ 
is said to be {\sc consistent w.r.t.~$\varrho$} if there 
exists a subset $\cal G$ of ${\frak E}(\HH)_{{\frak T}}$ 
such that ${\cal G}_0 = \widetilde{\oplus}_{\frak D} {\cal G}$ is 
consistent w.r.t~$\varrho$. \label{D22} \end{de}
\begin{rem} \label{R14} Above we have found two sorts of D-posets 
on which the \df can be unambiguously defined such that the \df is 
additive in both arguments. The \df induces a probability measure 
on consistent sub-D-posets of $({\frak E}(\HH)_{\frak T}, 
\widetilde{\oplus}_{\frak D})$ and on full D-posets of 
effect histories. Since effect histories represent the general 
physical properties of a quantum system, for a {\sf physical 
description} of a quantum system one needs a probability measure 
which is defined directly on some set of effect histories. This 
task has only be achieved in the latter case for the full D-posets 
of effect histories. The elements of ${\frak 
E}(\HH)_{\frak T}$ are not effect histories. Hence in the former 
case we have to single out those sets of effect histories for 
which the description in terms of elements of consistent 
sub-D-posets of $({\frak E}(\HH)_{\frak T}, 
\widetilde{\oplus}_{\frak D})$ can be lifted to an unambiguous 
description in terms of effect histories. 
Furthermore, it is possible to define a `reasoning' on a D-poset 
$D$ on which an additive \df $d$ is given. Let $a,b,c \in (D, 
\oplus)$ be pairwise orthogonal elements, then we say that $a 
\oplus c \Longrightarrow_d a \oplus b$ if $d(a \oplus c, a \oplus 
c) = d(a,a) \neq 0$ and $d(a \oplus b, a \oplus b) = d(a\oplus b 
\oplus c, a \oplus b \oplus c) \neq 0$. If $d$ induces a 
probability measure on $D$, then the second condition is 
redundant. Hence we also have to single out those sets of effect 
histories for which the reasoning in terms of elements of 
consistent sub-D-posets of $({\frak E}(\HH)_{\frak T}, 
\widetilde{\oplus}_{\frak D})$ can be lifted to an unambiguous 
reasoning in terms of effect histories. \end{rem} 
\begin{rem} If $\widetilde{\oplus}_{\frak D} {\cal G}$ is 
consistent w.r.t.~$\varrho$, then 
$d_{\varrho, {\frak T}}(a,a)$ can be interpreted as probability of 
$a$ in $\widetilde{\oplus}_{\frak D} {\cal G}$ . \end{rem}
\begin{rem} \label{frN} There is a canonical map ${\frak N} : 
{\Bbb E}_{fin}(\HH) \to {\frak E}(\HH)_{{\frak T}}$ defined by 
${\frak N}(u) := {\widetilde{\bigotimes}_{{\frak D}, t \in {\frak 
s}(u)}} u_t$. The decoherence functional $d_{\varrho, {\frak T}}$ 
induces a map $d_{\varrho, {\frak N}} : {\Bbb E}_{fin}(\HH) \times 
{\Bbb E}_{fin}(\HH) \to \CC, d_{\varrho, {\frak N}} (p_1, p_2) := 
d_{\varrho, {\frak T}} ({\frak N}(p_1), {\frak N}(p_2)).$ 
\end{rem} 
\begin{de} We say that a map ${\frak F} : \Delta \subset 
\widehat{\Bbb E}(\HH) \to {\frak E}(\HH)_{{\frak T}}$ {\sc 
extends} $\frak N$ if ${\frak F}(p) = {\frak N}(p)$, for all $p 
\in \Delta \cap {\Bbb E}_{fin}(\HH)$. \end{de} 
\begin{de} \label{CEH} A Boolean lattice $({\cal B}, 
\vee_{\cal B}, \wedge_{\cal B}, \neg_{\cal B})$ is said to be an 
{\sc allowed Boolean lattice of (inhomogeneous) effect histories} 
if the following conditions are satisfied \begin{itemize} \item 
${\cal B}$ is a Boolean sublattice of $\widehat{\Bbb 
E}_{fin}(\HH)$; the lattice-operations 
$\vee_{\cal B}$ and $\wedge_{\cal B}$ are the restrictions of 
$\vee_{\cal L}$ and $\wedge_{\cal L}$ to $\cal B$ respectively. 
The lattice-operations $\vee_{\cal B}$ and $\wedge_{\cal B}$ are 
such that a complementation $\neg_{\cal B}$ can be unambiguously 
defined on $\cal B$; 
\item $\cal B$ is atomic and the set 
of atoms consists of homogeneous elements; 
\item The canonical map ${\frak N} : 
{\cal B} \cap {\Bbb E}_{fin}(\HH) \to {\frak E}(\HH)_{{\frak T}}$ 
defined in Remark $\mbox{\em \ref{frN}}$ can be uniquely extended 
to a {\sf positive valuation} {\fraktur B} on $\cal B$ with values 
in ${\frak E}(\HH)_{{\frak T}}$, i.e., to a map {\fraktur 
B}$:{\cal B} \to {\frak E}(\HH)_{{\frak T}}$ satisfying {\fraktur 
B}$\left({b}_1 \vee_{{\cal B}} {b}_2 \right) 
\widetilde{\ominus}_{\frak D}${\fraktur B}$({b}_1)= $ {\fraktur 
B}$({b}_2) \widetilde{\ominus}_{\frak D}${\fraktur B}$({b}_1 
\wedge_{{\cal B}} {b}_2)$, for all 
${b}_1, {b}_2 \in {\cal B}$. This condition means in particular 
that the left hand side and the right hand side are well-defined 
for all ${b}_1, {b}_2 \in {\cal B}$. \end{itemize} 
An allowed Boolean 
sublattice of $\widehat{\Bbb E}_{fin}(\HH)$ will be briefly 
denoted by $({\cal B},$ {\fraktur B}$)$. \end{de} 
\begin{rem} The greatest element $1_{\cal B}$ and the least 
element $0_{\cal B}$ in $\cal B$ do not necessarily coincide with 
the greatest element $1$ and the least element 
$0$ in $\widehat{\Bbb E}_{fin}(\HH)$ respectively. If the set of 
atoms of $\cal B$ contains more than two different elements, then 
$0_{\cal B} =0$. Every inhomogeneous history in $\cal B$ is the 
join of disjoint homogeneous atoms of $\cal B$. \end{rem} 
\begin{rem} \label{rem11} The decoherence functional $d_{\varrho, 
{\frak T}}$ induces a \df on ${\cal B} \times 
{\cal B}$ by $d_{\varrho, {\cal B}} : {\cal B} \times 
{\cal B} \to \CC, d_{\varrho, {\cal B}}(p_1, p_2) := 
d_{\varrho, {\frak T}} (${\fraktur B}$(p_1),$ {\fraktur 
B}$(p_2))$, which is additive in both arguments with respect to 
the Boolean lattice structure on $\cal B$. \end{rem} 
\begin{lem} The value {\fraktur B}$(u)$ does not depend on the 
choice of the allowed Boolean lattice of effect histories ${\cal 
B} \ni u$. \end{lem}
{\bf Proof}: If $u$ is a homogeneous history, then the assertion 
is trivial. If $u$ is an inhomogeneous history belonging to two 
allowed Boolean lattices, say $({\cal B}, $ {\fraktur B}$)$ and 
$({\cal B}', $ {\fraktur B}'$)$, then 
it is easy to see that there exist homogeneous histories $h_1, 
..., h_n \in {\cal B} \cap {\cal B}'$ (not necessarily atoms of 
${\cal B}$ and ${\cal B}'$) such that $u = \bigvee_{i=1}^n h_i$. 
Thus {\fraktur B}$(u) = $ {\fraktur B}'$(u)$. \hfill $\Box$ \\ 
\begin{de} Let $u_1$ and $u_2$ denote two finite effect histories. 
Then we say that $u_1$ {\sc implies} $u_2$ 
{\sc in the state} $\varrho$ if $u_1$ and $u_2$ lie in 
a common allowed Boolean lattice $({\cal B},$ {\fraktur B}$)$ 
of effect histories and if $d_{\varrho, {\cal 
B}}(u_1 \wedge_{\cal B} u_2, u_1 \wedge_{\cal B} u_2) = 
d_{\varrho, {\cal B}}(u_1, u_1) \neq 
0$ and if $d_{\varrho, {\cal 
B}}(u_1 \vee_{\cal B} u_2, u_1 \vee_{\cal B} u_2) = 
d_{\varrho, {\cal B}}(u_2, u_2) \neq 
0$. \label{D23} In this case we write $u_1 
\Longrightarrow_{\varrho} u_2$. \\
\end{de} 
\begin{rem} It is easy to 
verify that if $p_1 \Longrightarrow_{\varrho} p_2$ is valid in one 
allowed Boolean lattice, then $p_1 \Longrightarrow_{\varrho} 
p_2$ is also valid in every other allowed Boolean lattice of 
$\widehat{\Bbb E}_{fin}(\HH)$ containing $p_1$ and $p_2$. 
\end{rem} 
{\bf Example:} If $u_1$ and $u_2$ are nonzero finite homogeneous 
effect histories such that $u_1 \leq u_2$ where $\leq$ denotes the 
partial order defined in Remark \ref{cgr}, then $u_1 
\Longrightarrow_{\varrho} u_2$ for all $\varrho$ for which 
$d_{\varrho, {\frak T}}(u_1, u_1) \neq 0$ and $d_{\varrho, {\frak 
T}}(u_2, u_2) \neq 0$. To see this, consider the trivial allowed 
Boolean lattice ${\cal B}_0$ containing only the two elements $u_1 
\equiv 0_{{\cal B}_0}$ and $u_2 \equiv 1_{{\cal B}_0}$. 
\begin{rem} Let $h_1$ and $h_2$ denote two nonzero ordinary 
homogeneous histories. If $h_1 \Longrightarrow_{\varrho} h_2$ is 
valid for some $\varrho$ in the sense of Section $\mbox{\em 3}$, 
then $h_1 \Longrightarrow_{\varrho} h_2$ is also valid in the 
sense of Definition $\mbox{\em \ref{D23}}$ and vice versa. To see 
this, consider the allowed Boolean lattice $\cal B$ of effect 
histories consisting of the effect histories $h_1; h_2; h_1 
\vee_{\cal B} h_2; h_1 \wedge_{\cal B} h_2$. ($h_1 \vee_{\cal B} 
h_2$ is a possibly inhomogeneous effect history. It 
is important not to confuse the notion of ordinary inhomogeneous 
history as defined in Section $\mbox{\em 2}$ and the notion of 
inhomogeneous effect history as defined in this section. In this 
section the term `inhomogeneous history' is always meant in the 
sense of Definition $\mbox{\em \ref{IH}}$ (unless explicitly 
otherwise stated).) The least element in $\cal B$ is $h_1 
\wedge_{\cal B} h_2$ and the greatest element in $\cal 
B$ is $h_1 \vee_{\cal B} h_2$. The atoms in $\cal B$ are $h_1$ and 
$h_2$. The map {\fraktur B} extends $\frak 
N$ and hence maps homogeneous histories to their corresponding 
projection operators on $\otimes_i \HH_i$ which in turn can be 
identified with the corresponding homogeneous elements in ${\frak 
E}(\HH)_{\frak T}$, e.g., $h_1$ is mapped to {\fraktur B}$(h_1) = 
\otimes_{t \in {\frak s}(h_1)} h_{1,t}$. The condition {\fraktur 
B}$\left({b}_1 \vee_{{\cal B}} {b}_2 \right) 
\widetilde{\ominus}_{\frak D}${\fraktur B}$\left({b}_1 
\right) = $ { \fraktur B}$({b}_2) \widetilde{\ominus}_{\frak 
D}${\fraktur B}$({b}_1 \wedge_{{\cal B}} {b}_2)$, for 
all ${b}_1, {b}_2 \in {\cal B}$ fixes {\fraktur B} on the 
inhomogeneous element $h_1 \vee_{\cal B} h_2$ of $\cal B$. \\ On 
the other hand consider the set ${\cal C} \subset 
\pout{fin}$ consisting of $\{h_1, h_2, h_1 \vee h_2, h_1 \wedge 
h_2 \}$ where $h_1 \vee h_2$ and $h_1 \wedge h_2$ are 
ordinary histories (possibly inhomogeneous in the sense of Section 
$\mbox{\em 2}$). The assertion of this remark is now an easy 
consequence of Definition $\mbox{\em \ref{D23}}$ and Corollary 
$\mbox{\em \ref{CSH}}$. \end{rem} 

\begin{de} An allowed Boolean lattice $({\cal B}, $ {\fraktur 
B}$)$ is called {\sc consistent w.r.t.~$\varrho$} if 
{\fraktur B}$({\cal B})$ is preconsistent w.r.t.~$\varrho$. 
\end{de} 

\begin{theo} Let $({\cal B}, $ {\fraktur B}$)$ be a consistent 
allowed Boolean lattice of effect histories.
Then the decoherence functional $d_{\varrho, {\frak T}}$ induces a 
probability functional on ${\cal B}$ by $b \mapsto 
\frac{d_{\varrho, {\frak T}}(\mbox{\small \fraktr B}(b), 
\mbox{\small \fraktr B}(b))}{d_{\varrho, {\frak T}}(\mbox{\small 
\fraktr B}(1_{\cal B}), \mbox{\small \fraktr B}(1_{\cal B}))}$. 
\end{theo}

Before formulating the generalized logical rule of interpretation 
we return briefly to the discussion of exceptional sets of 
effect histories. We have already seen above that the D-poset 
structure on ${\frak E}(\HH)$ is not unique. It is possible to 
define a countably infinite family of D-poset structures on 
${\frak 
E}(\HH)$. Let $\alpha$ be a rational number with $\alpha > 0$ 
and define \[ A \oplus_{\alpha} B := \left( A^{1/\alpha} + 
B^{1/\alpha} \right)^{\alpha}, \mbox{ for all } A, B \in 
{\frak E}(\HH) \mbox{ satisfying } A^{1/\alpha} + B^{1/\alpha} 
\leq 1. \] That these expressions are well-defined is a 
consequence of the work of Langer (1962). In particular it follows 
from Proposition 2 in (Langer, 1962) that $E^{\alpha}$ is 
well-defined and that $E^{\alpha}$ is itself an effect 
operator for all $E \in {\frak E}(\HH)$ and all $\alpha \in 
{\Bbb Q}, \alpha >0$. The pair $({\frak E}(\HH), 
\oplus_{\alpha})$ is a D-poset for every $\alpha > 0$. Clearly, 
$\oplus_1 = \oplus$ and $\oplus_2 = \widetilde{\oplus}$. 
Moreover, $({\frak E}(\HH), 
\oplus_{\alpha})$ is an ${\Bbb N}_{\alpha}$-semimodule. 
It is now possible to define the tensor product $E_1 
\widehat{\otimes}_{\alpha,{\frak D}} \cdot \cdot \cdot 
\widehat{\otimes}_{\alpha,{\frak D}} E_m$ in the category of 
${\Bbb N}_{\alpha}$-semimodules to be the D-poset 
consisting of all $\sim_{\alpha, \frak D}$-equivalence classes of 
$\oplus_{\alpha, \frak D}$-parts of $1 \otimes_{\Bbb N} \cdot 
\cdot \cdot \otimes_{\Bbb N} 1$ subject 
to the familiar three relations. The equivalence relation 
$\sim_{\alpha, \frak D}$ and the partial addition $\oplus_{\alpha, 
\frak D}$ are defined completely analogously to $\sim_{\frak D}$ 
and $\oplus_{\frak D}$. \\ Pick an 
arbitrary finite homogeneous effect history $w_0$, choose $k 
\in {\Bbb N}, k > 0$ and choose for all $r \in \{1, ..., m \}$ a 
$k$-tuple of times $(t_{r,1}^*, t_{r,2}^*, ..., t_{r,k}^*)$ with 
$t_{r,k}^* > ... > t_{r,2}^* > t_{r,1}^*$ such that for all $r$ 
there is no $t \in {\frak s}(w_0)$ such that $t_{r,l}^* \geq t 
\geq t_{r,l-1}^*$ for some $1< l \leq k$. Now we 
pick $m$ effect operators $E_1, ..., E_m$ and define 
$\sqrt{E_r(t)} := U(t, t_i(w_{E_1, ..., E_m})) \sqrt{E_r} U(t, 
t_i(w_{E_1, ..., E_m}))^{\dagger}$. Here $t_i(w_{E_1, ..., E_m})$ 
denotes the initial time of the effect history $w_{E_1, E_2, ..., 
E_m}$ defined by \\ ${\frak s}^*_{m,k}(w_0) 
:= {\frak s}(w_{E_1, ..., E_m}) := {\frak s}(w_0) \cup \{ 
t_{1,1}^*, ..., t_{1,k}^*, ..., t_{m,1}^*, ..., t_{m,k}^* \}$ and 
\[ (w_{E_1, ..., E_m})_t := \left\{ \begin{array}{r@{\quad : 
\quad}l} (w_0)_t & t \neq t_{r,1}^*, ..., t_{r,k}^* \\ E_r(t) & t 
\in \{ t_{r,1}^*, ..., t_{r,k}^* \} \end{array} \right. . \] That 
is, $w_{E_1, ..., E_m}$ is the extension of $w_0$ by the effects 
$E_r(t)$ at the intermediate times $t_{r,j}^*$, where $r \in \{1, 
..., m \}$ and $j \in \{ 1, ..., k \}$. 

Define $\widehat{\frak E}_{k,m} := {\{ w_{E_1, ..., 
E_m}}{\mid}{E_1, ..., E_m \in {\frak E}(\HH) \}}$. The above 
decoherence functional $\widehat{d}_{\varrho,{\frak 
s}_{m,k}^*(w_0)}$ restricted to the set $\Theta(\widehat{\frak 
E}_{k,m}) := {\{ \Theta(w_{E_1, ..., E_m})}{\mid}{E_1, ..., E_m 
\in {\frak E}(\HH) \}}$ is additive in both arguments, e.g., 
\begin{eqnarray*} 
\widehat{d}_{\varrho,{\frak s}_{m,k}^*(w_0)} \left(\Theta(w_{E_1 
\oplus_{2/k} D_1, ..., E_m}), \Theta(w_{F_1, ..., F_m}) \right) & 
= & \widehat{d}_{\varrho,{\frak s}_{m,k}^*(w_0)} 
\left(\Theta(w_{E_1, ..., E_m}), \Theta(w_{F_1, ..., F_m}) \right) 
\\ &  & + \widehat{d}_{\varrho,{\frak s}_{m,k}^*(w_0)} 
\left(\Theta(w_{D_1, ..., E_m}), \Theta(w_{F_1, ..., F_m}) 
\right), \end{eqnarray*} for arbitrary 
$E_1, ..., E_m, D_1, F_1, ..., F_m \in {\frak E}(\HH)$ for which 
$E_1 \oplus_{2/k} D_1$ is well-defined. 
$\Theta(w_{E_1, ..., E_m})$ 
depends upon $E_1, .., E_m$ only through the tensor product 
$E_1 \widehat{\otimes}_{2/k,{\frak D}} \cdot \cdot \cdot 
\widehat{\otimes}_{2/k,{\frak D}} E_m$ and hence there is an 
embedding ${\frak b}_k$ mapping $\Theta(\widehat{\frak E}_{m,k})$ 
injectively to ${\frak E}(\HH_1) \widehat{\otimes}_{2/k,{\frak D}} 
\cdot \cdot \cdot \widehat{\otimes}_{2/k,{\frak D}} {\frak 
E}(\HH_m)$. Thus $\Theta(\widehat{\frak E}_{m,k})$ can be 
identified with the set of all homogeneous elements in ${\frak 
E}(\HH_1) \widehat{\otimes}_{2/k,{\frak D}} \cdot \cdot \cdot 
\widehat{\otimes}_{2/k,{\frak D}} {\frak E}(\HH_m)$, i.e., 
elements of the form $E_1 \widehat{\otimes}_{2/k,{\frak D}} \cdot 
\cdot \cdot \widehat{\otimes}_{2/k,{\frak D}} E_m$. In this sense 
${\frak E}(\HH_1) \widehat{\otimes}_{2/k,{\frak D}} \cdot \cdot 
\cdot \widehat{\otimes}_{2/k,{\frak D}} {\frak E}(\HH_m)$ can be 
generated from $\Theta(\widehat{\frak E}_{m,k})$ by finitely many 
well-defined $\oplus_{2/k,{\frak D}}$-operations, that is, we 
write ${\bigoplus_{2/k,{\frak D}} {\frak b}_k 
\left(\Theta(\widehat{\frak E}_{m,k}) \right)}{\big/}{\sim_{2/k, 
\frak D}} = {\frak 
E}(\HH_1) \widehat{\otimes}_{2/k,{\frak D}} \cdot \cdot \cdot 
\widehat{\otimes}_{2/k,{\frak D}} {\frak E}(\HH_m)$. 
The \df $\widehat{d}_{\varrho,{\frak s}_{m,k}^*(w_0)}$ induces a 
\df $d^{{\frak b}_k}_{\varrho,{\frak s}_{m,k}^*(w_0)}$ on ${\frak 
b}_k \left( 
\Theta (\widehat{\frak E}_{m,k}) \right)$ in a trivial 
way by \[ d^{{\frak b}_k}_{\varrho,{\frak s}_{m,k}^*(w_0)} 
\left(a, b \right) 
:= \widehat{d}_{\varrho,{\frak s}_{m,k}^*(w_0)} \left({\frak 
b}_k^{-1}(a), {\frak b}_k^{-1}(b) \right). \] The right hand side 
is well-defined since ${\frak b}_k$ maps $\Theta 
\left(\widehat{\frak E}_{m,k} \right)$ bijectively to ${\frak b}_k 
\left(\Theta \left(\widehat{\frak E}_{m,k} \right) \right)$. 
It is now easy to see that for 
arbitrary $m \in \NN$ the \df $d^{{\frak b}_k}_{\varrho,{\frak 
s}_{m,k}^*(w_0)}$ can be extended to a 
$\oplus_{2/k,{\frak D}}$-additive functional on the D-poset 
${\bigoplus_{2/k,{\frak D}} {\frak b}_k \left( \Theta( 
\widehat{\frak E}_{m,k}) \right)}{\big/}{\sim_{2/k, \frak D}}$. 

Let ${\cal B}_k$ be a subset of the free lattice ${\cal L} 
\left(\widehat{\frak E}_{m,k} \right)$ generated by 
$\widehat{\frak E}_{m,k}$ (for $m \in \NN, m > 0$). We 
will say that $\widehat{\cal B}_k := \XX({\cal B}_k)$ is an {\sc 
allowed Boolean lattice of (inhomogeneous) effect histories of 
order $k$} if conditions analogous to those in Definition 
\ref{CEH} are satisfied, namely \begin{itemize} \item 
$\widehat{\cal B}_k$ is a Boolean sublattice of ${{\cal L} 
\left(\widehat{\frak E}_{m,k} \right)}{\big/}{\sim_{\Bbb E}}$; 
\item $\widehat{\cal B}_k$ is atomic and the set 
of atoms consists of homogeneous elements; 
\item the map ${\frak b}_k \circ \Theta$ can be uniquely 
extended to a positive valuation ${\frak I}_k$ on $\widehat{\cal 
B}_k$ with values in ${\bigoplus_{2/k,{\frak D}} {\frak b}_k 
\left(\Theta(\widehat{\frak E}_{m,k}) \right)}\big/{\sim_{2/k, 
\frak D}}$, i.e., to a map ${\frak I}_k : \widehat{\cal B}_k \to 
{\bigoplus_{2/k,{\frak D}} {\frak b}_k 
\left(\Theta(\widehat{\frak E}_{m,k}) \right)}\big/{\sim_{2/k, 
\frak D}}$ satisfying ${\frak I}_k(d_1 \vee_{k} d_2) 
{\ominus}_{2/k,{\frak D}} {\frak I}_k(d_1) = {\frak I}_k(d_2) 
{\ominus}_{2/k,{\frak D}} {\frak I}_k(d_1 \wedge_{k} d_2)$ for 
$d_1, d_2 
\in \widehat{\cal B}_k$. This condition means in particular that 
the left hand side and the right hand side are both well-defined 
for all $d_1, d_2 \in \widehat{\cal B}_k$. Here $\wedge_k$ and 
$\vee_k$ denote the lattice-operations on $\widehat{\cal B}_k$. 
\end{itemize} 
The decoherence functional $d^{{\frak 
b}_k}_{\varrho,{\frak s}_{m,k}^*(w_0)}$ and the map ${\frak I}_k$ 
induce a \df on $\widehat{\cal B}_k \times \widehat{\cal B}_k$ by 
\begin{equation} \label{bibabu}
d^{{\frak I}_k}_{\varrho,{\frak s}_{m,k}^*(w_0)} : \widehat{\cal 
B}_k \times \widehat{\cal B}_k \to \CC, 
d^{{\frak I}_k}_{\varrho,{\frak s}_{m,k}^*(w_0)}(u,v) := 
d^{{\frak b}_k}_{\varrho,{\frak s}_{m,k}^*(w_0)} \left({\frak 
I}_k(u), {\frak I}_k(v) \right). \end{equation} 
The \df $d^{{\frak I}_k}_{\varrho,{\frak s}_{m,k}^*(w_0)}$ is 
additive in both arguments with respect to the Boolean lattice 
structure on $\widehat{\cal B}_k$. 
Furthermore, we say that an allowed Boolean lattice 
$\widehat{\cal B}_k$ of effect histories of order $k$ is a {\sc 
consistent allowed Boolean lattice of effect histories of order 
$k$} if the \df $d^{{\frak I}_k}_{\varrho,{\frak s}_{m,k}^*(w_0)}$ 
defines a probability measure on the Boolean lattice 
$\widehat{\cal B}_k$ by \begin{equation} \label{E4} p_{\varrho,k} 
: \widehat{\cal B}_k \to \RR, p_{\varrho,k}(v) := d^{{\frak 
I}_k}_{\varrho,{\frak s}_{m,k}^*(w_0)}(v,v) = d^{{\frak 
b}_k}_{\varrho,{\frak s}_{m,k}^*(w_0)} 
\left( {\frak I}_k(v), {\frak I}_k(v) \right). \end{equation} It 
is clear that this $p_{\varrho,k}$ defines a probability measure 
on $\widehat{\cal B}_k$ if and only if $\mbox{\rm Re } d^{{\frak 
I}_k}_{\varrho,{\frak 
s}_{m,k}^*(w_0)}(u,v)$ = Re $d^{{\frak b}_k}_{\varrho,{\frak 
s}_{m,k}^*(w_0)} \left( {\frak I}_k(u), {\frak I}_k(v) \right) = 
0$, for all $u, v \in \widehat{\cal B}_k$ for which ${\frak 
I}_k(u) \oplus_{2/k, {\frak D}} {\frak I}_k(v)$ is well-defined. 
\\ 

\begin{rem} It is possible to define the direct limit $({\frak 
E}(\HH)_{2/k, {\frak T}}, \oplus_{2/k, {\frak D}})$ for 
arbitrary $k > 0$ and fixed $w_0$ and to extend the \df $d^{{\frak 
I}_k}_{\varrho,{\frak s}_{m,k}^*(w_0)}$ to a \df on ${\frak 
E}(\HH)_{2/k, {\frak T}}$. We omit the details. \end{rem}
\begin{rem} \label{imp} The implication 
$\Longrightarrow_{\varrho}$ has been 
defined above in Definition $\mbox{\em \ref{D23}}$ only for pairs 
of effect histories belonging to an allowed Boolean lattice of 
effect histories (defined with respect to $({\frak E}(\HH)_{\frak 
T}, \widetilde{\oplus}_{\frak D})$). \begin{itemize} \item 
Analogously we say for two effect histories $w_1$ and $w_2$ 
belonging to some common allowed Boolean lattice $\widehat{\cal 
B}_k$ of effect 
histories of order $k>0$ that $w_1$ {\sc implies $w_2$ 
in the state $\varrho$} if $d^{{\frak I}_k}_{\varrho, {\frak 
s}^*_{m,k}(w_0)}(w_1 \wedge_{k} w_2, w_1 \wedge_{k} w_2) = 
d^{{\frak I}_k}_{\varrho, {\frak s}^*_{m,k}(w_0)}(w_1, w_1) \neq 
0$ and if $d^{{\frak I}_k}_{\varrho, {\frak s}^*_{m,k}(w_0)}(w_1 
\vee_{k} w_2, w_1 \vee_{k} w_2) = d^{{\frak I}_k}_{\varrho, 
{\frak s}^*_{m,k}(w_0)}(w_2, w_2) \neq 0$. We write $w_1 
\Longrightarrow_{\varrho} w_2$. \item For two histories $u_1$ and 
$u_2$ belonging to a full set $\widetilde{\frak E}$ of effect 
histories, we say that $u_1$ {\sc implies} $u_2$ {\sc in the state 
$\varrho$} if $u_1 \wedge u_2$ is well-defined 
in $\widetilde{\frak E}$ and if $p_{\varrho}(u_1 \wedge u_2) = 
p_{\varrho}(u_1) \neq 0$. Here $\wedge$ denotes the 
partially defined meet operation on $\widetilde{\frak E}$ induced 
by the partially defined meet operation on ${\Bbb E}_{fin}(\HH)$. 
We write $u_1 \Longrightarrow_{\varrho} u_2$. \end{itemize} 
\end{rem}

The universal rule of interpretation of quantum mechanics can now 
be generalized
\begin{rle} \label{rle2}
Propositions about quantum mechanical systems should solely be 
expressed in terms of 
inhomogeneous effect histories which represent the {\sf general 
physical properties} of a quantum mechanical system. 
\begin{itemize} \item Every description of a quantum mechanical 
system (i.e., 
probabilistic or predictive statements) should be 
expressed either \begin{itemize} \item solely in terms of effect 
histories belonging to 
a common consistent allowed Boolean lattice $({\cal B}, 
${\fraktur B}$)$ of 
effect histories. The probability measure on ${\cal B}$ 
is induced by the decoherence functional $d_{\varrho, {\frak T}}$ 
on ${\cal B}$; \item[or] \item solely in 
terms of a consistent allowed Boolean lattice $(\widehat{\cal 
B}_k, {\frak I}_k)$ of effect histories of order $k>1$. The 
probability measure on $\widehat{\cal B}_k$ is 
defined by Equation $\mbox{\em \ref{E4}}$; \item[or] \item solely 
in terms of effect histories belonging to a 
full D-poset $\widetilde{\frak E}$ of effect histories. 
The probability measure on $\widetilde{\frak E}$ is defined by 
Equation $\mbox{\em \ref{E7}}$. \end{itemize}
\item Every reasoning relates solely effect histories 
\begin{itemize} \item belonging to a common allowed Boolean 
lattice $({\cal B},$ {\fraktur B}$)$ of effect histories; 
\item[or] \item belonging to a common allowed Boolean lattice 
$(\widehat{\cal B}_k, {\frak I}_k)$ of effect histories of order 
$k>1$; \item[or] \item belonging to a full D-poset of effect 
histories. \end{itemize}
Every reasoning relating histories belonging to a common allowed 
Boolean lattice (of order $0$) should solely be expressed in terms 
of the logical relations induced by the functional $d_{\varrho, 
{\cal B}} : {\cal B} \times {\cal B} \to \CC$ defined in Remark 
$\mbox{{\em \ref{rem11}}}$ and Definition $\mbox{{\em 
\ref{D23}}}$. \\ Every reasoning 
relating histories belonging to a common allowed Boolean lattice 
of order $k>1$ should solely be expressed in terms of the 
logical relations induced by the functional $d^{{\frak 
I}_k}_{\varrho, {\frak s}^*_{m,k}(w_0)}$ defined in Equation 
$\mbox{{\em \ref{bibabu}}}$ (see also Remark $\mbox{{\em 
\ref{imp}}}$). \\ Every reasoning 
relating histories belonging to a full D-poset of 
effect histories should solely be expressed in terms of the 
logical relations induced by the probability functional defined in 
Equation $\mbox{\em \ref{E7}}$ (see also Remark $\mbox{\em 
\ref{imp}}$). \end{itemize} \end{rle} 

\begin{rem} It is easy to 
verify that if $p_1 \Longrightarrow_{\varrho} p_2$ is valid in one 
allowed Boolean lattice, then $p_1 \Longrightarrow_{\varrho} 
p_2$ is also valid in every other allowed Boolean lattice of 
$\widehat{\Bbb E}_{fin}(\HH)$ containing $p_1$ and $p_2$. 
\end{rem} 

\subsection{The algebraic structure of generalized 
history theories}

We now summarize our discussion by stating the general axioms for 
a generalized quantum theory based on our generalized history 
concept. This subsection parallels the discussion in 
(Isham, 1994). However, it contains only a rough summary of the 
main concepts and structures. In every particular history theory 
one has to {\sf show} that everything is well-defined and 
consistent and if necessary to modify the concepts and structures. 
\begin{enumerate}
\item The space $\cal U$ of history filters or homogeneous 
histories. \begin{itemize} \item $\cal U$ is the space of the 
basic physical properties of a physical system. An element of 
$\cal U$ represents the equivalence classes of (operationally 
undistinguishable) basic entities (propositions) in the 
interpretation of the theory. 
There exists a map $F$ 
mapping the elements of $\cal U$ to a D-poset $\frak E$. $\frak E$ 
can be interpreted as the set of (equivalence classes of) 
one-time propositions. Moreover, 
there is a canonical map ${\frak N}$ imbedding ${\cal U}$ in a 
D-poset $({\frak M}, \oplus)$. [in Section 4.3: $\cal U$ equals 
the space of homogeneous effect histories ${\cal U} = 
{\Bbb E}_{fin}(\HH)$, cf.~Definition \ref{b2}; ${\frak E}$ is 
given by ${\frak E}(\HH)$ and $F$ is given by $F(u) = 
C_{t_0}(u)^{\dagger}C_{t_0}(u)$. ${\frak M} = {\frak 
E}(\HH)_{{\frak T}}$]. \item $\cal U$ 
is a partially ordered set with unit history 1 and null history 0. 
\item $\cal U$ is a partial semigroup with composition law 
$\circ$, cf.~(Isham, 1994). $a \circ b$ is well-defined if $t_f(a) 
< t_i(b)$. In this case we say that $a$ {\sc proceeds} $b$ or that 
$b$ {\sc follows} $a$. Further, $1 \circ a = a \circ 1 = a$ and $a 
\circ 0 = 0 \circ a =0$. If $a \circ b$ is defined, then $a\circ 
b = a \wedge b$, in particular the right hand side is 
well-defined. \item The partial ordering on $\cal U$ induces a 
partial unary operation $\neg$ (complementation) and two 
partial binary operations $\wedge$ and $\vee$ (meet and join) on 
${\cal U}$. \end{itemize} \item The space of 
decoherence functionals. \\ 
A decoherence functional is a map $d : {\frak M} \times 
{\frak M} \to \CC$ which satisfies for all $\alpha, 
\alpha',\beta \in {\frak M}$
\begin{itemize} \item $d(\alpha,\alpha) \in \RR$ and 
$d(\alpha,\alpha) 
\geq 0$. \item $d(\alpha,\beta) = d(\beta,\alpha)^*$. \item 
$d(1,1) =1$. \item $d(0,\alpha) =0$, for all 
$\alpha$. \item $d(\alpha_1 \oplus 
\alpha_2, \beta) = d(\alpha_1,\beta) + d(\alpha_2,\beta)$ for all 
$\alpha_1, \alpha_2, \beta \in {\frak M}$ for which 
$\alpha_1 \oplus \alpha_2$ is well-defined. \end{itemize}
\item The space $\frak U$ of general history propositions.
\begin{itemize} \item $\frak U$ is the quotient space of the free 
lattice generated by $\cal U$ by the congruence 
relation induced by the partial ordering on $\cal U$ [see Remark 
\ref{cgr}]. \item 
There exists an embedding $\tau : {\cal U} \to \frak U$, i.e., 
$\tau({\cal U}) \subset \frak U$. \end{itemize} \pagebreak[3] 
\item The physical interpretation.
\begin{itemize} \item $\frak U$ cannot globally be mapped to 
${\cal U}$ or to ${\frak M}$ respectively. 
The physically interesting subsets of $\frak U$ are the 
`allowed' Boolean sublattices ${\frak U}_0$ of ${\frak U}$ (see 
Definition \ref{CEH}) on which the canonical map $\frak N$ can be 
uniquely extended to a valuation {\fraktur B} on ${\frak U}_0$ 
with values in $\frak M$ such that for every $u \in {\frak U}_0'$ 
the value {\fraktur B}$(u)$ of this extension does not depend upon 
the particular `allowed' Boolean lattice ${\frak U}_0'$ chosen. 
\item The \df induces a probability measure on the 
consistent (w.r.t.~the decoherence functional) `allowed' Boolean 
sublattices of ${\frak U}$. \item On the `allowed' Boolean 
sublattices of $\frak U$ the decoherence functional defines a 
partial logical implication which allows to make logical 
inferences. \item The D-poset $({\frak M}, \oplus)$ 
may be not unique. There may be other D-posets 
$({\frak M}', \oplus')$ containing ${\frak N}({\cal U})$ 
such that the decoherence functionals can be extended to 
${\oplus}'$-additive functionals on ${\frak M}' \times {\frak 
M}'$. It is possible to define `allowed' Boolean sublattices 
${\frak U}_0'$ of $\frak U$ with respect to ${\frak M}'$. 
The \df induces a probability measure on the 
consistent `allowed' Boolean sublattices ${\frak 
U}_0'$ of $\frak U$. Moreover, the \df defines a partial logical 
implication on the Boolean sublattices ${\frak U}_0'$ of $\frak 
U$. \item there exist D-posets $\cal EU$ of effect 
histories isomorphic to $\frak E$ [e.g., 
the full D-poset of effect histories] on which every 
decoherence functional induces a probability measure, i.e., for 
which there exists a D-poset isomorphism ${\frak I} : {\cal EU} 
\to {\frak E} \subset {\frak N}({\cal U})$ mapping the space 
${\cal EU}$ bijectively to the 
D-poset ${\frak E}$ such that the \df can be extended to 
${\frak I}({\cal EU})$ and such that \[ d \left({\frak I}(a) 
\oplus {\frak I}(b), {\frak I}(a) 
\oplus {\frak I}(b) \right) = d \left({\frak I}(a), {\frak I}(a)) 
+ d({\frak I}(b), {\frak I}(b) \right), \mbox{ for all } a, b \in 
{\cal EU}, \] whenever the left hand side is well-defined. In 
particular, no consistency condition is required. Under certain 
additional conditions it is also 
possible to define an unambiguous partial implication on $\cal 
EU$. \item All probability measures defined on some consistent 
allowed Boolean sublattices of ${\frak U}$ defined with respect to 
some of the various D-posets ${\frak M, \frak M}'$ 
or on the D-posets of histories ${\cal EU}$ have 
equal physical status in the theory and have to be 
treated egalitarianly. There seems to be no reason to prefer one 
class over the others. \end{itemize} \end{enumerate}
\section{Discussion and Conclusion}

Above, we have defined the decoherence functional 
$\widehat{d}_{\varrho,S}$ on pairs of homogeneous elements in 
$\left( {\frak E}(\HH)_{t_1} \otimes_{\Bbb N} \cdot \cdot \cdot 
\otimes_{\Bbb N} {\frak E}(\HH)_{t_n} \right) \times 
\left( {\frak E}(\HH)_{t_1} \otimes_{\Bbb N} \cdot \cdot 
\cdot \otimes_{\Bbb N} {\frak E}(\HH)_{t_n} \right)$ 
by \[ \widehat{d}_{\varrho,S}(a,b) := {\tr} \left( C_{t_0}' \left( 
\sqrt{a} \right) \varrho C_{t_0}' \left( \sqrt{b} 
\right)^{\dagger} \right), \] and noticed that it 
cannot be extended to a 
$\oplus_{\frak D}$-additive functional on $\left( {\frak 
E}(\HH)_{t_1} \widehat{\otimes}_{\frak D} \cdot \cdot \cdot 
\widehat{\otimes}_{\frak D} {\frak E}(\HH)_{t_n} \right) \times \\ 
\times \left( {\frak E}(\HH)_{t_1} \widehat{\otimes}_{\frak D} 
\cdot \cdot \cdot \widehat{\otimes}_{\frak D} {\frak E}(\HH)_{t_n} 
\right)$. However, if we define a \df $\widetilde{d}_{\varrho,S}$ 
on pairs of square roots of homogeneous elements in $\left( {\frak 
E}(\HH)_{t_1} \otimes_{\Bbb N} \cdot \cdot 
\cdot \otimes_{\Bbb N} {\frak E}(\HH)_{t_n} \right) \times 
\left( {\frak E}(\HH)_{t_1} \otimes_{\Bbb N} \cdot \cdot 
\cdot \otimes_{\Bbb N} {\frak E}(\HH)_{t_n} \right)$ 
by \[ \widetilde{d}_{\varrho,S} \left(\sqrt{a}, \sqrt{b} \right) 
:= {\tr} \left( C_{t_0}' \left( \sqrt{a} \right) \varrho C_{t_0}' 
\left( \sqrt{b} \right)^{\dagger} \right), \]
then $\widetilde{d}_{\varrho,S}$ can straightforwardly be extended 
to a $\oplus_{\frak D}$-additive functional on \\ $\left( {\frak 
E}(\HH)_{t_1} \widehat{\otimes}_{\frak D} \cdot \cdot \cdot 
\widehat{\otimes}_{\frak D} {\frak E}(\HH)_{t_n} \right) \times 
\left( {\frak E}(\HH)_{t_1} \widehat{\otimes}_{\frak D} \cdot 
\cdot \cdot \widehat{\otimes}_{\frak D} {\frak E}(\HH)_{t_n} 
\right)$. However, this approach and the one discussed in Section 
4.3 are mathematically equivalent 
since $\sqrt{a} \oplus_{\frak D} \sqrt{b}$ is well-defined 
if and only if $\sqrt{a \widetilde{\oplus}_{\frak D} b}$ is 
well-defined for arbitrary $a, b \in {\frak 
E}(\HH)_{t_1} \widetilde{\otimes}_{\frak D} \cdot \cdot \cdot 
\widetilde{\otimes}_{\frak D} {\frak E}(\HH)_{t_n}$. \\ 

We add a few remarks about 
the physical significance of the sets of effect 
histories of order $k>0$. In such effect histories the physical 
interesting physical qualities are always repeated at $k$ 
successive times. If we restrict ourselves to ordinary physical 
qualities represented by projection operators, then repetition of 
some physical quality adds nothing new. This fact is 
mathematically expressed through the equation $P_1 \oplus_{\alpha} 
P_2 = P_1 \oplus P_2$ for all $\alpha \in {\Bbb Q}, \alpha > 0$ 
and all projection operators $P_1$ and $P_2$ for which $P_1 \oplus 
P_2$ is well-defined. Hence, for histories of ordinary physical 
qualities all different D-poset structures coincide. The situation 
that for effect histories there are different inequivalent 
algebraic structures (D-poset structures) which have to be 
dealt with on the same footing is a new aspect of our 
generalized history approach. \\ 

We have already mentioned above that the set of effects does not 
fulfil some requirements which are habitually associated with the 
notion of property. In particular effects can in 
general neither be measured ideally nor repeatedly. Moreover, in 
general the effect $1-F$ cannot be interpreted as the 
property complementary to the property represented by the effect 
$F$. The same is true for, 
e.g., $(1- F^{1/2})^2$. Of course, in mathematical terms 
$1-F$ and $(1- F^{1/2})^2$ simply {\sf are} the complements of $F$ 
in the D-posets $({\frak E}(\HH), {\oplus})$ or $({\frak 
E}(\HH), \widetilde{\oplus})$ respectively. However, if in a 
measurement of the effect $F$ the measuring apparatus intended to 
measure $F$ is not triggered, then we can say that the effect $F$ 
has not occurred (trivial) but in general we cannot say that there 
is another complementary effect which has occurred instead. This 
is in particular true for effects $F$ satisfying $F \leq 
\frac{1}{2}$ or $\frac{1}{2} \leq F$. Therefore some authors claim 
that only {\em regular} effects whose spectrum extends both below 
as well as above the value $\frac{1}{2}$ represent the (unsharp) 
properties of a quantum mechanical system (Busch et al., 1995). 
However, in the present work all effects are treated on the same 
footing and no ad hoc assumptions are added to single out effects 
representing the `well-behaved' properties of a quantum mechanical 
system. \\ 

In the literature of the standard consistent histories approach 
some authors find it necessary to consider various approximate 
consistency conditions in order to describe the classical or 
the quasiclassical properties of quantum mechanical systems. 
For instance, Omn\`es associates `quasi-projectors' with 
macroscopic regular cells in classical phase space. However, 
strictly speaking histories involving quasi-projectors lie outside 
the mathematical framework of the standard consistent histories 
approach and it is a little bit disconcerting that the discussion 
of the (semi)classical limit of quantum mechanics in the 
consistent histories approach involves approximations of the 
fundamental concepts of the theory, even when it is of `no 
practical importance.' This point may seem a bit pedantic at 
first, but this author nevertheless believes that it is also a 
quite unsatisfactory state of affairs in a physical theory. The 
solution of this problem should be clear now. Of course, 
quasi-projectors are only a special sort of effects. Thus 
histories of quasi-projectors fit perfectly well in our 
generalized history theory and no ad hoc modifications of the 
basic concepts and principles are necessary in the discussion of 
the (semi)classical limit of quantum mechanics. \\ Gell-Mann and 
Hartle encounter in their discussions of 
quasiclassical domains and measurement processes the situation 
that the consistency conditions and the probability sum rules for 
physically interesting histories are only approximately satisfied 
(see in particular section II.11 in (Hartle, 1991)). The standard 
`for 
all practical purposes' argument to justify the use of approximate 
probabilities and approximate consistency conditions is that if 
the violation is small enough, then no experiment could detect the 
discrepancy. Dowker and Kent (1996) stress that `{\em \sf this 
seems a rather casual disruption of the mathematical structure of 
a fundamental theory.}' Indeed, this question is not a question of 
practicability but rather a question of principle. While 
admittedly approximations of any (reasonable) kind may be used for 
the computation of the (numerical) predictions of a theory, the 
basic principles and concepts of a theory should be formulated 
without invoking any kind of approximation. 
Otherwise the result would be a theory with semantically slippery 
rules of interpretation or without firm mathematical foundations. 
\\
In any case our generalized history approach allows to incorporate 
any kind of `unsharp' physical quality into the theory at a 
fundamental level, regardless whether it describes the 
quasiclassical domain of the universe or a measurement situation 
with limited resolution or other. Histories of generalized 
physical qualities and effect histories have a clear-cut status in 
the theory and thus concepts of 
approximate consistency are superfluous. 
Since these questions are not the main topic of the present work, 
we postpone a thorough discussion to the future. \\ 

\indent The problem of interpreting a physical theory is of 
foremost 
importance in the physical science but especially in the case of 
quantum theory difficult. The axioms of every 
interpretation of a physical theory 
are always introduced by fiat. They are independent of and cannot 
be derived from the formalism. Thus it comes as no surprise 
that no agreement about the interpretation of quantum mechanics 
has been achieved yet. Nevertheless even nowadays most physicists 
do not care too much about interpretational issues and simply rely 
on one of the diverse versions of the so-called Copenhagen 
interpretation of quantum mechanics. This interpretation is (as is 
well-known) plagued with paradoxes and puzzles such as the 
subject$-$object muddle and the actualization of facts as a 
consequence of our measurements. One of its basic claims is that 
main aspects of the microscopic world are in principle 
unanalyzable. Furthermore, it has been stressed by Omn\`es 
(1992) that the Copenhagen interpretation is incomplete. 
The paradoxes of the 
Copenhagen interpretation of quantum mechanics cannot be resolved 
but only be removed by another interpretation of quantum 
mechanics. \\ 

\indent In the last decades there have been several 
attempts to construct a realistic and individual interpretation of 
quantum mechanics. In this work we studied two such attempts, 
namely the generalized operational quantum theory and the logical 
interpretation based on the concept of histories. We investigated 
their interrelationship and constructed a generalized logical 
interpretation. Both the generalized operational interpretation 
and the old logical interpretation can be viewed as special cases 
of our generalized logical interpretation. Our generalized Rule 2 
is, however, only a first tentative step towards a {\em consistent 
quantum reasoning} involving generalized properties. 
Unfortunately, there is no general theory of coexistent sets of 
effects or generalized observables in quantum mechanics. With such 
a theory it would presumably be possible to considerably 
improve and generalize our approach and find a more natural 
definition of the notion of allowed Boolean lattice of 
effect histories than the one given in Definition \ref{CEH}. \\ 

\indent The interpretation of the consistent histories formalism 
underlying the present work differs significantly from the 
wide-spread interpretation succinctly summarized under the name 
{\em Unknown Set Interpretation} by Kent (1995). Firstly $-$ and 
perhaps most importantly $-$ we do not claim that nonrelativistic 
Hamiltonian quantum mechanics can be applied to the `whole 
universe,' see Note 5. Hence, we 
also do not adopt the view that exactly one history is {\em 
realized} and describes all of physics. Whether or not the 
classical and quasiclassical features of the observable world can 
be fully understood in the consistent histories formalism is an 
open problem. The most promising and most concrete description of 
the semiclassical limit of quantum mechanics in the 
framework of consistent histories has been given by Omn\`es 
(1994, Chapter 6). Secondly, we assert that different 
consistent sets of histories are complementary in the sense 
described above and should be treated on the same footing (this is 
completely analogous to the assertion 
that position and momentum are complementary variables in the 
description of an elementary particle). 
Quantum theory ascribes probabilities to different 
possible events. The propensity that some particular event occurs 
depends upon the quantum system itself and upon the 
integral physical situation. Particularly, in measurement 
situations the result of the measurement depends upon the object 
under study and upon the mode of observation. At present no rule 
determining a preferred consistent set of histories for the 
universe (the Unknown Set) is known; in our interpretation such a 
rule will hardly be needed: instead for every quantum 
system and every `physical situation' one needs a rule fixing the 
`correct' consistent set. In measurement situations the choice of 
the consistent set is determined by the measuring apparatuses. 

There are three methodical arguments supporting the 
logical interpretation. Since according to the logical rule 
any allowed reasoning 
relates only properties belonging to some `allowed' Boolean 
algebra and since there can as a matter of principle never arise a 
paradoxical or inconsistent situation in a Boolean logic, all 
paradoxes and inconsistencies have been expelled from the language 
of quantum mechanics by one single simple rule. 
Thus the three arguments supporting the logical interpretation are 
in brief \begin{itemize} \item Nonrelativistic quantum 
mechanics with the generalized logical interpretation (and 
together with decoherence) gives unambiguous 
predictions for every conceivable experimental situation in 
nonrelativistic quantum mechanics; \item Simplicity and economy of 
principles; \item The logical interpretation is free of logical 
paradoxes. \end{itemize} Examples illustrating the third point 
have been discussed by Omn\`es (1994). However, at least one 
mystery has remained. The microscopic world is unanalyzable. 
Quantum physics as viewed by the logical interpretation provides 
no model for what `actually happens.'  

\subsubsection*{Acknowledgments}
The author wants to thank Frank Steiner for encouraging and 
supporting this work and Christian Grosche for drawing his 
attention upon the work of Karl Popper. Financial support given by 
Deutsche Forschungsgemeinschaft (Graduiertenkolleg f\"ur 
theoretische Elementarteilchenphysik) is gratefully acknowledged. 
\begin{appendix}
\end{appendix} 
\end{document}